\documentclass[a4paper,11pt]{article}
\pdfoutput=1 

\usepackage{jheppub} 

\usepackage[T1]{fontenc} 
\usepackage[cc]{titlepic}
\usepackage[figuresright]{rotating}
\usepackage{float}
\usepackage{lineno}

\newcommand{\zc}{Z_{c}(3900)}
\newcommand{\zpm}{Z_{c}^{\pm}(3900)}
\newcommand{\zz}{Z_{c}^{0}(3900)}

\newcommand{\p}{\psi_{2}(3823)}
\newcommand{\x}{X(3900)}
\newcommand{\pp}{\pi^+\pi^-}
\newcommand{\pip}{\pi^+}
\newcommand{\pim}{\pi^-}

\newcommand{\LL}{\ell^+\ell^-}
\newcommand{\EE}{e^+e^-}
\newcommand{\ee}{e^+e^-}

\newcommand{\mm}{\mu^+\mu^-}
\newcommand{\GG}{\gamma\gamma}
\newcommand{\etap}{\eta^\prime}
\newcommand{\psip}{\psi(3686)}

\newcommand{\jpsi}{J/\psi}
\newcommand{\piz}{\pi^0}

\newcommand{\BESIIIorcid}[1]{\href{https://orcid.org/#1}{\hspace*{0.1em}\raisebox{-0.45ex}{\includegraphics[width=1em]{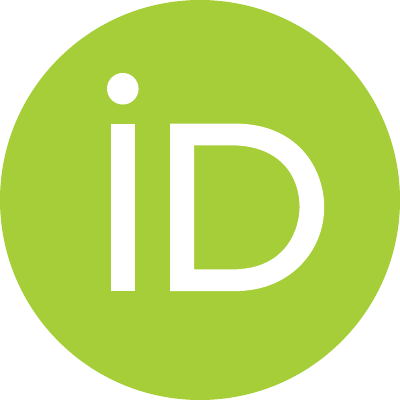}}}}

\title{\boldmath Search for an isoscalar partner of the $\zc$ in $\ee\to\pp\eta\jpsi$}

\collaboration{The BESIII Collaboration}
\collaborationImg{\includegraphics[height=0.5in]{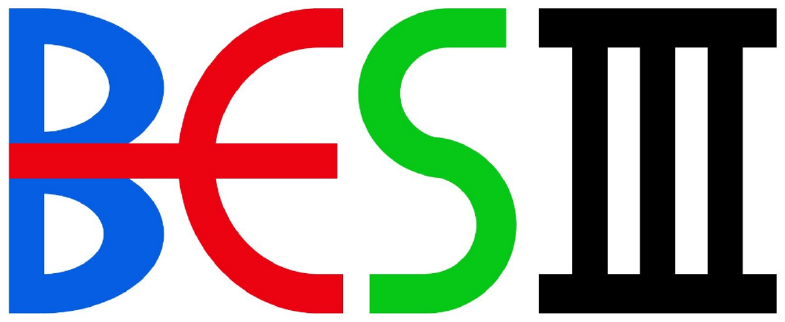}}
\author{
M.~Ablikim$^{1}$\BESIIIorcid{0000-0002-3935-619X},
M.~N.~Achasov$^{4,c}$\BESIIIorcid{0000-0002-9400-8622},
P.~Adlarson$^{81}$\BESIIIorcid{0000-0001-6280-3851},
X.~C.~Ai$^{86}$\BESIIIorcid{0000-0003-3856-2415},
R.~Aliberti$^{39}$\BESIIIorcid{0000-0003-3500-4012},
A.~Amoroso$^{80A,80C}$\BESIIIorcid{0000-0002-3095-8610},
Q.~An$^{64,77,\dagger}$,
Y.~Bai$^{62}$\BESIIIorcid{0000-0001-6593-5665},
O.~Bakina$^{40}$\BESIIIorcid{0009-0005-0719-7461},
Y.~Ban$^{50,h}$\BESIIIorcid{0000-0002-1912-0374}
% ...
} 
\affiliation{
$^{1}$ Institute of High Energy Physics, Beijing 100049, People's Republic of China\\
% ...
$^{4}$ Budker Institute of Nuclear Physics SB RAS (BINP), Novosibirsk 630090, Russia\\
% ...
$^{39}$ Johannes Gutenberg University of Mainz, Johann-Joachim-Becher-Weg 45, D-55099 Mainz, Germany\\
$^{40}$ Joint Institute for Nuclear Research, 141980 Dubna, Moscow region, Russia\\
% ...
$^{50}$ Peking University, Beijing 100871, People's Republic of China\\
% ...
$^{62}$ Southeast University, Nanjing 211100, People's Republic of China\\
% ...
$^{64}$ State Key Laboratory of Particle Detection and Electronics, Beijing 100049, Hefei 230026, People's Republic of China\\
% ...
$^{77}$ University of Science and Technology of China, Hefei 230026, People's Republic of China\\
% ...
$^{80}$ University of Turin and INFN, (A)University of Turin, I-10125, Turin, Italy; (B)University of Eastern Piedmont, I-15121, Alessandria, Italy; (C)INFN, I-10125, Turin, Italy\\
$^{81}$ Uppsala University, Box 516, SE-75120 Uppsala, Sweden\\
% ...
$^{86}$ Zhengzhou University, Zhengzhou 450001, People's Republic of China\\

\vspace{0.2cm}
$^{\dagger}$ Deceased\\
% ...
$^{c}$ Also at the Novosibirsk State University, Novosibirsk, 630090, Russia\\
% ...
$^{h}$ Also at State Key Laboratory of Nuclear Physics and Technology, Peking University, Beijing 100871, People's Republic of China\\
% ...
}
\emailAdd{besiii-publications@ihep.ac.cn}

\abstract{  
Using a data sample corresponding to an integrated luminosity of 20.3 fb$^{-1}$ 
collected at center-of-mass energies from 4.18 to 4.95 GeV with the BESIII detector, we observe the process $\ee\to\pp\eta\jpsi$ with a statistical significance of $6.0 \sigma$, including systematic uncertainties. The isoscalar partner of the $Z_c(3900)$, denoted $X(3900)$, is searched for in the $\eta J/\psi$ final state, and no significant signal is observed. The upper limits on the product of the Born cross section $\sigma^{\rm Born}[e^{+}e^{-}\to\pi^{+}\pi^{-} X(3900)$] and the branching fraction $\mathcal{B}[X(3900)\to\eta J/\psi]$ are given with various assumptions for the mass and width of the $X(3900)$.
}

\begin{document} 
\maketitle
\flushbottom

\section{Introduction}
\label{sec:intro}
In 2013, the charged charmonium-like state $\zpm$ was observed through its decay to $\pi^{\pm}\jpsi$ by the BESIII experiment~\cite{zc-BESIII}, and then confirmed by Belle~\cite{zc-Belle}, CLEO-c data~\cite{zc-cleoc} and D0~\cite{zc-D0}. Two years later, the neutral partner $\zz$ was observed by BESIII in the process $\ee\to\piz\piz\jpsi$~\cite{zc0-BESIII}. The spin-parity
of the $\zc$ has been determined to be $J^{P}=1^{+}$~\cite{zcjp}, and the isospin triplet states of the $\zc$ have been experimentally established.  BESIII also reported the charged and neutral structures $Z_{c}(3885)$ close to the production threshold of $D\bar{D}^*$ in the process $\ee\to\pi(D\bar{D}^*)$, with a mass close to the $\zc$~\cite{DD-BESIII, D0D0-BESIII}, which indicates they might correspond to the same hadronic state. Currently, the well-established decay channels of the $\zc$ primarily include $\pi\jpsi$ and $D\bar{D}^*$.
\par The discovery of the $\zc$ has sparked a great deal of interest in the hadron physics community. 
To conserve electric charge, the charged $Z_c^{\pm}(3900)$ must contain a pair of light quarks in addition to the $c\overline{c}$ pair~\cite{zc-nature-rev}, making it a popular tetraquark candidate. However, its internal structure remains unclear. Possible interpretations of the $\zc$ mainly include hadronic molecules~\cite{the-zc-mole} and tetraquark states~\cite{the-zc-tetra}. Therefore, experimental study is still needed to deepen our understanding of this state.
\par Searching for an isoscalar partner of the $\zc$ decaying to $\eta\jpsi$ could help  establish the family of $Z_{c}$ states under SU(3) symmetry, which would greatly expand our understanding of the nature of the $Z_{c}$ states. In 2015, the search for the isospin violating process $\ee\to\piz\eta\jpsi$ was reported by BESIII, but no $\zc$ signal was found~\cite{pi0etajpsi}.  For the isoscalar $\x$ state, more decay channels must be considered, such as the isospin-conserving process $\ee\to\pp\eta\jpsi$. In the $\ee\to\pp\eta\jpsi$ process, the $\pp$ can come from an isospin singlet $\sigma(500)$. Therefore, isospin is conserved in the $\ee\to\sigma(500)\x$ process, and its production cross section should be enhanced with respect to $\ee\to\piz\eta\jpsi$. Recently, BESIII also searched for the isoscalar partner of the $\zc$ in $\ee\to\eta\eta\jpsi$, and no obvious resonance signal was observed in the $\eta\jpsi$-subsystem~\cite{etaeta}. 

\par In this article, we report an observation of the process $\ee\to\pp\eta\jpsi$ and perform the first search for the $\zc$ isoscalar partner $\x$ in its decay to $\eta\jpsi$. The $\eta$ and $\jpsi$ candidates are reconstructed via their decays to $\gamma\gamma$ and $\LL$ ($\ell=e$ or $\mu$), respectively. The data samples are taken with the BESIII detector~\cite{bes3-detector} at thirty-five  center-of-mass (c.m.) energies ranging from $\sqrt{s}=4.18$ to $4.95$~GeV~\cite{4.6-4.9 lum and ecm, 4.1-4.4 ecm},
corresponding to an integrated luminosity of 20.3 $\rm fb^{-1}$~\cite{4.6-4.9 lum and ecm, 4.0-4.6 lum}.

\section{BESIII detector and Monte Carlo simulation}


The BESIII detector~\cite{bes3-detector} records symmetric $e^+e^-$ collisions 
provided by the BEPCII storage ring~\cite{Yu:IPAC2016-TUYA01}
in the c.m.~energy region from 1.84 to 4.95~GeV,
with a peak luminosity of $1.1 \times 10^{33}\;\text{cm}^{-2}\text{s}^{-1}$ 
achieved at $\sqrt{s} = 3.773\;\text{GeV}$. BESIII has collected large data samples in this energy region~\cite{BESIII:2020nme, EcmsMea, Zhang:2022bdc}. 
The cylindrical core of the BESIII detector covers 93\% of the full solid angle and consists of a helium-based
multilayer drift chamber~(MDC), a time-of-flight
system~(TOF), and a CsI(Tl) electromagnetic calorimeter~(EMC),
which are all enclosed in a superconducting solenoidal magnet
providing a 1.0~T magnetic field. The solenoid is supported by an
octagonal flux-return yoke with resistive plate counter muon
identification modules (MUC) interleaved with steel~\cite{detect}. 
The charged-particle momentum resolution at $1~{\rm GeV}/c$ is
$0.5\%$, and the 
${\rm d}E/{\rm d}x$
resolution is $6\%$ for electrons
from Bhabha scattering. The EMC measures photon energies with a
resolution of $2.5\%$ ($5\%$) at $1$~GeV in the barrel (end cap)
region. The time resolution in the  plastic scintillator TOF barrel region is 68~ps, while
that in the end cap region is 110~ps. The end cap TOF
system was upgraded in 2015 using multigap resistive plate chamber
technology, providing a time resolution of
60~ps, which benefits 83\% of the data used in this analysis~\cite{etof}.   

Simulated data samples produced with a {\sc geant4}-based~\cite{geant4} Monte Carlo~(MC) simulation 
software package, which includes
the geometric description of the BESIII detector and the detector response,
are used to optimize the event selection criteria,
determine the detection efficiency, and estimate the backgrounds.  The simulation models the beam
energy spread and initial state radiation (ISR) in the $e^+e^-$
annihilations with the generator {\sc kkmc}~\cite{kkmc, kkmc_2}.
For the signal process, we generate 100,000 signal MC events $\ee\to\pp\x$ with $\x\to\eta\jpsi$  at each c.m.~energy with a phase-space (PHSP) model using {\sc evtgen}~\cite{evtgen}. In addition, the signal MC events without the isoscalar $\x$ as an intermediate state, $\ee\to\pp\eta\jpsi$, are also generated using the PHSP model.
ISR is simulated by incorporating the $\sqrt{s}$-dependent production cross section into the program, including the $\sqrt{s}$-dependent three-body PHSP factor~\cite{pdg}.
The maximum ISR photon energy is set according to  the production threshold of the $\pp\x$ system.
Final-state-radiation is simulated with the {\sc photos} package~\cite{photos}.


Background contributions are investigated using inclusive MC samples, which include open-charm processes, the ISR process of vector charmonium(-like) states, and the continuum processes incorporated in {\sc kkmc}.
All particle decays are modelled with the {\sc evtgen}~\cite{evtgen} using branching fractions taken from the Particle Data Group (PDG)~\cite{pdg} when available, and otherwise modelled with {\sc lundcharm}~\cite{lundcharm}. The luminosity of inclusive MC samples are ten times that of the data samples. A generic event-type analysis tool~\cite{topo} is employed to study the backgrounds.

\section{Event selection and background study}

The final-state particles in this analysis include $\pp\gamma\gamma\mm$ or $\pp\gamma\gamma\ee$. Charged tracks detected in the MDC are required to be within a polar angle ($\theta$) range of $|\rm{\cos\theta}|<0.93$ (the coverage of the MDC),
where $\theta$ is defined with respect to the $z$-axis, which is the symmetry axis of the MDC. 
For each charged track, the distance of the closest approach to the interaction point (IP) must be less than 10\,cm along the $z$-axis, $|V_z|<10$\,cm, and less than 1\,cm
in the transverse plane, $|V_{xy}|<1$\,cm.  
\par The pions produced alongside the $\jpsi$ and the leptons from the $\jpsi$ decay are kinematically well-separated.
Charged tracks with momenta greater than 1.0~GeV/$c$ in the laboratory frame are taken as lepton candidates, while those with momenta less than 1.0~GeV/$c$ are taken as pion candidates.
For the candidate events of interest, the total number of charged tracks ($\pp\LL$) must be at least four, 
with exactly two oppositely-charged leptons, and at least two oppositely-charged pions. 
The energy deposition in the EMC for a lepton candidate is used to separate electrons from muons. 
Muon candidates are required to have an energy deposition less than 0.4 GeV, while electron candidates must have more than $1.1~$GeV.

\par Photon candidates are identified using isolated showers in the EMC. 
The deposited energy of each shower must be more than 25~MeV in the barrel region ($|\cos \theta|< 0.80$) and more than 50~MeV in the end cap region ($0.86 <|\cos \theta|< 0.92$). 
To exclude showers that originate from charged tracks, the angle subtended by the EMC shower and the position of the closest charged track extrapolated to the EMC must be greater than 10 degrees as measured from the IP. To suppress electronic noise and showers unrelated to the event, the difference between the EMC time of a shower 
and the event start time is required to be within 
[0, 700]\,ns.  
At least two good photon candidates are required in each event.

To improve the resolution and suppress background events, a four-constraint (4C) kinematic fit, constraining the total four-momentum of the reconstructed final state particles to that of the initial $\ee$ system, is performed to each event. If there is more than one combination satisfying the 4C kinematic fit due to extra pion or photon candidates, 
the one with the minimum $\chi^2$ value from the 4C kinematic fit ($\chi_{\rm 4C}^2$) is retained. Events with $\chi_{\rm 4C}^2<80$ are kept for further analysis.

To veto background events from $\ee\to\piz\piz\psi'$, $\ee\to\gamma_{\rm ISR}\psi'$ and $\ee\to\eta\psi'$, with $\psi'\to\pp\jpsi$, $|M(\pp\jpsi)- m_{\psip}|>0.01$~GeV/$c^2$ is required, and $M(\pp\jpsi)\equiv M(\pp\LL)-M(\LL)+m_{\jpsi}$ is used to improve the resolution, where $m_{\psip}$ and $m_{\jpsi}$ are the nominal masses of the $\psip$ and $\jpsi$~\cite{pdg}, respectively. 
To eliminate backgrounds from $\gamma\to\ee$ conversions in the beam pipe or inner wall of the MDC, where the $\ee$ pair is misidentified as the $\pp$ pair, the opening angle of the pion pair ($\theta_{\pip\pim}$) is required to satisfy $\cos\theta_{\pip\pim} < 0.98$. In order to suppress background from $\ee\to\gamma_{\rm ISR}\psip$ with $\psip\to\eta\jpsi\to\pp\piz\jpsi$, $\chi^{2}_{\rm 4C}(3\gamma)>\chi^{2}_{\rm 4C}$ and $\frac{E_{\gamma}^H-E_{\gamma}^L}{E_{\gamma}^H+E_{\gamma}^L}< 0.62$ are applied. Here $\chi^{2}_{\rm 4C}(3\gamma)$ is the $\chi^2$ from the 4C kinematic fit with a $\pp\gamma\gamma\gamma\LL$ hypothesis, $E_{\gamma}^H$ and $E_{\gamma}^L$ are the energies of the higher and lower energy photons in an event passing the 4C kinematic fit with a $\pp\gamma\gamma\LL$ hypothesis, respectively. To reduce $\mu/\pi$ misidentification backgrounds in the $\jpsi\to\mm$ channel, e.g. $\ee\to\omega\pp$ events, at least one of the muon candidates must have a hit depth larger than 38 cm in the MUC.
\par 
To select the signal candidates that contain $\jpsi$ and $\eta$ resonances, we define
$M(\LL)\in[3.05, 3.14]$~GeV/$c^2$ and $M(\gamma\gamma)\in[0.5, 0.6]$~GeV/$c^2$ as the $\jpsi$ and $\eta$ mass windows, respectively. The mass resolutions $\sigma$ of the $\eta$ and $\jpsi$ are 14.7 MeV and 15.0 MeV, respectively, as determined by the fit to the signal MC sample. The signal region is defined as $\pm 3\sigma$ around the nominal mass value. The non-$\jpsi$ background contribution is estimated using the events in the $\jpsi$ mass sideband regions, which are defined as $M(\LL)\in$ [2.91, 3.00]~GeV/$c^2 \cup$ [3.19, 3.28]~GeV/$c^2$, and are twice as wide as the signal region.

After imposing the above event selection, the remaining background contributions mainly come from $\ee \to \pip\pim \psi_2(3823) \to \pp \gamma\chi_{c1}$,
$\ee \to \eta^\prime J/\psi \to \pip\pim\eta\jpsi$ and $\ee\to\pp\psip$ with $\psip\to\eta\jpsi/\gamma\chi_{cJ}$, whose final states are also $\pp\GG\jpsi$.

\begin{figure*}
	\centering
	\includegraphics[height=2.2in]{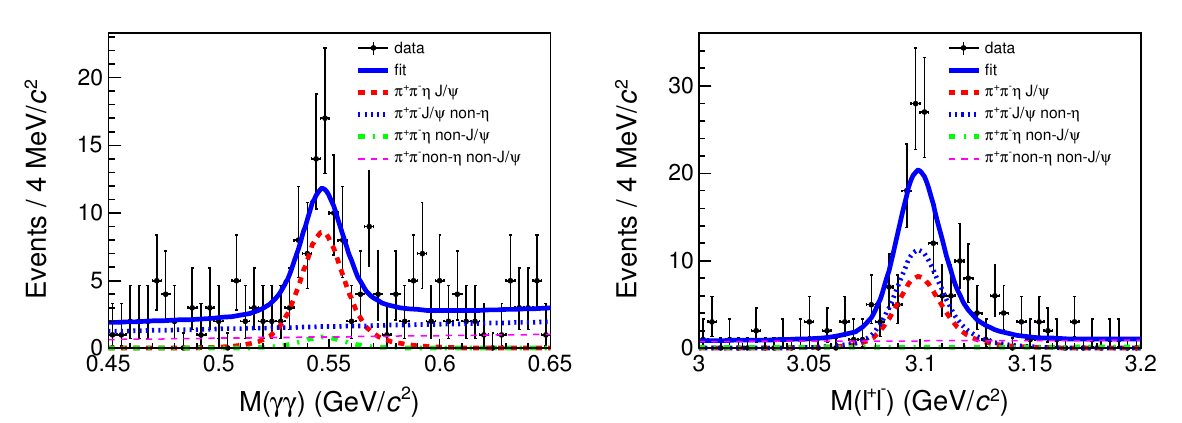}
	\vspace{-5mm}
	\caption{Projections of the 2D fit onto the $M(\gamma\gamma)$ and $M(\LL)$ distributions. Dots with error bars are data samples from all c.m.~energy points, the blue solid curves are the fit results and the red dashed curves are $\pp\eta\jpsi$ events ($S_{\eta}S_{\jpsi}$). The blue dotted, green dash-dotted and thin dashed curves represent $\pp\jpsi$ non-$\eta$ ($B_{\gamma\gamma}S_{\jpsi}$), $\pp\eta$ non-$\jpsi$ ($S_{\eta}B_{\LL}$) and $\pp$ non-$\eta$ non-$\jpsi$ ($B_{\gamma\gamma}B_{\LL}$), respectively.}
	\label{2d-fit}
\end{figure*}

\begin{figure*}
	\centering
	\includegraphics[height=2.2in]{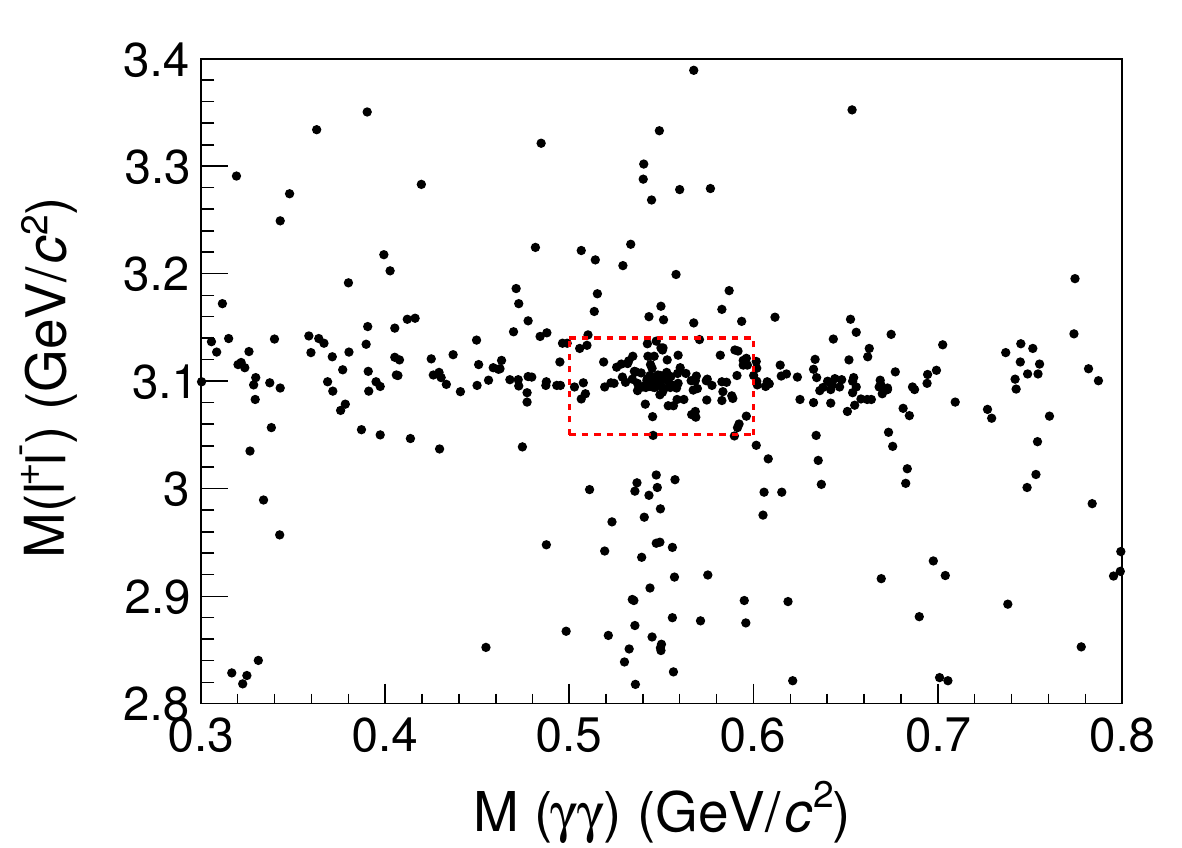}
	\vspace{-5mm}
	\caption{The 2D distribution of $M(\gamma\gamma)$ versus $M(\LL)$ for the data samples from all c.m.~energy points. The area bounded by the red dashed line is the signal region.}
	\label{2d dis}
\end{figure*}



\section{The yield of $\ee\to\pp\eta\jpsi$}

The signal process $\ee\to\pp\eta\jpsi$ includes both an 
$\eta$ and a $J/\psi$ in the final state.
The $\ee \to \pp \psi_2(3823)$ background forms a peak only in the $\jpsi$ signal region but not in the $\eta$ signal region. Therefore, it does not need to be removed. The backgrounds of $\ee\to\pp\psip$ with $\psip\to\eta\jpsi$ and $\ee \to \etap \jpsi$ with $\etap\to\pp\eta$ form peaks in both the $\eta$ and $\jpsi$ signal regions and have to be rejected.  These two backgrounds can be removed by requiring $|M^{\rm recoil}(\pp)-m_{\psip}|>0.044$ GeV/$c^2$
and $|M(\gamma\gamma\pp)-m_{\etap}|>0.04$~GeV/$c^2$, where $M^{\rm recoil}(\pp)\equiv\sqrt{(P_{\EE}-P_{\pp})^2}$ 
is the recoil mass of the $\pp$ system, and $m_{\etap}$ is the nominal mass of the $\etap$~\cite{pdg}. Here $P_{\EE}$ and $P_{\pp}$ are the four-momenta of the initial colliding beams and the $\pp$ pair, respectively. The $M(\LL)$, $M(\gamma\gamma)$ and the corresponding 2D distributions from all data samples are shown in Fig.~\ref{2d-fit} and Fig.~\ref{2d dis}. Significant $\eta$ and $\jpsi$ signals are observed. 

To extract the number of $\pp\eta\jpsi$ events $N$, a 2D fit is performed to the $M(\gamma\gamma)$ versus $M(\LL)$ distributions. The 2D PDF $f_{2D}$ is constructed with the
following four components:
\begin{equation}
	\begin{aligned}
		&f_{2D} = S_{\eta}S_{\jpsi}+B_{\gamma\gamma}S_{\jpsi}+S_{\eta}B_{\LL}+B_{\gamma\gamma}B_{\LL}\label{2d pdf}, 	
	\end{aligned}
\end{equation}
where $S_{\eta}S_{\jpsi}$, $B_{\gamma\gamma}S_{\jpsi}$, $S_{\eta}B_{\LL}$ and $B_{\gamma\gamma}B_{\LL}$ correspond to the $\pp$($\eta\jpsi$, $\jpsi$ non-$\eta$, $\eta$ non-$\jpsi$, and non-$\eta$ non-$\jpsi$) components, respectively. The symbols $S$ and $B$ stand for the signal and background shapes, respectively. The signal shape is described by the MC-simulated shape convolved with a Gaussian resolution. Here, the signal MC shape is extracted from combined MC samples by summing the signal simulations at each energy point, with each scaled according to its integrated luminosity and detection efficiency.
The Gaussian parameters are obtained from the $\ee\to\etap\jpsi$ with $\etap\to\pp\eta\to\pp\GG$ control sample. The background shapes of $B_{\gamma\gamma}$ and $B_{\LL}$ are described by 1st-order Chebychev functions. Figure~\ref{2d-fit} shows the fit result, yielding $N=58.2\pm10.2\pm3.4$, where the first uncertainty is statistical and the second is systematic. The signal significance is estimated by comparing the log-likelihood values with and without the $S_{\eta}S_{\jpsi}$ component included in the fit, which gives $\Delta(-2\ln\mathcal{L})=52.2$. Considering the change in the number of degrees of freedom $\Delta ndf=1$ between the fits with and without the $\pp\eta\jpsi$ component, the significance is calculated to be 7.2$\sigma$. The significance after considering the systematic uncertainties, including varying the resolution of the convolved Gaussian within $\pm 1\sigma$, and changing the background shapes from first-order to second order and the fit range by $\pm 10$ MeV, is 6.0$\sigma$. And the total systematic uncertainty from these three sources is 5.9\%. Possible intermediate resonance states have also been checked, as shown in Appendix~\ref{inter}, Fig.~\ref{inter state}. Due to the current statistics, it is not possible to establish the existence of intermediate resonance states.

\section{The average cross section of $\ee\to\pp\x$}\label{ave xs}
\begin{figure}
	\centering
	\includegraphics[height=2.2in]{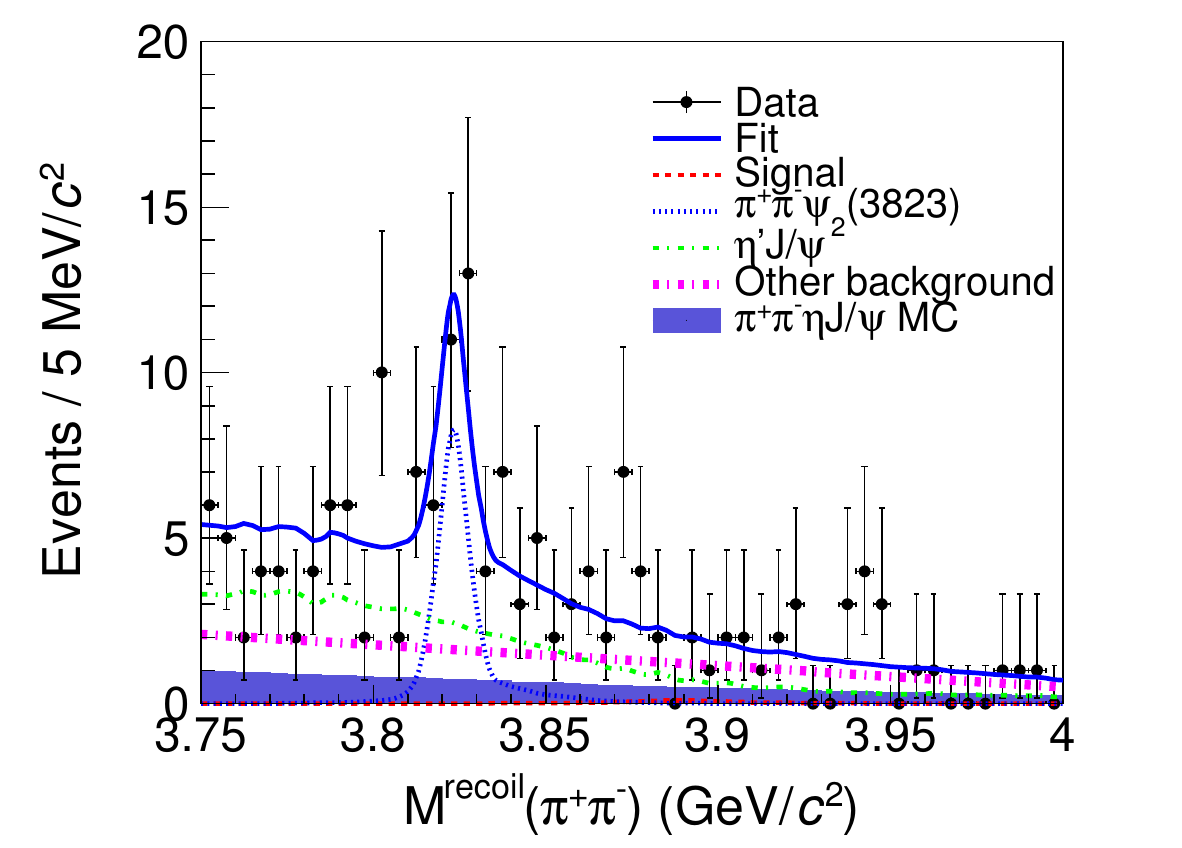}\\
	\caption{Result of the fit to the $M^{\rm recoil}(\pp)$ distribution.
		The dots with error bars are the data samples from all c.m.~energy points, the blue solid curve represents the total fit, and the blue dotted, green dash-dotted and pink thick dash-dotted curves are different background contributions. The shaded area is the contribution from the $\pp\eta\jpsi$ process and the red dashed curve is the signal contribution, which is not visible. }
	\label{fit}
\end{figure}

When considering the process $\ee\to\pp\x$ with $\x\to\eta\jpsi$, the backgrounds from $\ee\to\pp\psip$ and $\pp\psi_2(3823)$ form peaks around 3.686~GeV/$c^2$ and 3.823~GeV/$c^2$ in the $M^{\rm recoil}(\pp)$ spectrum, respectively, which do not affect the $\x$ search. The background from $\ee\to\etap\jpsi$ forms a wide envelope in the $M^{\rm recoil}(\pp)$ spectrum. These three background contributions have been well-studied by BESIII~\cite{etapjpsi, ppx3823, pppsip}
and can be reliably simulated. Thus these backgrounds are not removed when searching for $\ee\to\pp\x$ with $\x\to\eta\jpsi$. Figure~\ref{fit} shows the distribution of $M^{\rm recoil}(\pp)$ from the full data set. No obvious $\x$ signal is observed. Here the $M^{\rm recoil}(\pp)$ range is set to be greater than $3.75$~GeV/$c^2$ as the contribution in the range less than $3.75$~GeV/$c^2$ is mainly from the $\pp\psip$ process. To obtain the $\x$ signal yield, an unbinned maximum likelihood fit is performed,
as shown in Fig~\ref{fit}. The signal probability density function in the fit is represented by the MC-simulated $\x$ shape convolved with a Gaussian function,
which accounts for the difference in mass resolution between data and MC simulation.
The parameters of the $\x$ resonance in the simulation are the same as those of the $\zc$~\cite{pdg}, and the parameters of
the convolved Gaussian are fixed according to a study of the
$\ee\to\pp\psip$ control sample. The background in the fit consists of two components.
One is the simulated contribution from the $\ee \to \eta' J/\psi$ and $\ee \to \pip\pim \psi_2(3823)$ processes, 
which is normalized according to their measured cross sections~\cite{etapjpsi, ppx3823}. 
The other background is described by a first-order Chebychev polynomial, which represents the contributions from the $\ee\to\pp\eta\jpsi$ and continuum background events. 

Since no $\pp X(3900)$ signal is observed, the upper limit (U.L.) of the average Born cross section times the branching fraction $\sigma^{\rm U.L.}_{\rm Avg.}[\ee\to\pp X(3900)]\cdot\mathcal{B}[X(3900)\to\eta\jpsi]$ for all c.m.~energy points can be calculated by:
\begin{equation}
	\begin{aligned}
		&\sigma^{\rm U.L.}_{\rm Avg.}[\ee\to\pp X(3900)]\cdot\mathcal{B}[X(3900)\to\eta\jpsi] =&\frac{N^{\rm U.L.}_{\rm tot.}}{\mathcal{L}_{\rm int}^{\rm tot.}[(1+\delta)\frac{1}{|1-\Pi^2|}\epsilon_{\rm sig}]^{\rm Avg.}\mathcal{B}} \label{sigma}, 	
	\end{aligned}
\end{equation}
where $N^{\rm U.L.}_{\rm tot.}$ is the number of $\x$ signal events corresponding to an U.L. with 90$\%$ C.L., $\mathcal{L}_{\rm int}^{\rm tot.}$ is the total integrated luminosity, $[(1+\delta)\frac{1}{|1-\Pi^2|}\epsilon_{\rm sig}]^{\rm Avg.}$ is the average of the product of the $\x$ signal MC selection efficiencies, radiative correction factors and vacuum polarization factors (VP)~\cite{vacuum} for all c.m.~energies from 4.18 and 4.95~GeV. It is defined as $[(1+\delta)\frac{1}{|1-\Pi^2|}\epsilon_{\rm sig}]^{\rm Avg.}=\sum_{i}\mathcal{L}_{i}\sigma_{i}(1+\delta)_{i}\frac{1}{|1-\Pi^2|}_{i}\epsilon_{i}/\sum_{i}\mathcal{L}_{i}\sigma_{i}$, where $\mathcal{L}_{i}$, $\sigma_{i}$, $(1+\delta)_{i}$, $\frac{1}{|1-\Pi^2|}_{i}$, $\epsilon_{i}$ are the luminosity, cross section, radiative correction factors, vacuum polarization factors and selection efficiency at the $i$th c.m.~energy. The radiative correction factor is calculated by the {\sc kkmc} program with an accuracy of 0.1\%~\cite{kkmc, kkmc_2}. $\mathcal{B}\equiv\mathcal{B}(\eta\to\gamma\gamma) \, \mathcal{B}(\jpsi\to\LL)$ is a product of the corresponding branching fractions. 
\par The value of $N^{\rm U.L.}_{\rm tot.}$ is obtained by scanning the normalized likelihood value distribution using the Bayesian approach~\cite{Bayes} with all the additive systematic uncertainties taken into account. The normalized likelihood value distribution is obtained by fitting the $M^{\rm recoil}(\pp)$ mass spectrum (shown in Fig.~\ref{fit}) through a scan over the number of signal events. 
To incorporate the multiplicative systematic uncertainty into the U.L., the likelihood curve is further convolved by a Gaussian with a width parameter equal to the average multiplicative systematic uncertainty $\delta_{\rm sys}^{\rm Avg.}$, 9.3\%. The average systematic uncertainty $\delta_{\rm sys}^{\rm Avg.}$ is defined as $\delta_{\rm sys}^{\rm Avg.}=\sum_{i}\mathcal{L}_{i}\sigma_{i}\epsilon_{i}\delta_{i}/\sum_{i}\mathcal{L}_{i}\sigma_{i}\epsilon_{i}$, where $\delta_{i}$ is the systematic uncertainty at the $i$th c.m.~energy and listed in the last column of Table~\ref{systematical error for xs} in Appendix~\ref{sys-xs}. The systematic uncertainties will be discussed in detail in Sec.~\ref{sys error}.
\par Since the mass and width of the isoscalar partner $\x$ are not well determined, we give the U.L. with different assumptions of the mass and width. Three assumptions of mass (3.867 GeV/$c^2$, 3.887 GeV/$c^2$ and 3.907 GeV/$c^2$) and width (8.4 MeV, 28.4 MeV and 48.4 MeV) are used, resulting in 9 combinations of mass and width for the $\x$.  We use $X_1$ to $X_9$ to represent the 9 combinations. The mass, width, the U.L. of signal yield and the U.L. of average cross section of $X_1$ to $X_9$ are shown in Table~\ref{mgx1-x9}.

\section{The energy-dependent cross section of $\pp\x$}

The U.L.s for the product of the Born cross section  $\sigma^{\rm Born}[\ee\to\pp \x]$ and the branching fraction $\mathcal{B}[\x\to\eta\jpsi]$ are also measured at each energy point.
Due to the limitation of statistics, the U.L. of the signal yield at a $90\%$ C.L. for each c.m.~energy, $N^{\rm U.L.}$, is determined by counting the number of events in the $\x$ signal and sideband regions, where the sideband is 
used to estimate the background level in the signal region. The definitions of signal and sideband regions for $X_{1}$ to $X_{9}$ are shown in Table~\ref{mgx1-x9}. The distributions of $M^{\rm recoil}(\pp)$ at each c.m.~energy are shown in Fig.~\ref{rmpp1} and Fig.~\ref{rmpp2} of Appendix~\ref{rmpp-all}. 

\begin{figure*}[ht]
	\centering
	\includegraphics[height=2.0in]{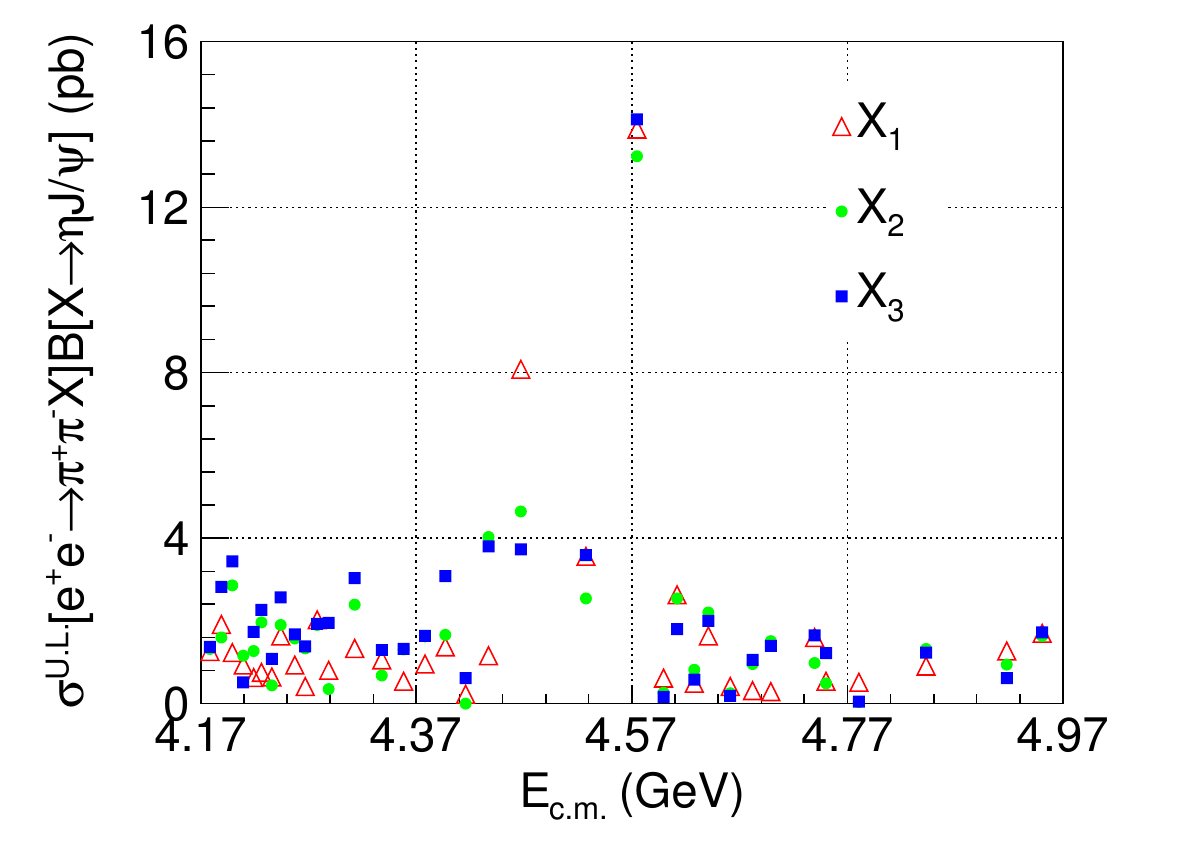}\hspace{5pt}
	\includegraphics[height=2.0in]{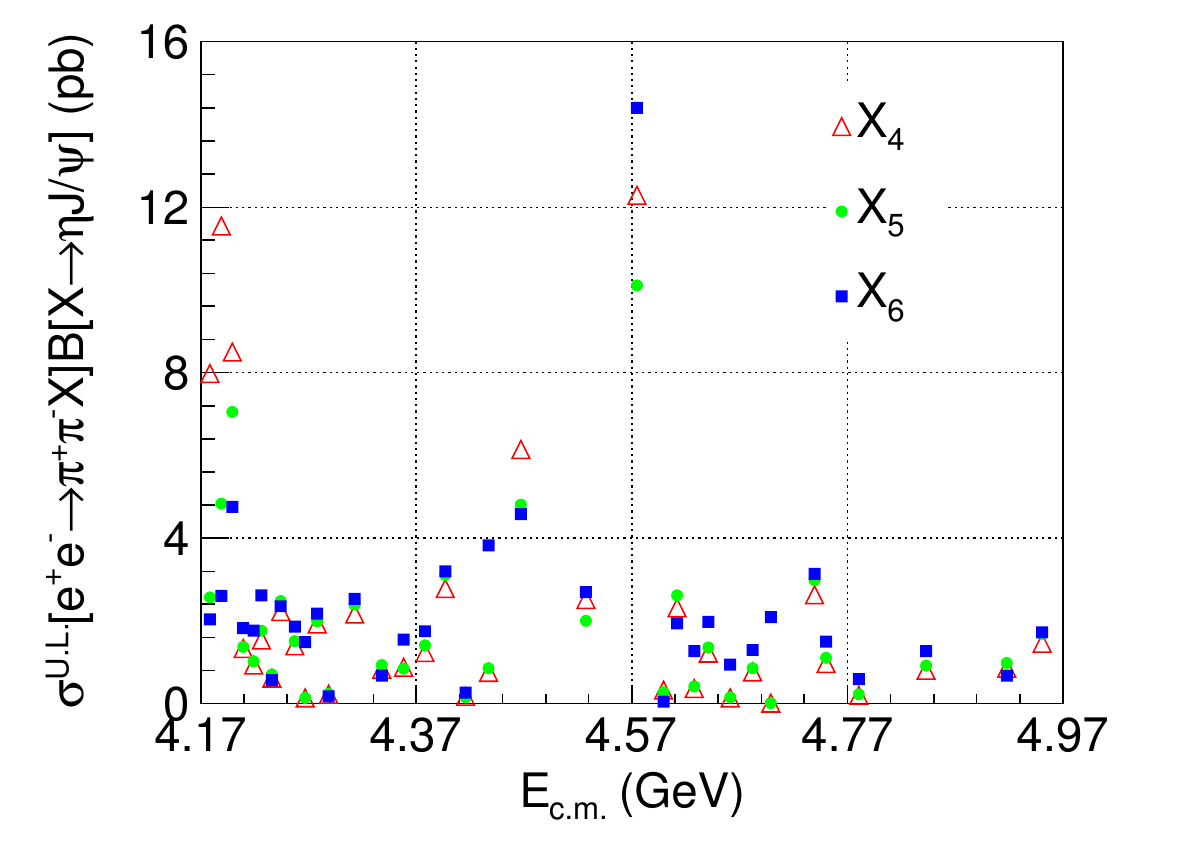}\hspace{5pt}
	\includegraphics[height=2.0in]{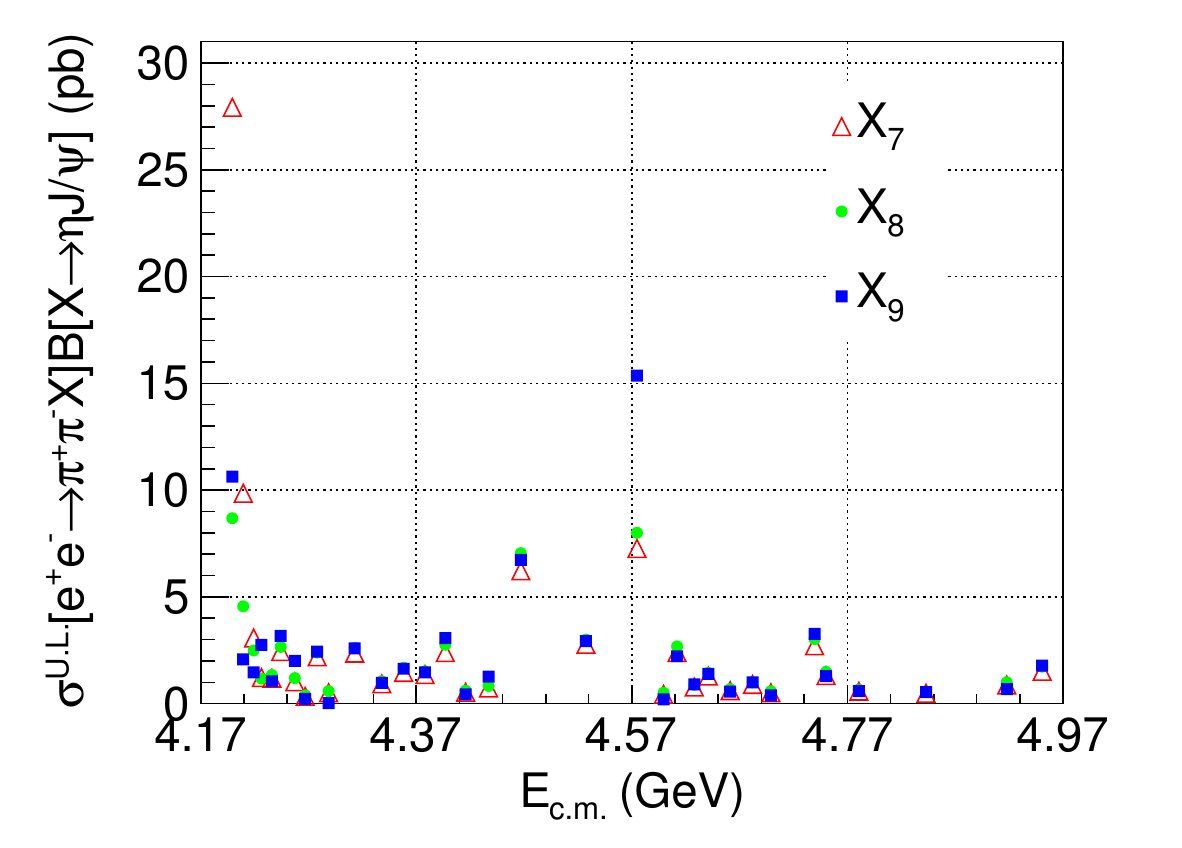}\hspace{5pt}
	\caption{The obtained U.L. $\sigma^{\rm U.L.}[\ee\to\pp \x] \cdot \mathcal{B}[\x\to\eta\jpsi]$ at 90\% C.L. for each c.m.~energy. The first plot is for $X_1$ to $X_3$, the second plot is for $X_4$ to $X_6$ and the third plot is for $X_7$ to $X_9$.}
	\label{fig_ul_x19}
\end{figure*}

The $N^{\rm U.L.}$ at each c.m.~energy is calculated using a frequentist method with an unbounded profile likelihood treatment~\cite{TRolke}, and we assume the numbers of observed events in the $\x$ signal and sideband regions both follow Poisson distributions and the signal efficiency follows a Gaussian distribution with a standard deviation equal to the systematic uncertainty $\delta_{i}$. 
Since the background distribution is not flat in the $\x$ signal and sideband regions according to the MC simulation, 
the ratio of the number of background events between the two regions is calculated using the MC simulation of $\etap\jpsi$, $\pp \p$, $\pp\eta\jpsi$ and the data sample from $\jpsi$ sideband as shown in Fig.~\ref{rmpp1} and Fig.~\ref{rmpp2}. 
Tables~\ref{count_UL_x13},~\ref{count_UL_x46} and~\ref{count_UL_x79} summarize the results related to the $\sigma^{\rm U.L.}[\ee\to\pp\x] \cdot \mathcal{B}[\x\to\eta\jpsi]$ measurement. The obtained U.L.s at 90$\%$ C.L. for $X_1$ to $X_9$ at each c.m.~energy are shown in Fig.~\ref{fig_ul_x19}. 

\begin{sidewaystable}
\begin{center}
	\caption{The mass and width of $X_1$ to $X_9$, the corresponding U.L. on the number of signal events $N^{\rm U.L.}_{\rm tot.}$, the U.L. on the average of the product $\sigma^{\rm U.L.}_{\rm Avg.}[\ee\to\pp X(3900)]\cdot\mathcal{B}[X(3900)\to\eta\jpsi]$, denoted as $(\sigma\cdot\mathcal{B})^{\rm U.L.}_{\rm Avg.}$, and the signal and sideband regions.}

	\begin{tabular}{c c c c c c c} 
		\hline 
		\hline
		Label & Mass~(GeV/$c^2$) & Width~(MeV) & $N^{\rm U.L.}_{\rm tot.}$ & $(\sigma\cdot\mathcal{B})^{\rm U.L.}_{\rm Avg.}$(pb) & Signal region~(GeV/$c^2$) & Sideband region~(GeV/$c^2$)\\
		\hline
		$X_1$ & 3.867 & 8.4 & 22.2 & 0.13 & [3.85, 3.89] & [3.80, 3.84] $\cup$[3.90,3.94]\\
		$X_2$ & 3.867 & 28.4 & 31.3 & 0.18 & [3.82, 3.92] & [3.72, 3.81]$\cup$[3.93, 4.04]\\
		$X_3$ & 3.867 & 48.4 & 42.9 & 0.26 & [3.80, 3.93]  & [3.72, 3.79]$\cup$[3.94, 4.13]\\
		$X_4$ & 3.887 & 8.4 & 8.5 & 0.05 & [3.84, 3.94] & [3.73, 3.83]$\cup$[3.95, 4.05]\\
		$X_5$ & 3.887 & 28.4 & 19.8 & 0.12 & [3.84, 3.94] & [3.73, 3.83]$\cup$[3.95, 4.05]\\
		$X_6$ & 3.887 & 48.4 & 34.7 & 0.22 & [3.82, 3.95] & [3.72, 3.81]$\cup$[3.96, 4.13]\\
		$X_7$ & 3.907 & 8.4 & 11.0 & 0.08 &  [3.86, 3.96] & [3.75, 3.85]$\cup$[3.97, 4.07]\\
		$X_8$ & 3.907 & 28.4 & 18.6 & 0.14 & [3.86, 3.96] & [3.75, 3.85]$\cup$[3.97, 4.07]\\
		$X_9$ & 3.907 & 48.4 & 27.1 & 0.21 & [3.84, 3.97] & [3.72, 3.83]$\cup$[3.98, 4.13]\\ 
		\hline 
		\hline
	\end{tabular}
	\label{mgx1-x9}  
\end{center}
\end{sidewaystable}

\section{Systematic uncertainty}\label{sys error}

The systematic uncertainties in the cross section measurement mainly come from the multiplicative and additive systematic uncertainties. The multiplicative part is from the luminosity, photon reconstruction, tracking efficiency, branching fraction, 4C kinematic fit, MC model, $\gamma_{\rm ISR}\psip$ cut, muon hit depth in the MUC, and radiative correction. The additive part is from the uncertainties of background, fit range and signal resolution.  

The integrated luminosity is measured using large angle Bhabha  scattering events, with an uncertainty of 0.7\% and 0.5\% for $\sqrt{s}<4.6$~GeV and $\sqrt{s}>4.6$~GeV, respectively~\cite{4.0-4.6 lum, 4.6-4.9 lum and ecm}. The systematic uncertainty of the photon reconstruction is 1.0\% per photon~\cite{uncertaintyofphoton} based on studies of the processes $\chi_{c0,2}\to\piz\piz$/$\eta\eta$. Because there are two photons in the final state, the systematic uncertainty is assigned as 2.0\%. The uncertainty of pion tracking is estimated to be 1.0\% per pion based on the study of $\jpsi\to p\bar{p}\pp$~\cite{syserrorofpi}.  The uncertainty of the tracking for high momentum leptons is assigned to be 1.0\% per track according to the study of $\ee\to\pp\jpsi$~\cite{erroroflep}. The total systematic uncertainty from tracking is assigned as 4.0\%. 

The branching fractions of $\eta\to\gamma\gamma$ and $\jpsi\to\LL$ are taken from the PDG~\cite{pdg}, and their uncertainties are propagated to the Born cross section measurement. The resulting systematic uncertainties are 0.5\% and 0.4\%, respectively. The systematic uncertainty associated with the 4C kinematic fit is estimated by comparing the efficiency difference with and without the correction of the helix parameters of charged tracks in the MC simulations~\cite{kf-correction}. The result after correction is taken as the nominal value.

The radiative correction factor $(1+\delta)$ and efficiency $\epsilon$ depend on the input cross section line shape in  {\sc kkmc}. In the nominal analysis, the cross section based on the three-body PHSP is used. To estimate the systematic uncertainty from radiative corrections, the measured $\ee\to\eta\eta\jpsi$ cross section from the BESIII measurement~\cite{etaeta} is used as input. Because the threshold of $\eta\eta\jpsi$ is greater than 4.2~GeV, the $\eta\eta\jpsi$ cross section at $\sqrt{s}<4.2$~GeV region is replaced by that of $\piz \zc$ from the BESIII measurement~\cite{pizZc}. For the MC model, in the nominal analysis, the $\ee\to\pp\x$ signal MC events are generated using a PHSP model with the three-body PHSP as production cross section. Assuming the $\pip\pim$ system is dominated by a $\sigma(500)$ resonance, the process $\ee\to\sigma(500)\x\to\pp\x$ with  two-body PHSP as the production cross section is used as an alternative model to estimate the systematic uncertainties. Since the mass and width of the $\x$ are unknown, we compare $\ee\to\pp \eta\jpsi$ and $\ee \to \sigma(500) \eta\jpsi$ as an alternative way to estimate the uncertainty. The difference of $\epsilon(1+\delta)$ is taken as the systematic uncertainty. 

The systematic uncertainty from the requirement of the muon hit depth in the MUC is studied with the control sample of $\ee\to\mm$, and the difference in efficiency between the data and MC simulation due to the requirement is taken as the systematic uncertainty. The systematic uncertainty from the cut, $\chi^{2}_{4\rm C}(3\gamma)>\chi^{2}_{4\rm C}$ and $\frac{E_{\gamma}^H-E_{\gamma}^L}{E_{\gamma}^H+E_{\gamma}^L}< 0.62$, are studied with the control sample of $\ee\to\pp\psip$.  The difference in efficiency between the data and MC simulation due to the requirements is taken as the systematic uncertainty.

When measuring the average cross section of $\pp\x$, the uncertainties of background shape, fit range, and signal resolution affect the $N^{\rm U.L.}_{\rm tot.}$ directly. Thus we change the background shape from a first-order polynomial to second-order, vary the resolution of the convolved Gaussian within $\pm 1\sigma$, and change the fit range by $\pm$ 10 MeV. The most conservative $N^{\rm U.L.}_{\rm tot.}$ from the combined effects of these additive sources is taken as the final result. When calculating the energy-dependent cross section, the ratio between the background events in the signal and sideband regions is calculated as the input to determine $N^{\rm U.L.}$ according to the MC simulation and $\jpsi$ sideband. Thus the uncertainty on the cross sections for the processes $e^+e^-\to \etap\jpsi$ and $\pp\p$ will introduce systematic uncertainty. To estimate this uncertainty,  we change the cross sections of $e^+e^-\to \etap\jpsi$ and $\pp\p$ by $\pm1\sigma$ and use the most conservative U.L..


Assuming that all the multiplicative sources are independent, the total systematic uncertainty for the $\sigma^{\rm U.L.}[\ee\to\pp\x] \cdot \mathcal{B}[\x\to\eta\jpsi]$ measurement at each c.m.~energy is calculated by adding them in quadrature.
Table~\ref{systematical error for xs} in Appendix~\ref{sys-xs} summarizes the systematic uncertainties 
for the $\sigma^{\rm U.L.}[\ee\to\pp\x] \cdot \mathcal{B}[\x\to\eta\jpsi]$ measurement.

\section{Summary}
In summary, with a data sample corresponding to an integrated luminosity of 20.3~fb$^{-1}$ collected with the BESIII detector at c.m.~energies ranging from 4.18 to 4.95~GeV,  the process $\ee\to\pp\eta\jpsi$ is observed with a statistical significance of $6.0 \sigma$.  
The isoscalar $\x$ is searched for through the process $\ee\to\pp\x$ with $\x\to\eta\jpsi$. No significant signal is found. The U.L.s on the average cross section $\sigma^{\rm Born}[\EE\to\pp\x] \cdot \mathcal{B}[\x\to\eta\jpsi]$ at 90\% C.L. are measured to be between 0.05 pb and 0.26 pb by assuming the $\x$ mass is in the range of 3.867 GeV/$c^2$ to 3.907~GeV/$c^2$ and the width is 8.4 MeV to 48.4 MeV. With the upgrade of the BEPCII~\cite{BESIII:2020nme} project, more data in this energy region is expected and a more comprehensive study of $\x$ production is anticipated, which will offer a promising chance to pin down the existence of this state, complete the family of $\zc$ states and understand their nature~\cite{Wang:2025dur}.

\acknowledgments
The BESIII Collaboration thanks the staff of BEPCII (https://cstr.cn/31109.02.BEPC) and the IHEP computing center for their strong support. This work is supported in part by National Key R\&D Program of China under Contracts Nos. 2025YFA1613900, 2023YFA1606000, 2023YFA1606704; National Natural Science Foundation of China (NSFC) under Contracts Nos. 12575090, 11635010, 11935015, 11935016, 11935018, 12025502, 12035009, 12035013, 12061131003, 12192260, 12192261, 12192262, 12192263, 12192264, 12192265, 12221005, 12225509, 12235017, 12361141819; the Chinese Academy of Sciences (CAS) Large-Scale Scientific Facility Program; the Strategic Priority Research Program of Chinese Academy of Sciences under Contract No. XDA0480600; CAS under Contract No. YSBR-101; 100 Talents Program of CAS; Project Nos. ZR2022JQ02, ZR2024QA151 supported by Shandong Provincial Natural Science Foundation; Supported by the China Postdoctoral Science Foundation under Grant No. 2023M742100; The Institute of Nuclear and Particle Physics (INPAC) and Shanghai Key Laboratory for Particle Physics and Cosmology; ERC under Contract No. 758462; German Research Foundation DFG under Contract No. FOR5327; Istituto Nazionale di Fisica Nucleare, Italy; Knut and Alice Wallenberg Foundation under Contracts Nos. 2021.0174, 2021.0299; Ministry of Development of Turkey under Contract No. DPT2006K-120470; National Research Foundation of Korea under Contract No. NRF-2022R1A2C1092335; National Science and Technology fund of Mongolia; Polish National Science Centre under Contract No. 2024/53/B/ST2/00975; STFC (United Kingdom); Swedish Research Council under Contract No. 2019.04595; U. S. Department of Energy under Contract No. DE-FG02-05ER41374

\clearpage
\appendix
\section{The invariant mass distribution for possible intermediate resonance states}\label{inter}
\begin{figure}[htbp]
	\centering
	\includegraphics[width=0.35\textwidth]{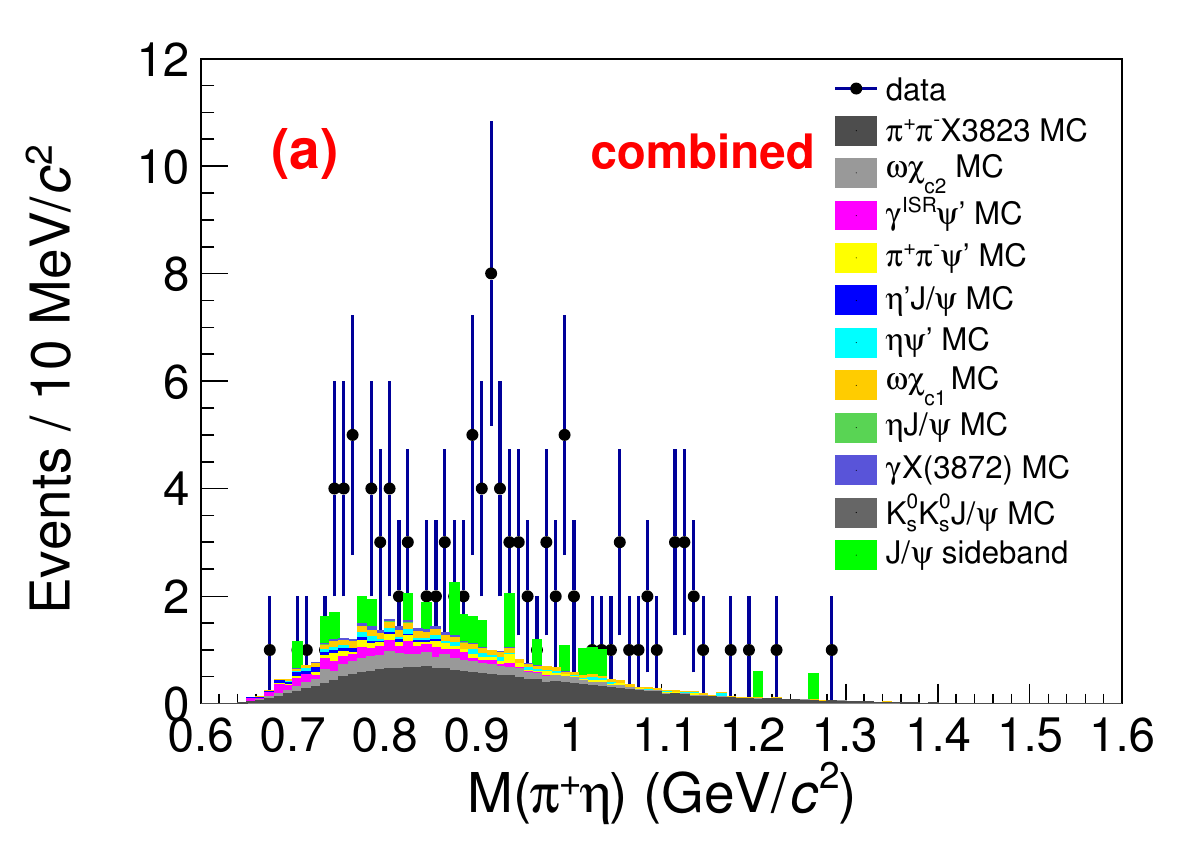}\hspace{5pt}
	\includegraphics[width=0.35\textwidth]{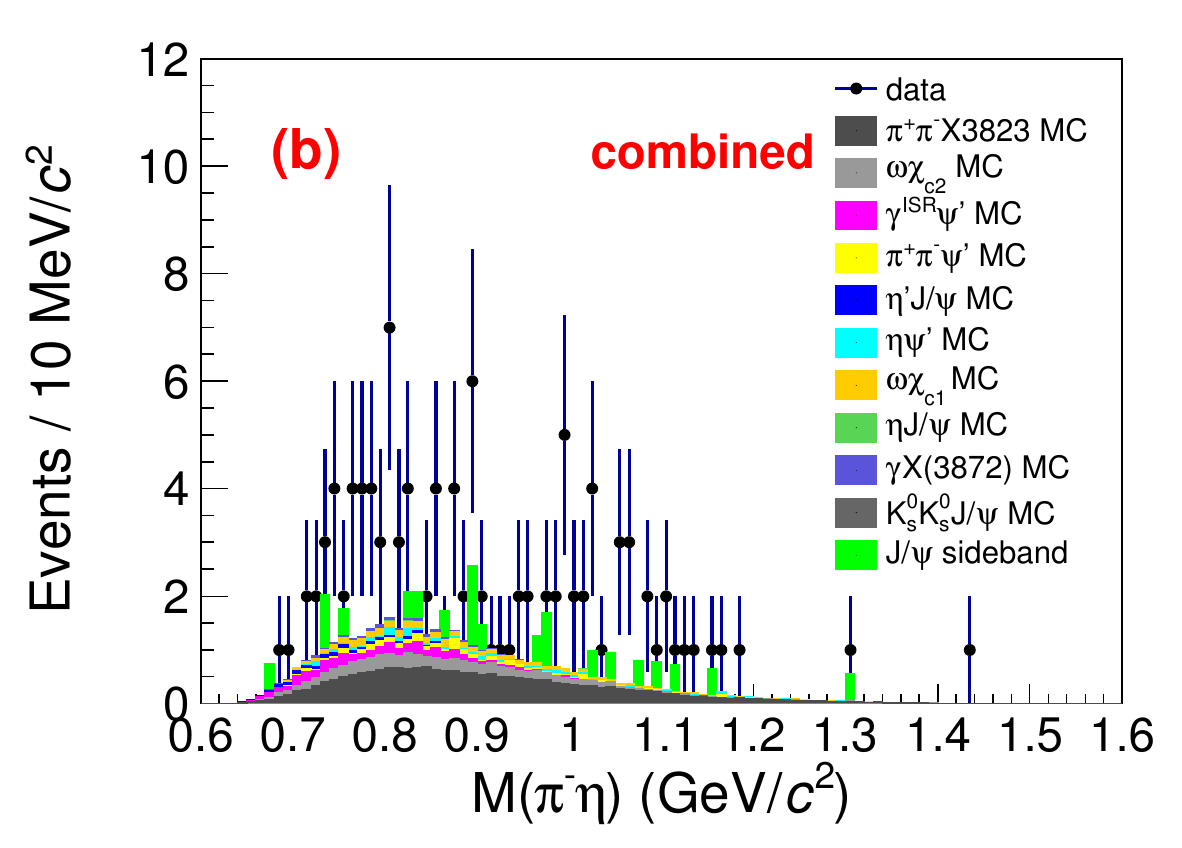}\hspace{5pt}
	\includegraphics[width=0.35\textwidth]{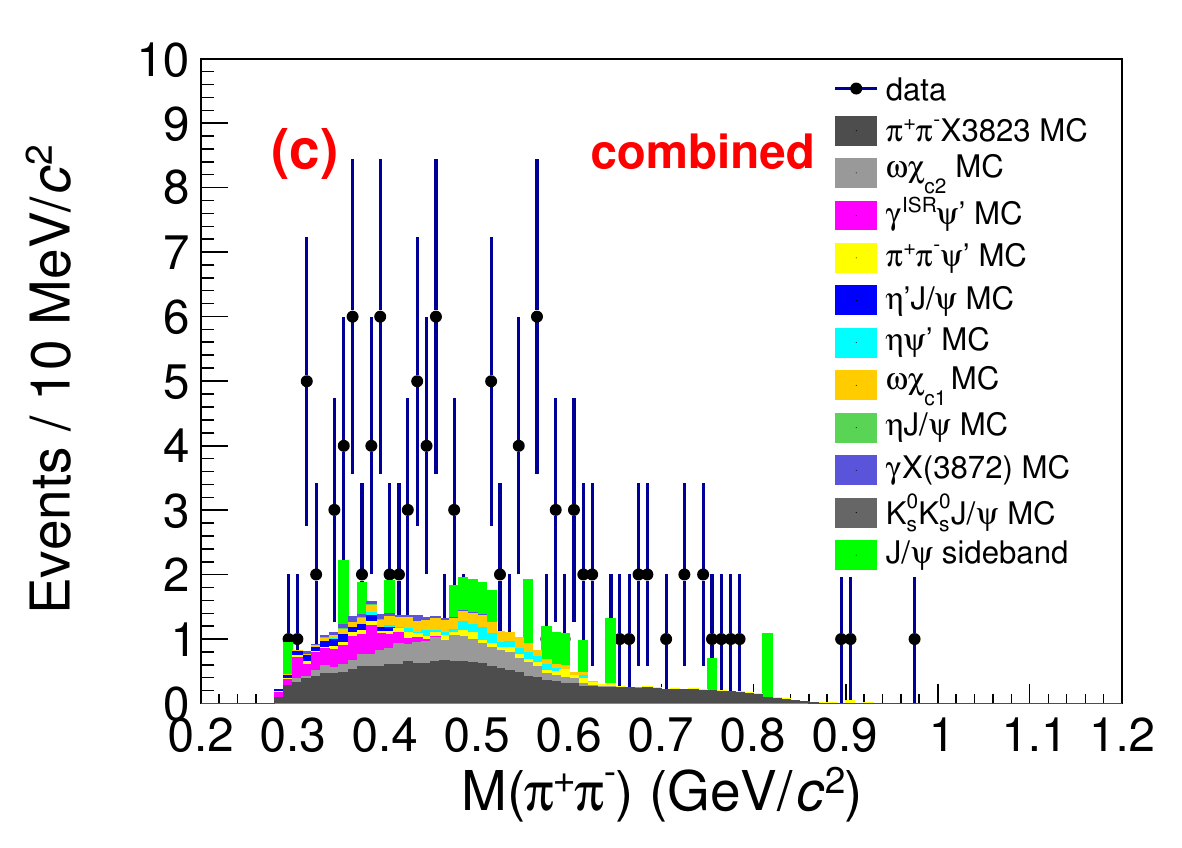}\hspace{5pt}
	\includegraphics[width=0.35\textwidth]{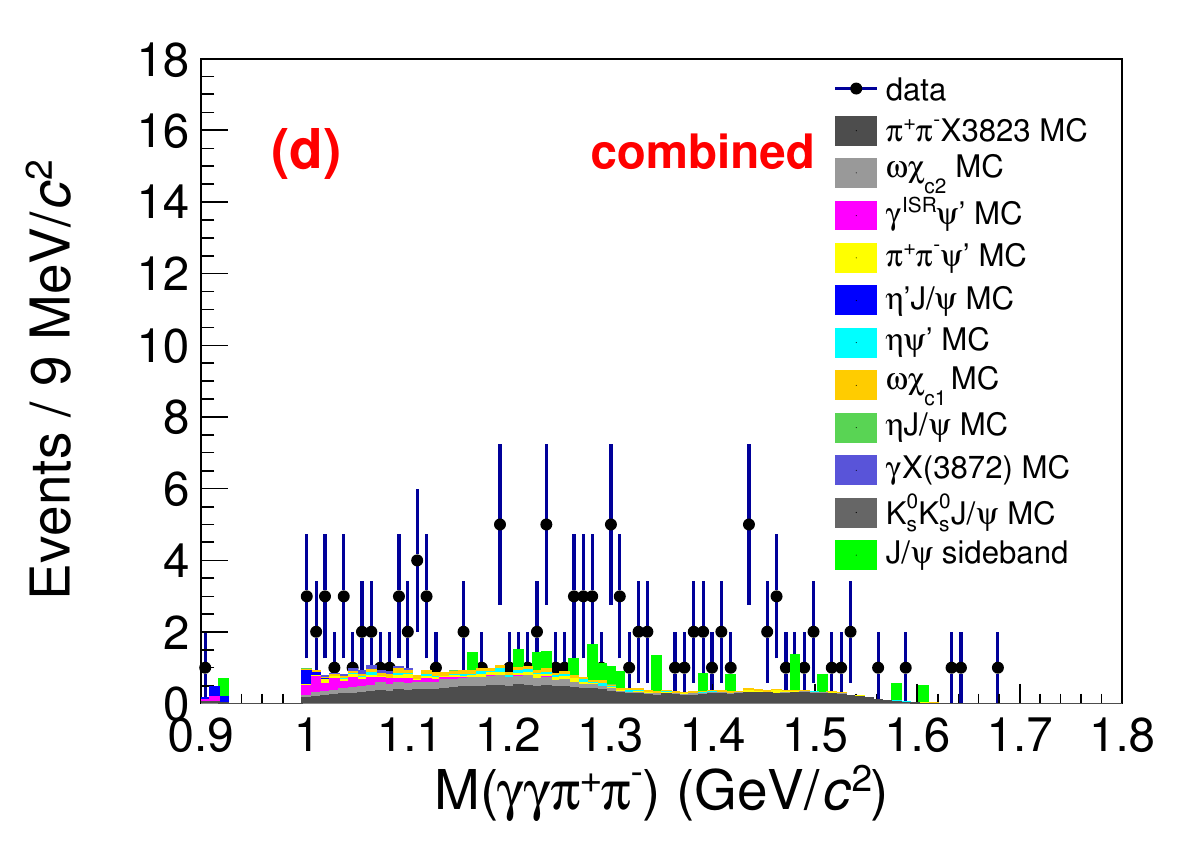}\hspace{5pt}
	\caption{The distributions of $M(\pip\eta)$ (a), $M(\pim\eta)$ (b), $M(\pp)$ (c) and $M(\gamma\gamma\pp)$ (d). The dots with error bars are the data samples from all c.m.~energy points, and the histograms are the background from MC simulation and $\jpsi$ sideband. }
	\label{inter state}
\end{figure}

\section{The distribution of $M^{\rm recoil}(\pp)$ at each center-of-mass energy}
\label{rmpp-all}
\vspace{-6mm}
\begin{figure}[H]
	\centering
	\includegraphics[width=0.3\textwidth]{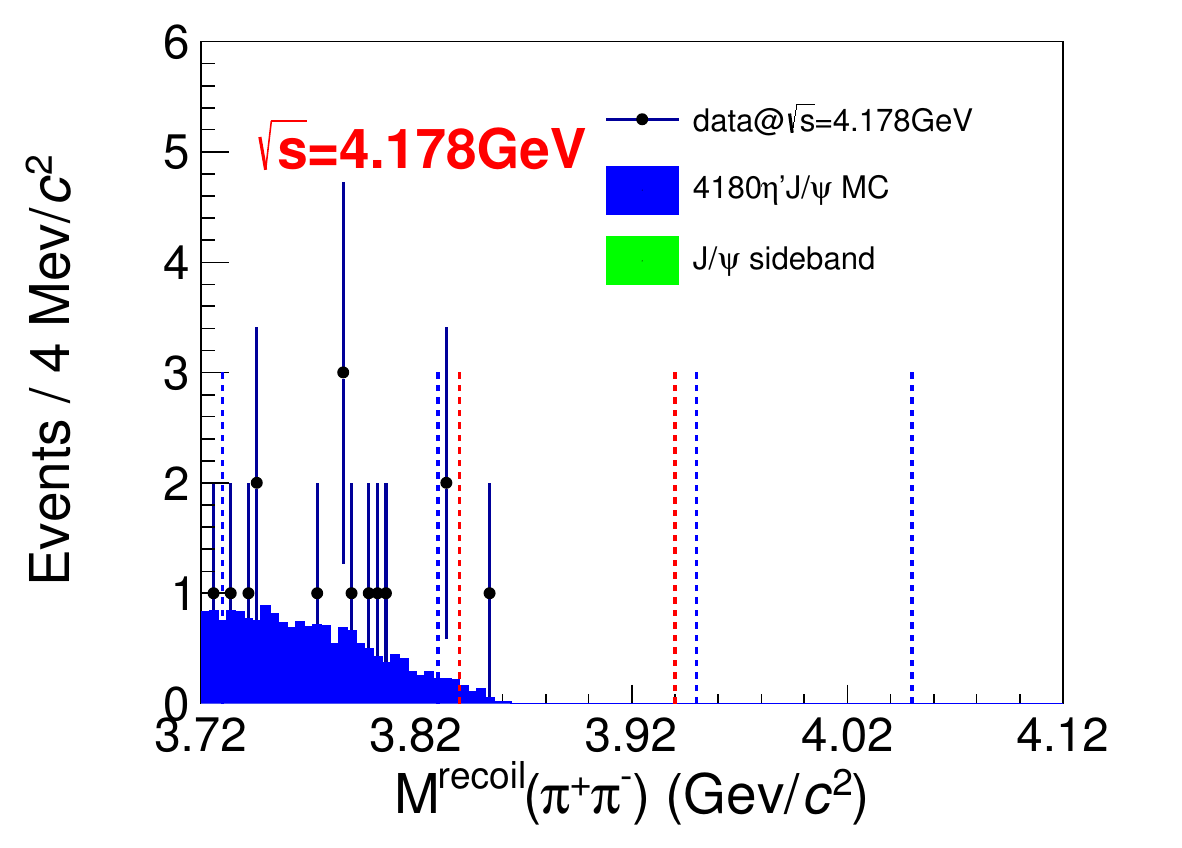}\hspace{5pt}
	\includegraphics[width=0.3\textwidth]{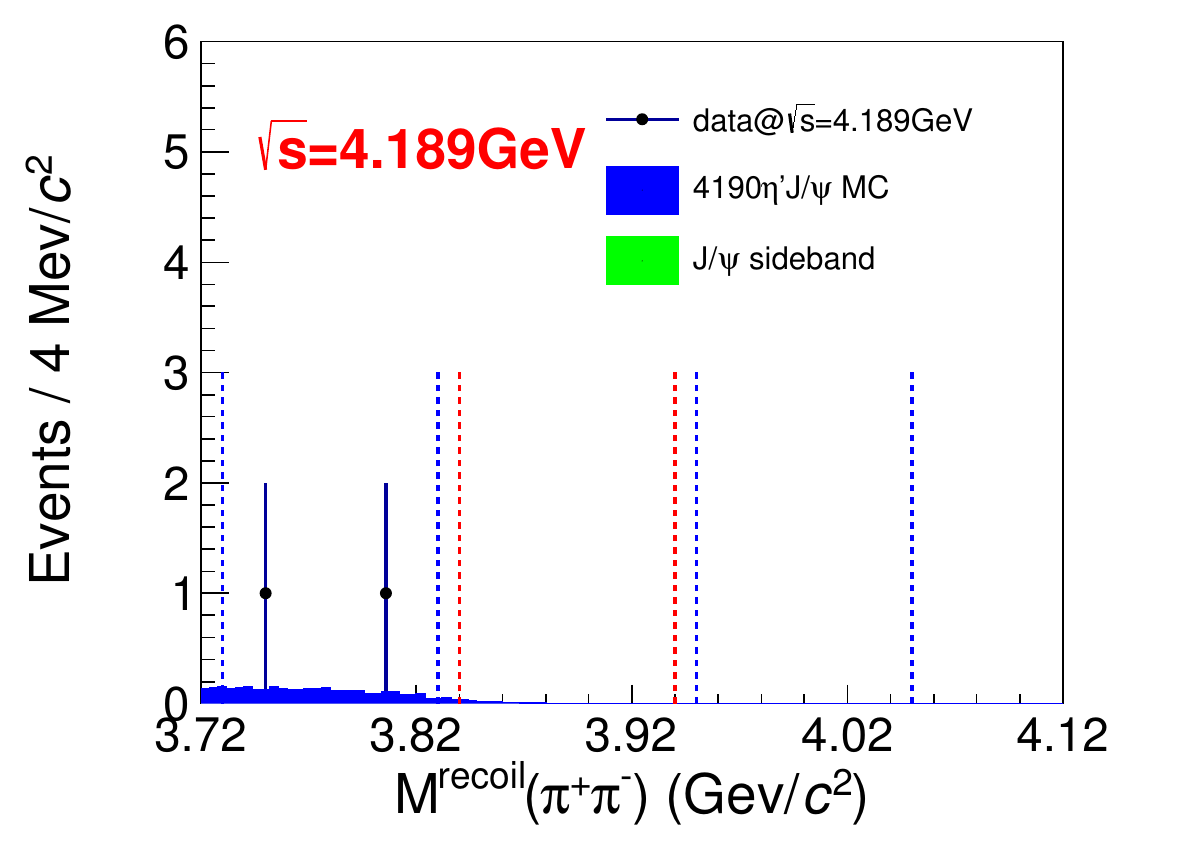}\hspace{5pt}
	\includegraphics[width=0.3\textwidth]{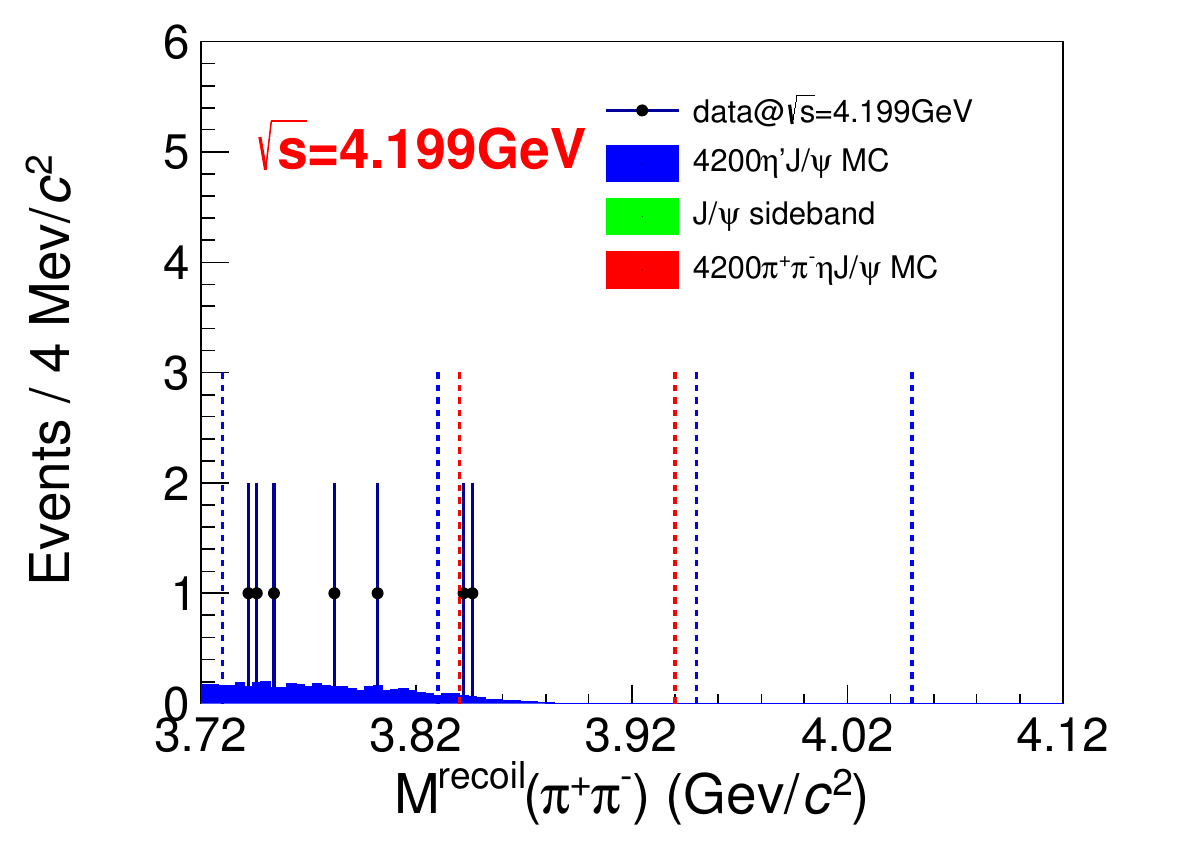}\hspace{5pt}
	\includegraphics[width=0.3\textwidth]{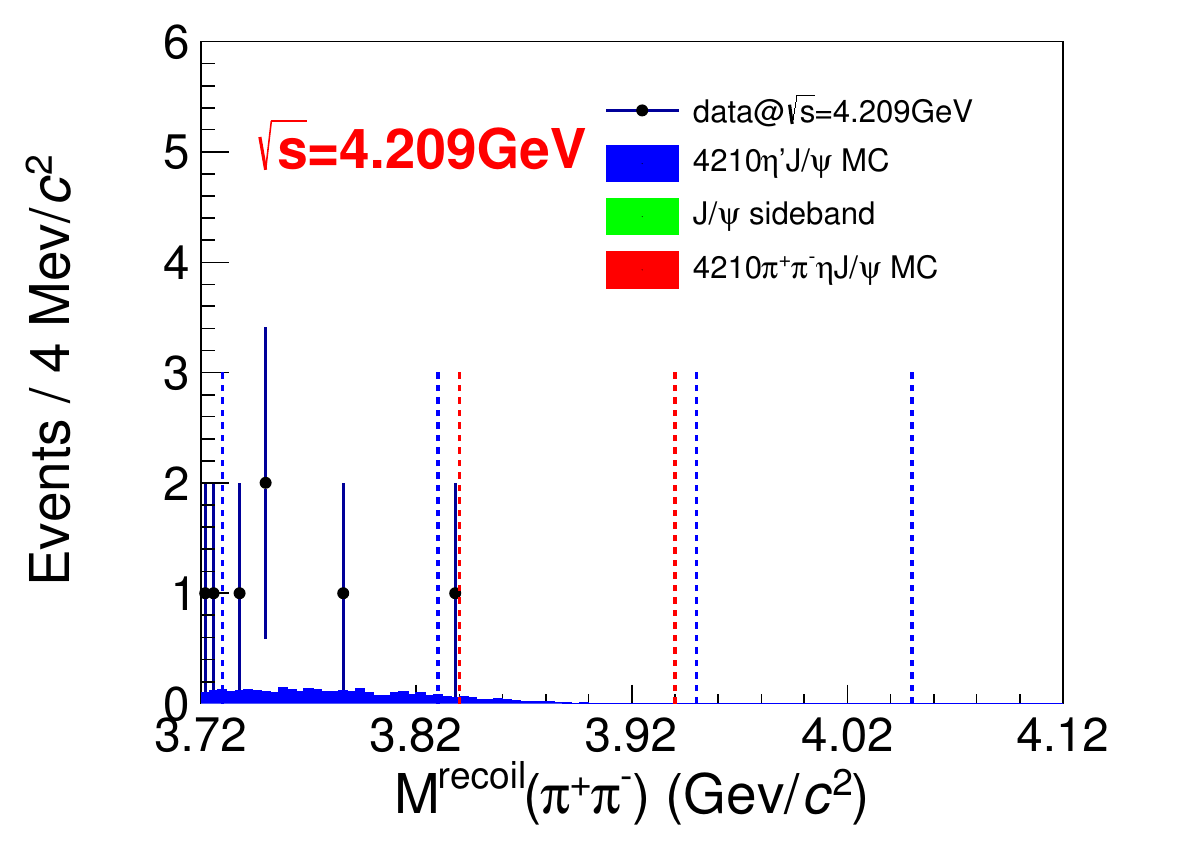}\hspace{5pt}
	\includegraphics[width=0.3\textwidth]{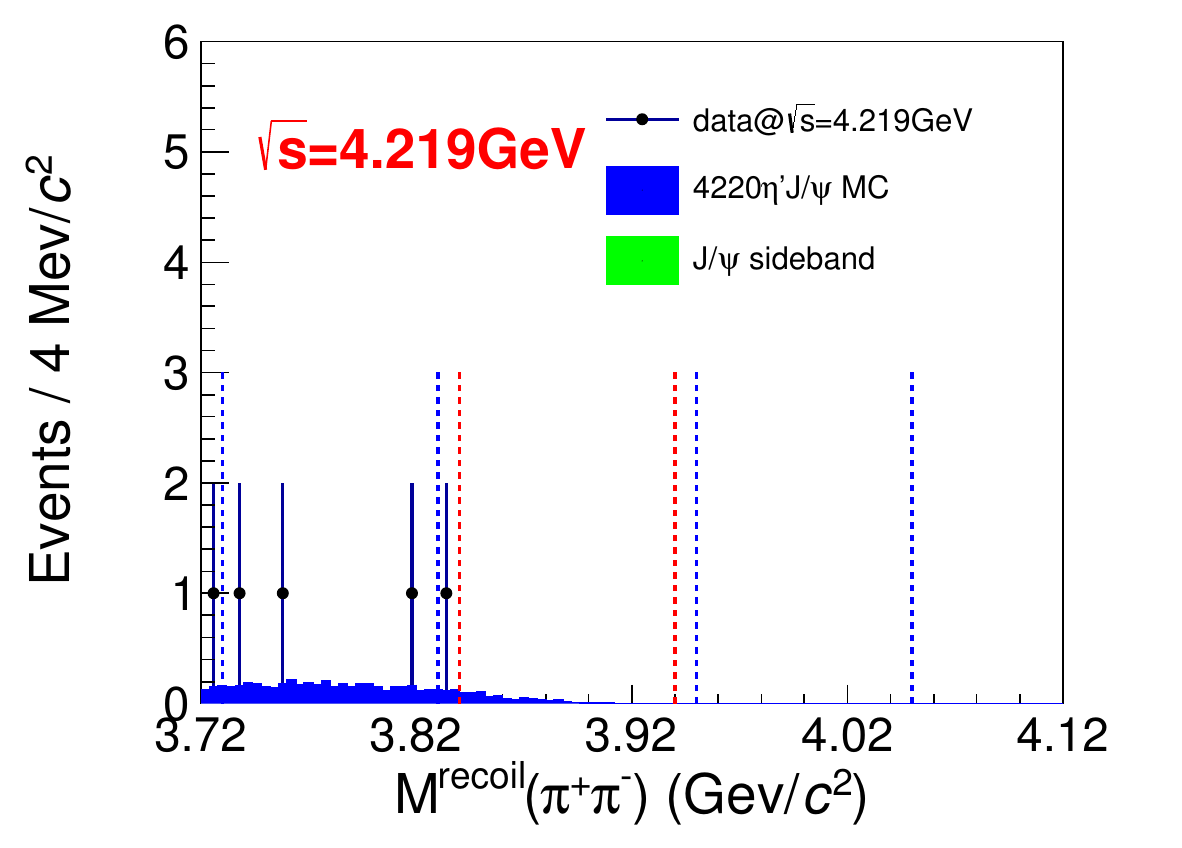}\hspace{5pt}
	\includegraphics[width=0.3\textwidth]{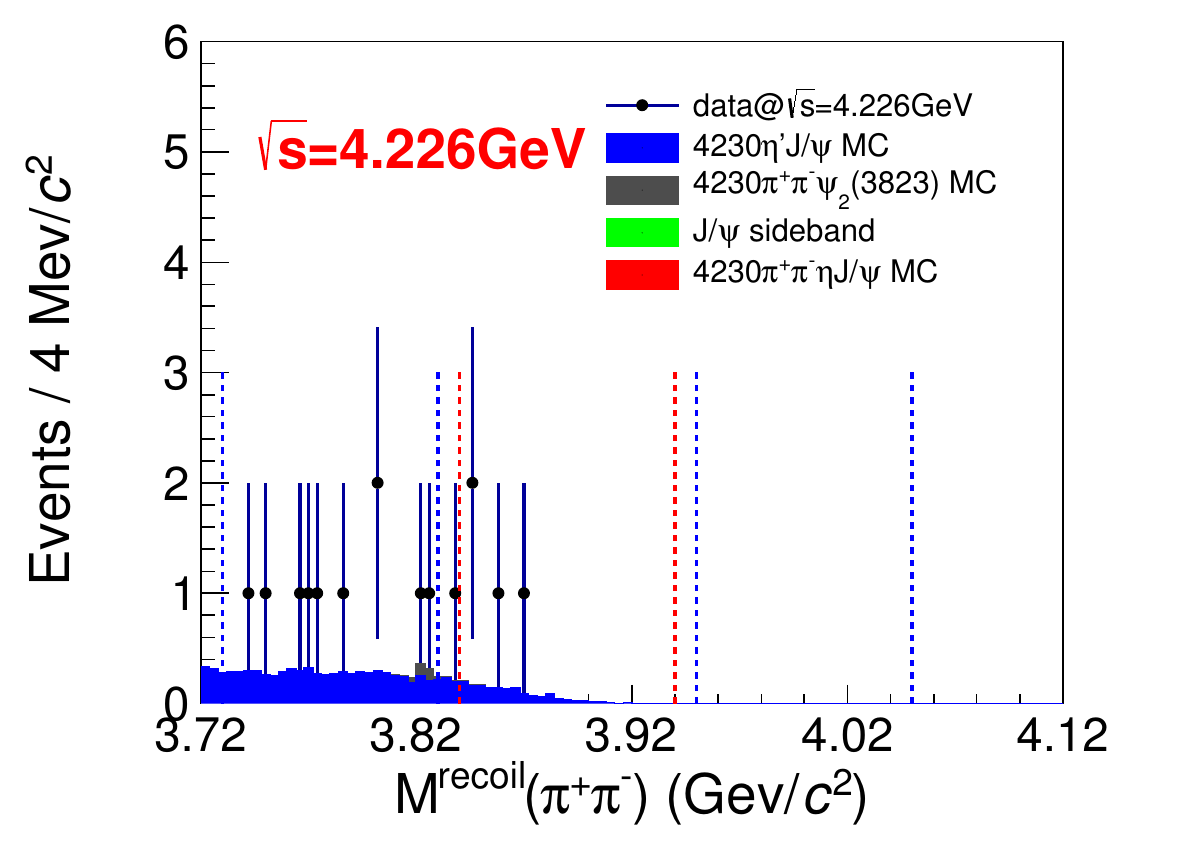}\hspace{5pt}
	\includegraphics[width=0.3\textwidth]{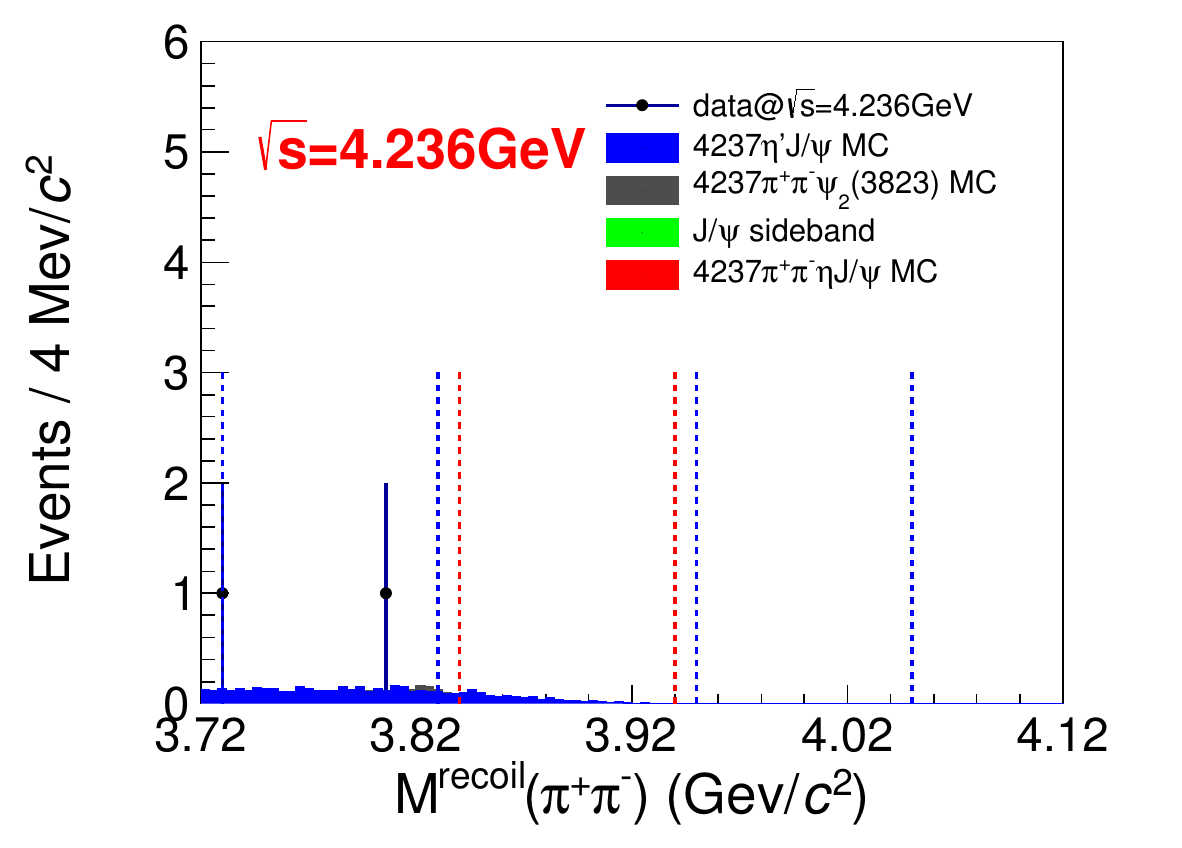}\hspace{5pt}
	\includegraphics[width=0.3\textwidth]{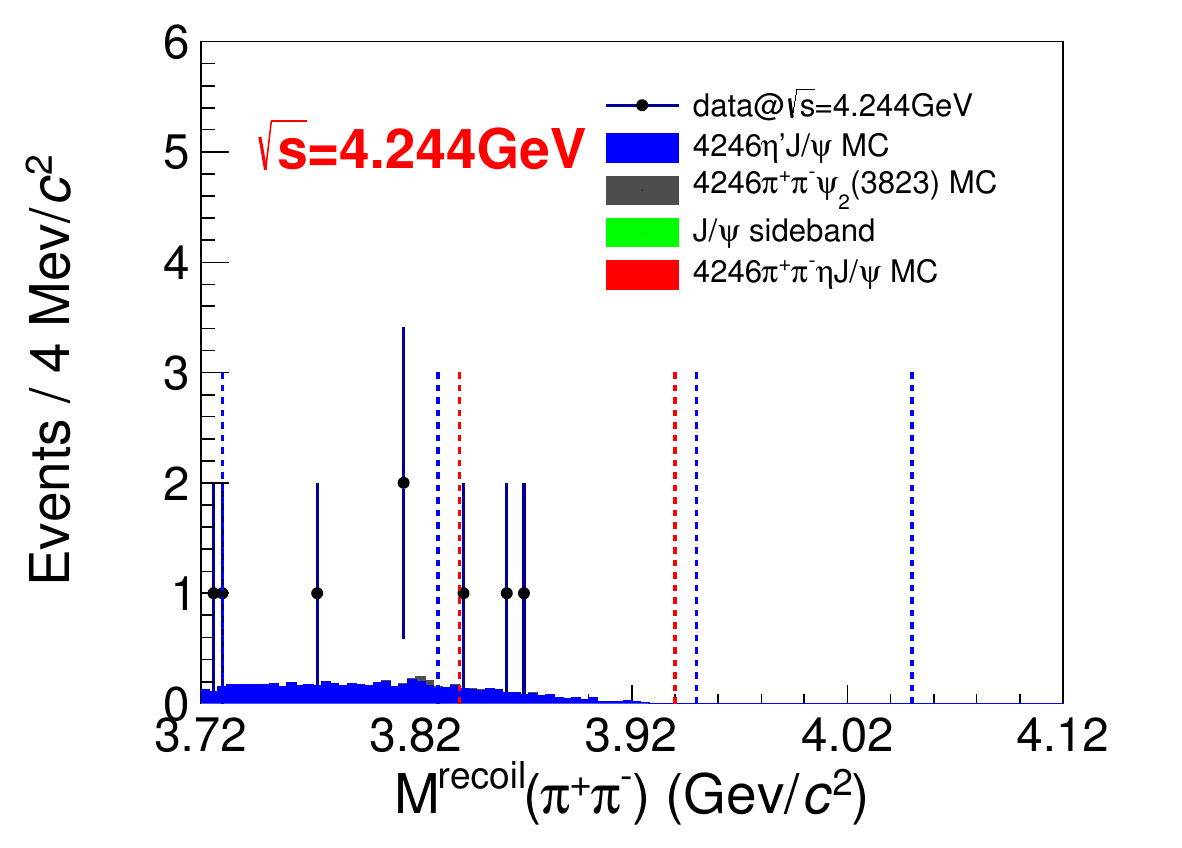}\hspace{5pt}
	\includegraphics[width=0.3\textwidth]{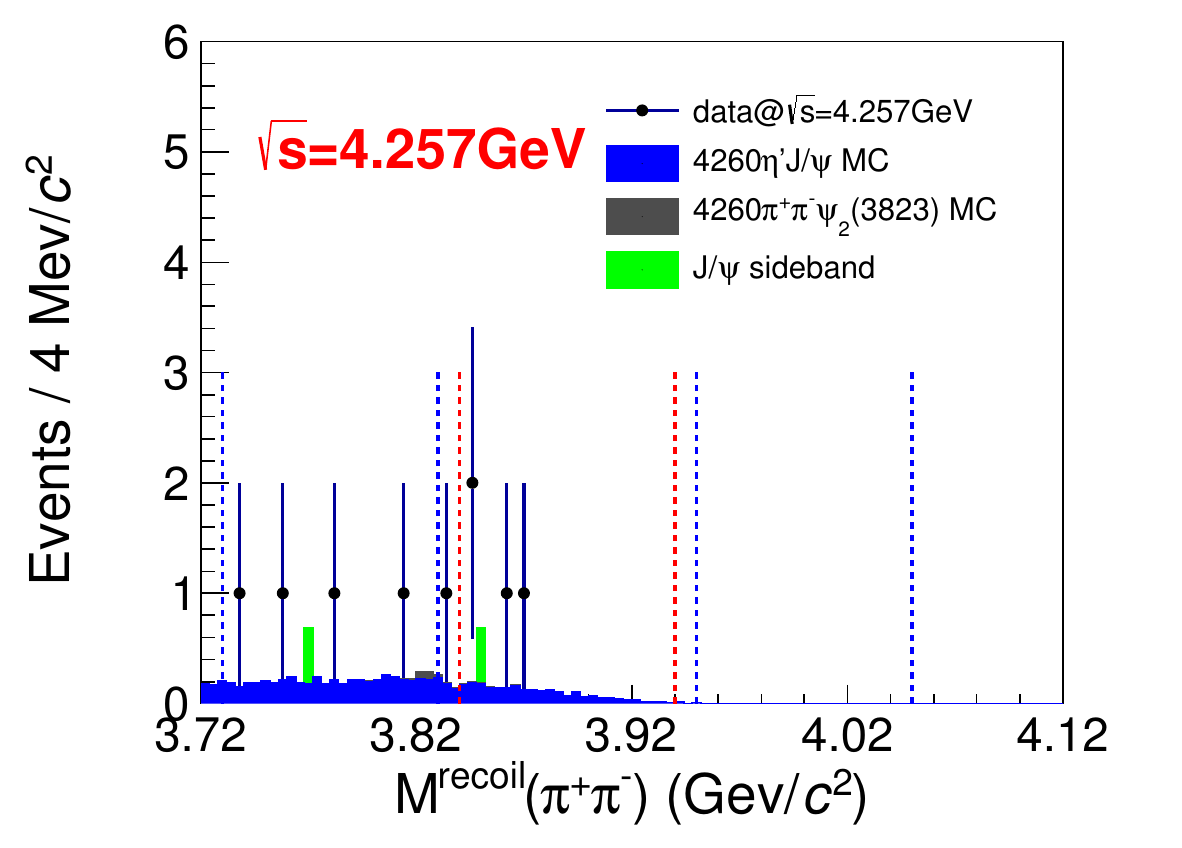}\hspace{5pt}
	\includegraphics[width=0.3\textwidth]{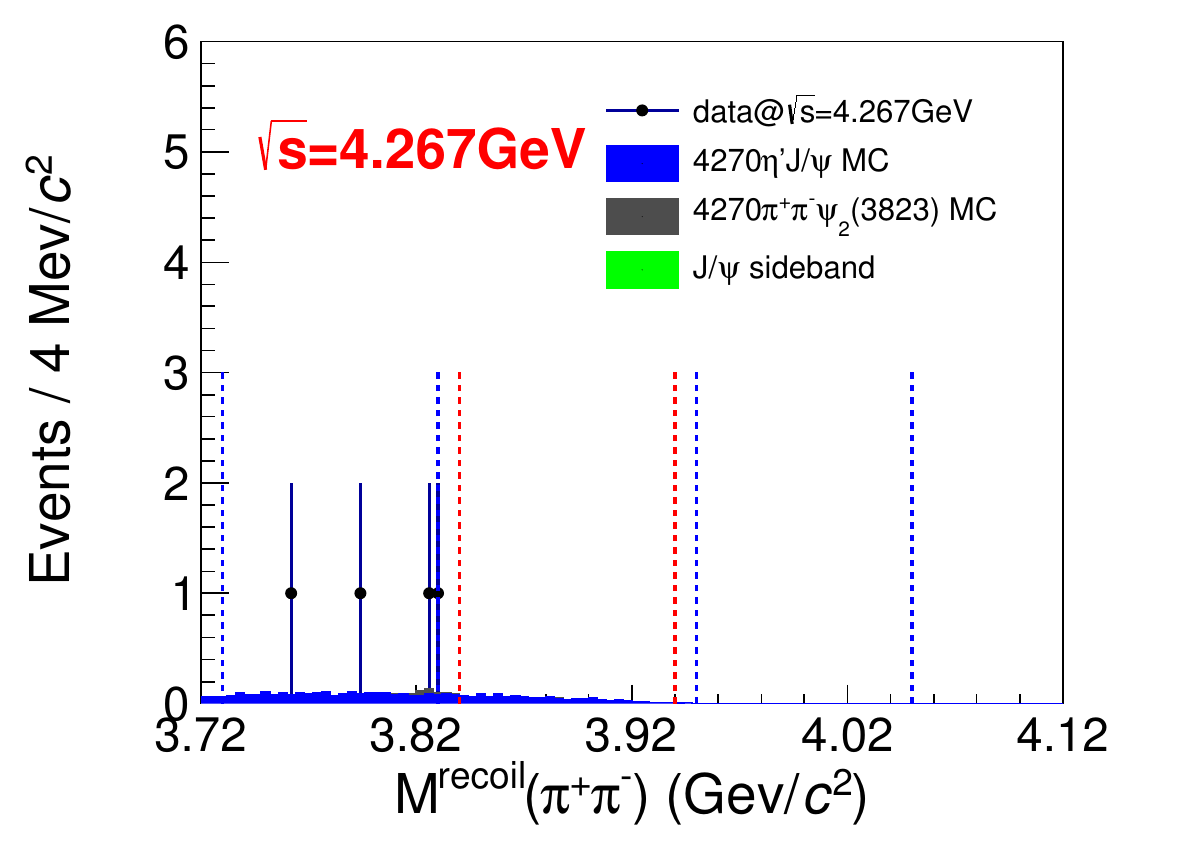}\hspace{5pt}
	\includegraphics[width=0.3\textwidth]{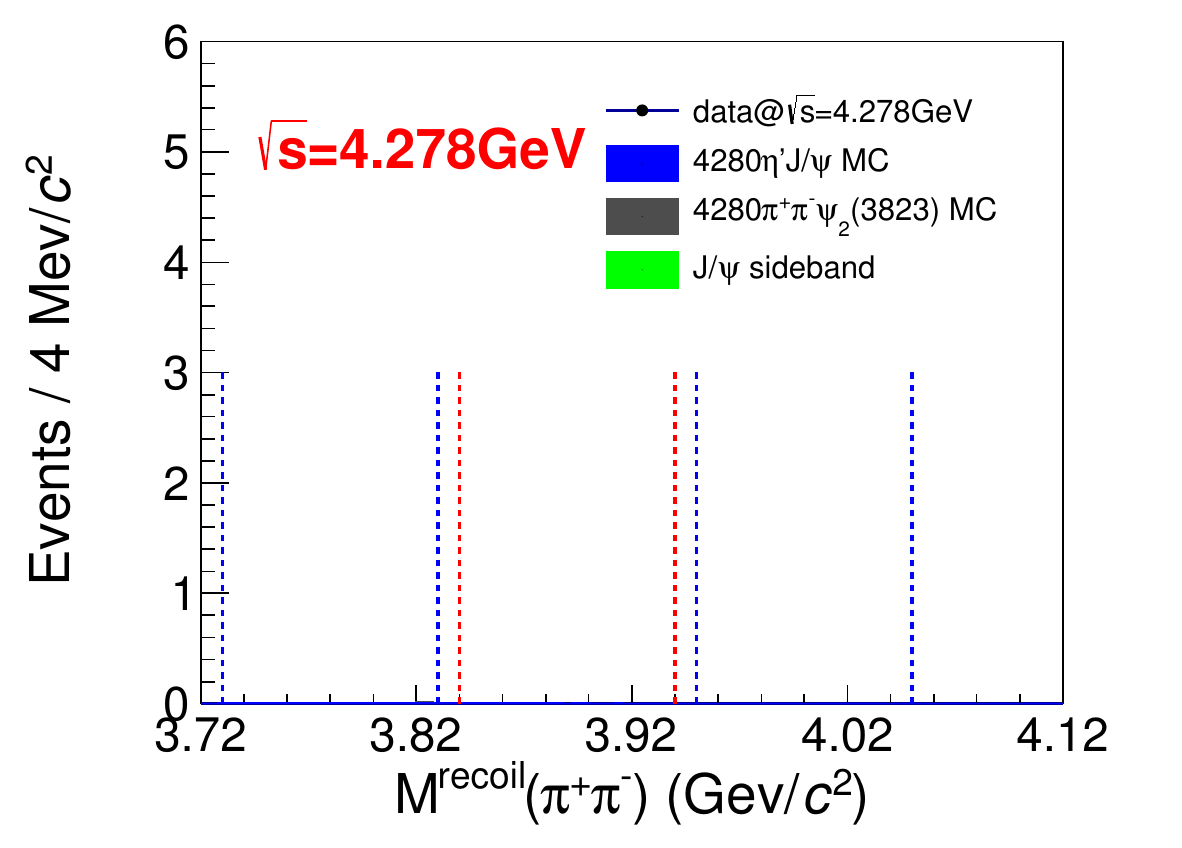}\hspace{5pt}
	\includegraphics[width=0.3\textwidth]{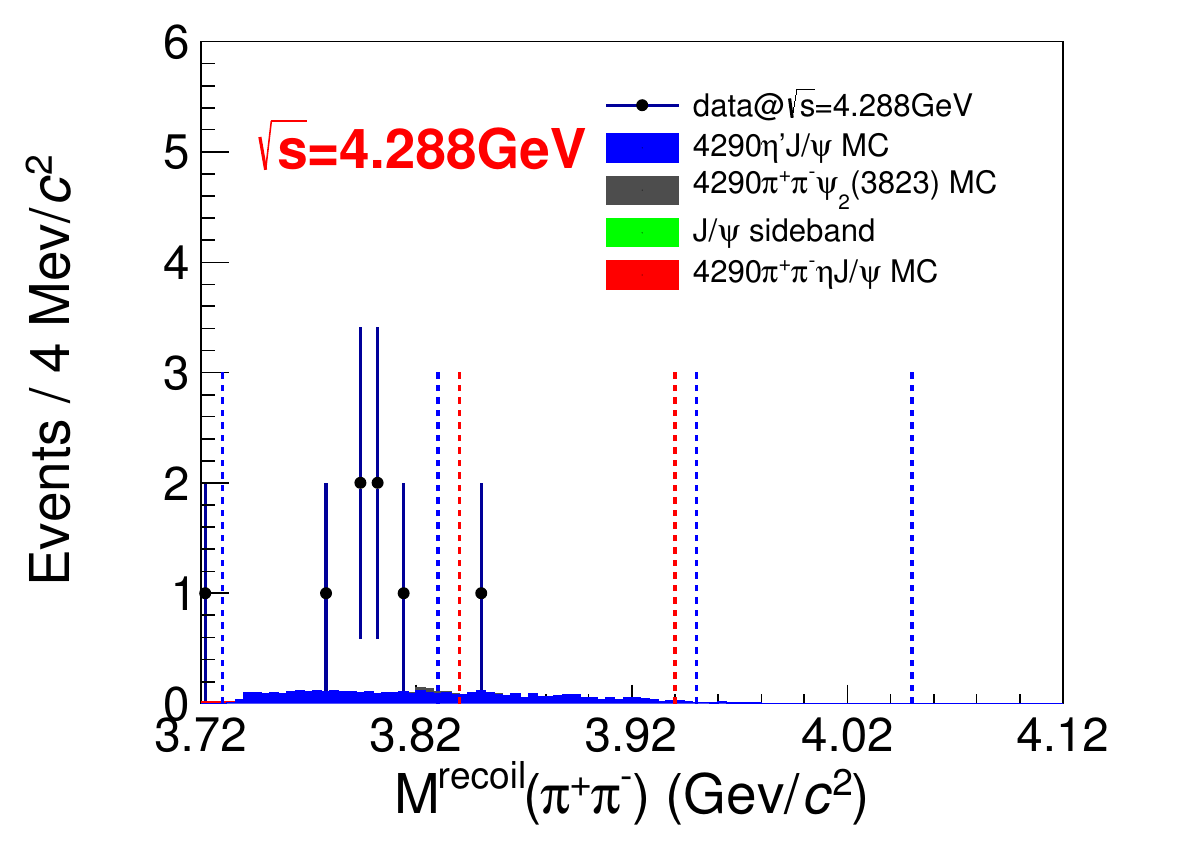}\hspace{5pt}
	\includegraphics[width=0.3\textwidth]{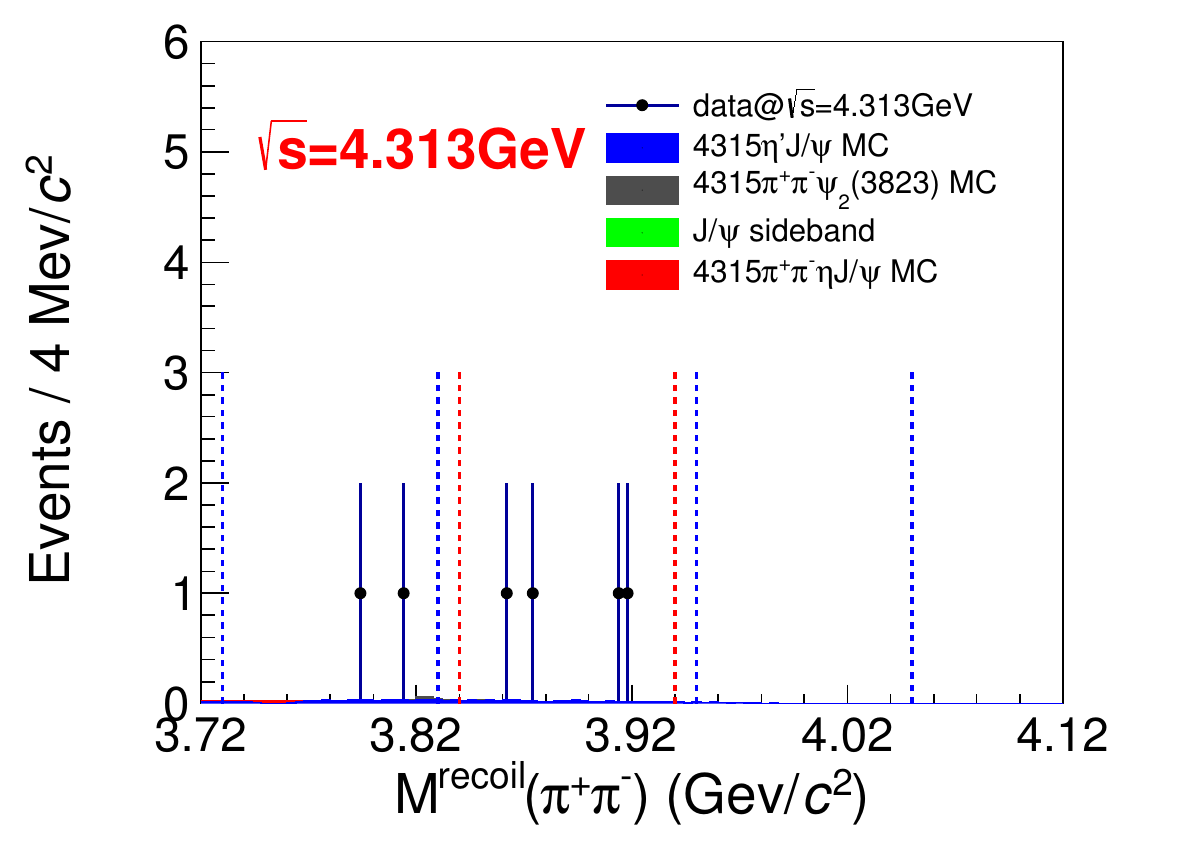}\hspace{5pt}
	\includegraphics[width=0.3\textwidth]{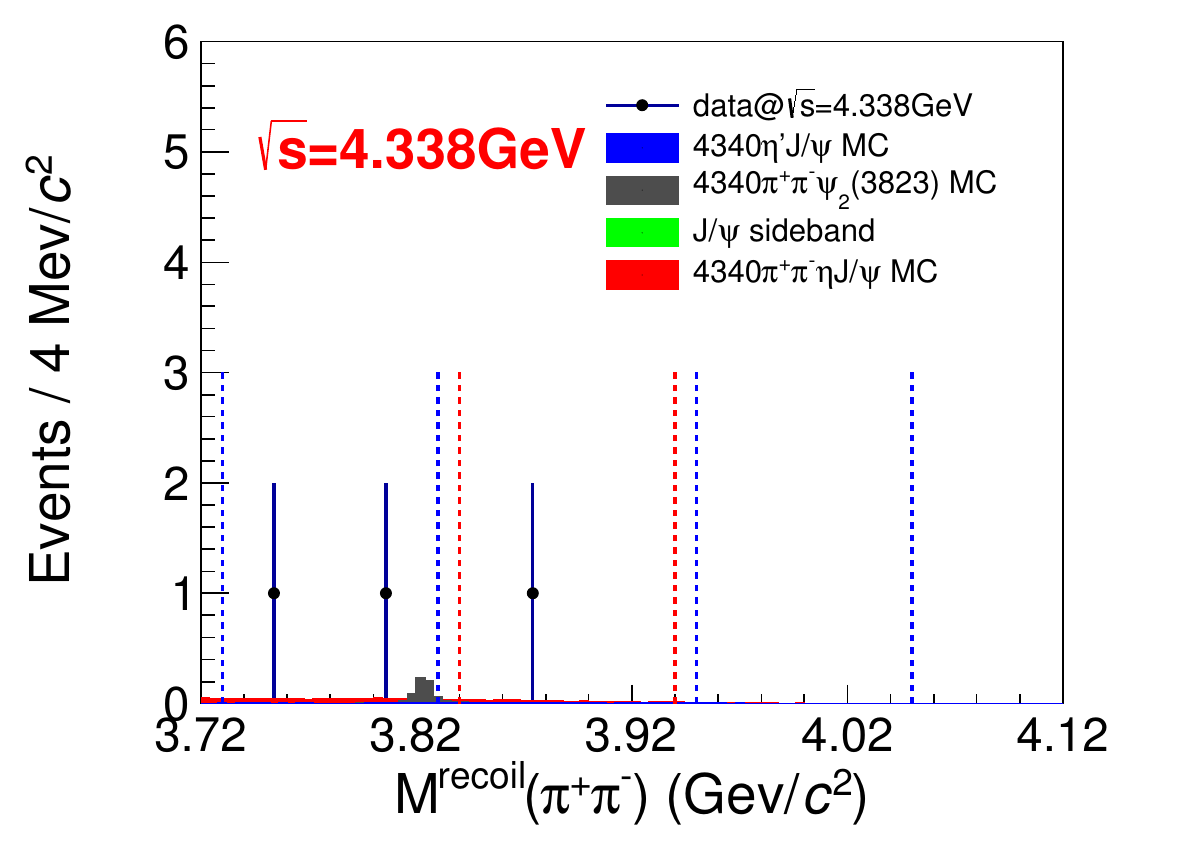}\hspace{5pt}
	\includegraphics[width=0.3\textwidth]{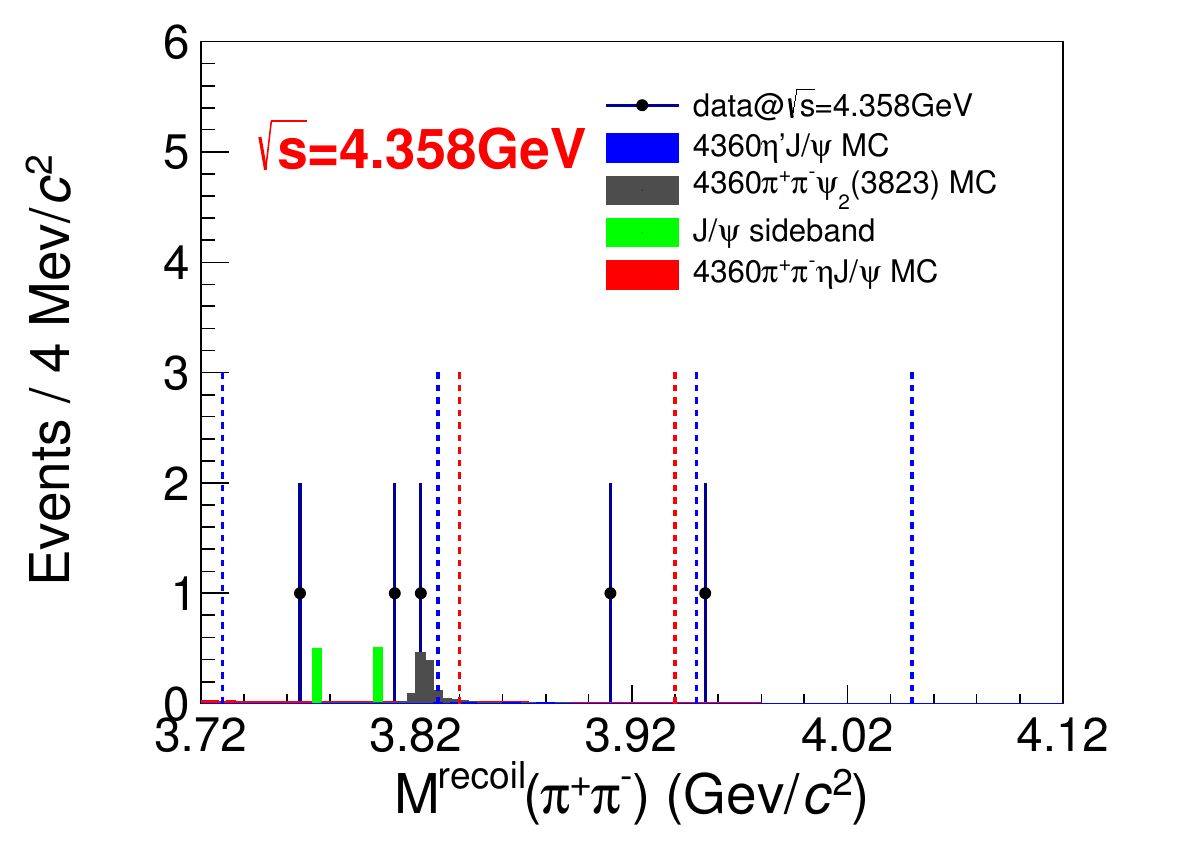}\hspace{5pt}
	\includegraphics[width=0.3\textwidth]{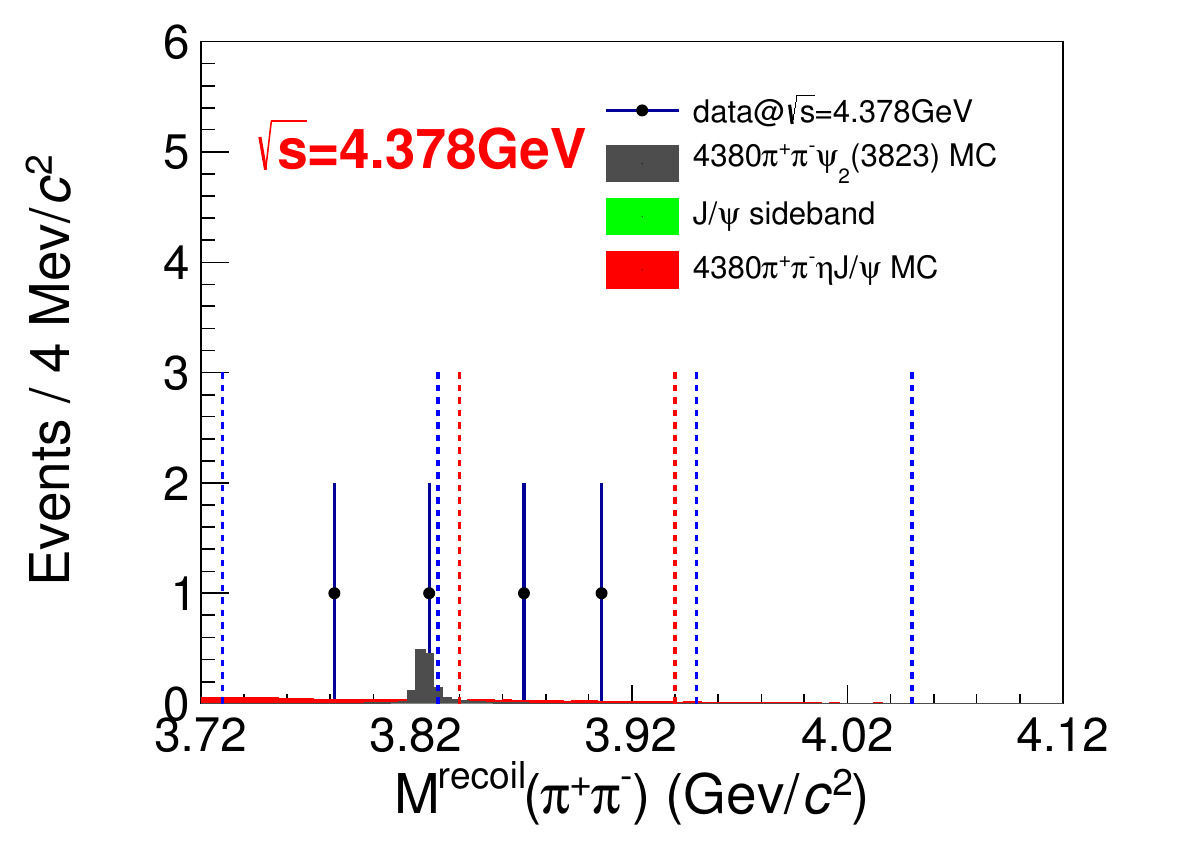}\hspace{5pt}
	\includegraphics[width=0.3\textwidth]{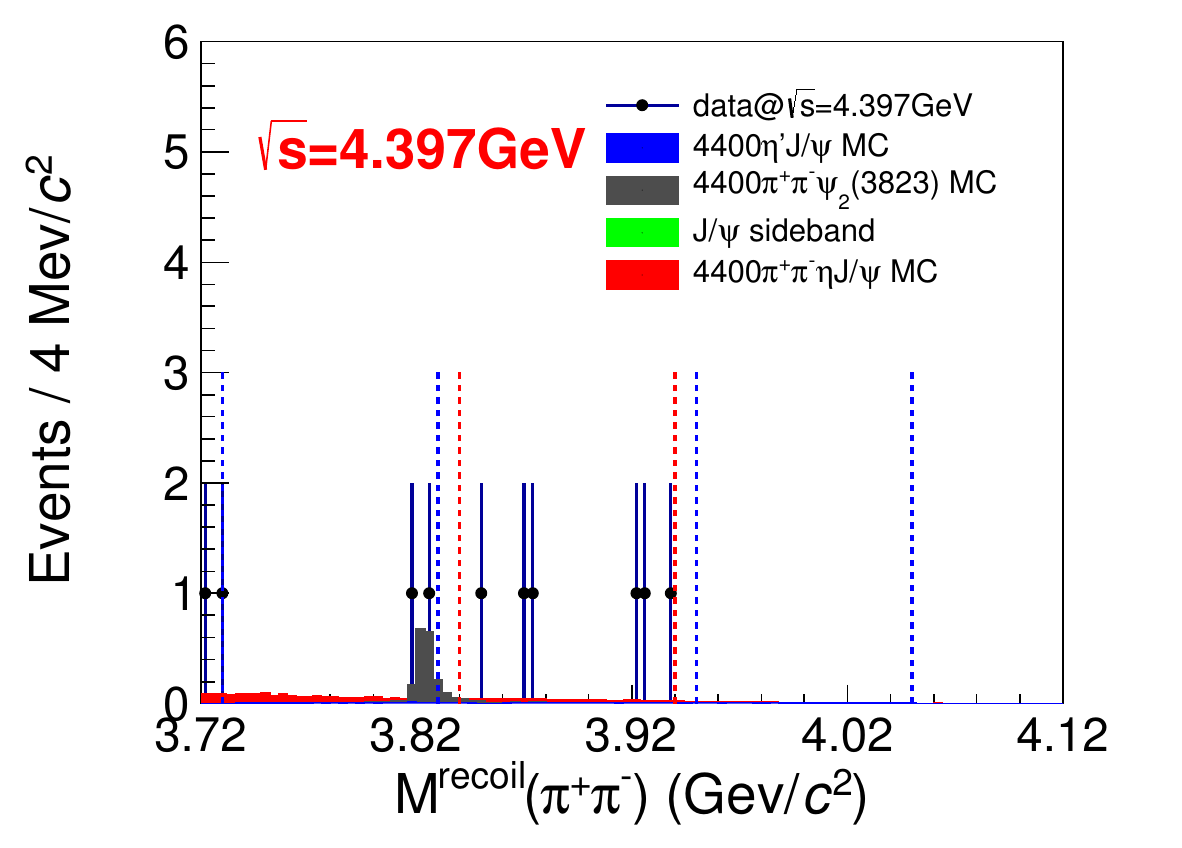}\hspace{5pt}
	\includegraphics[width=0.3\textwidth]{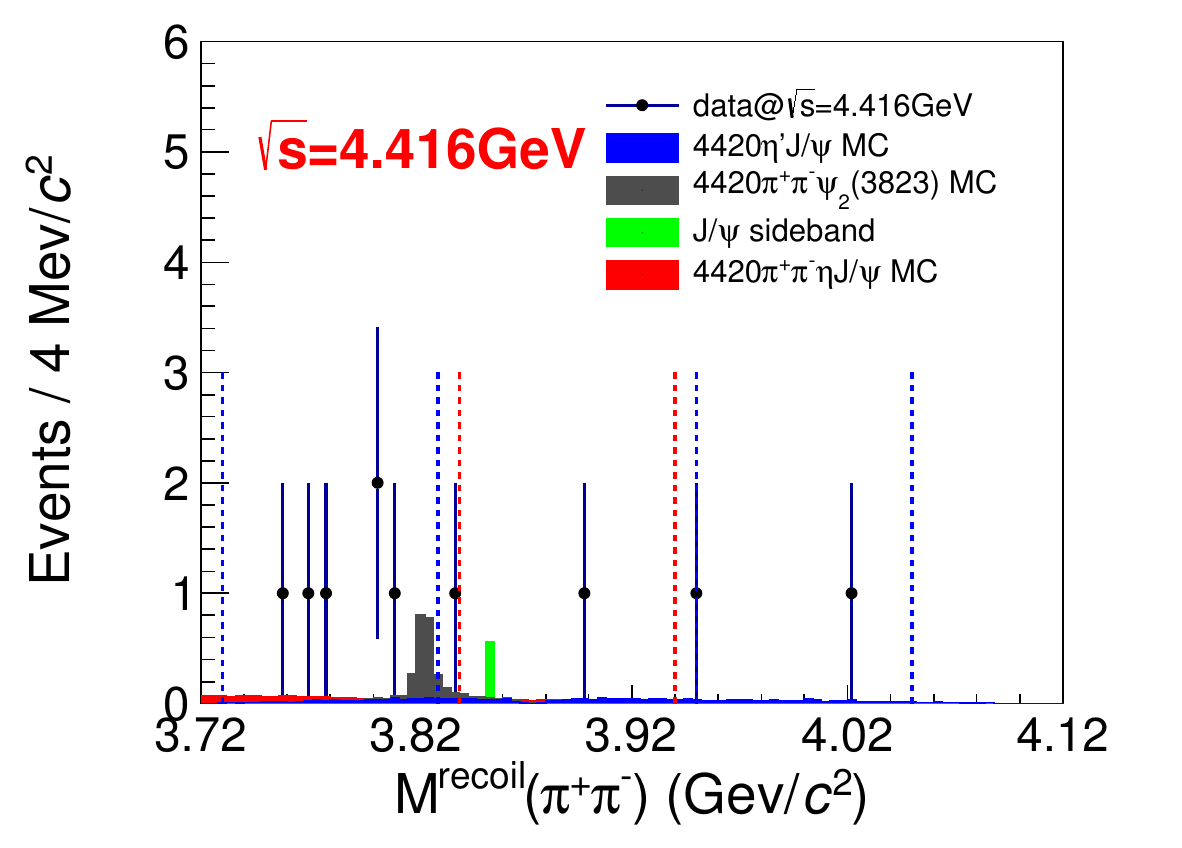}\hspace{5pt}
	
	\caption{The distributions of $M^{\rm recoil}(\pp)$ at each c.m.~energy from 4.18 GeV to 4.42 GeV. The black dots with error bars are data, the blue histogram is $\eta'\jpsi$ MC, the black histogram is $\pp \p$ MC, the red histogram is $\pp\eta\jpsi$ MC and the green histogram is $\jpsi$ sideband. The red dotted and blue dotted lines represent the signal and sideband regions of $X_5$, respectively.}
	\label{rmpp1}
\end{figure} 

\begin{figure}[H]
	\centering
	
	\includegraphics[width=0.3\textwidth]{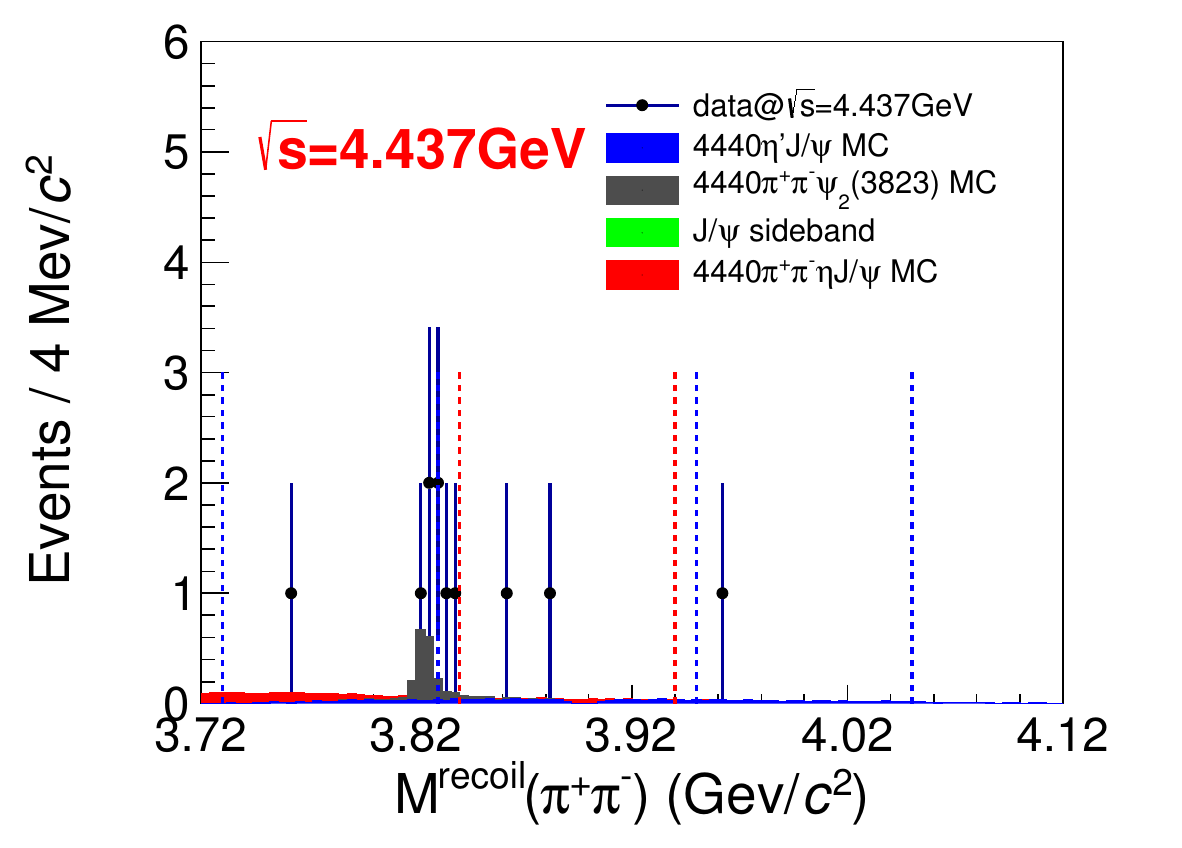}\hspace{5pt}
	\includegraphics[width=0.3\textwidth]{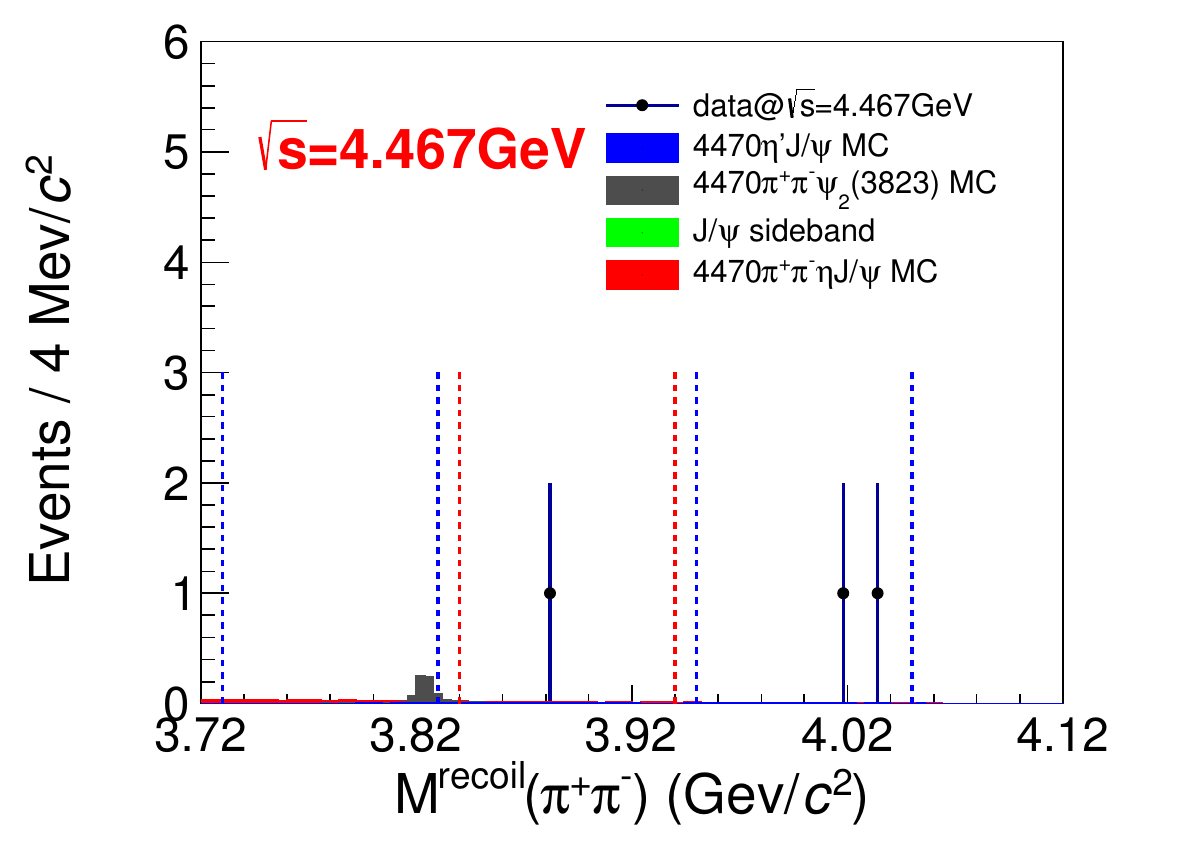}\hspace{5pt}
	\includegraphics[width=0.3\textwidth]{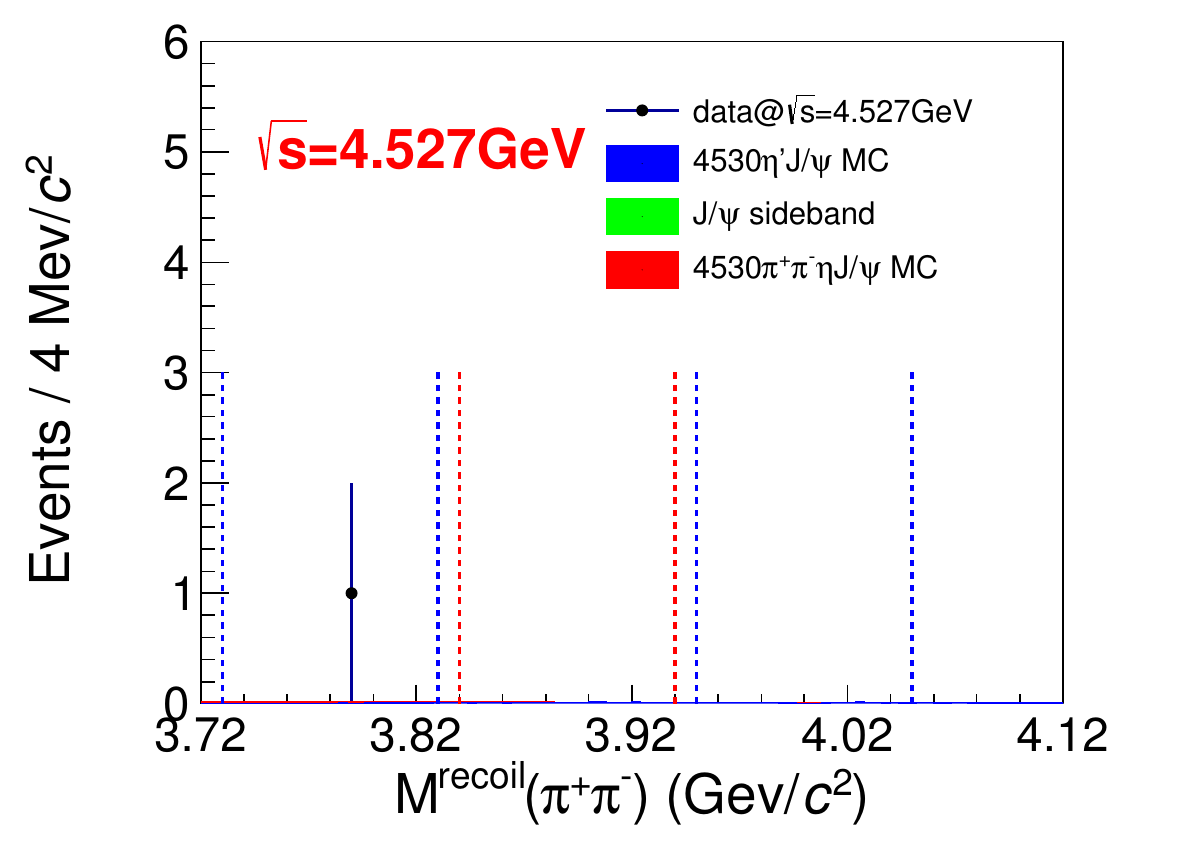}\hspace{5pt}
	\includegraphics[width=0.3\textwidth]{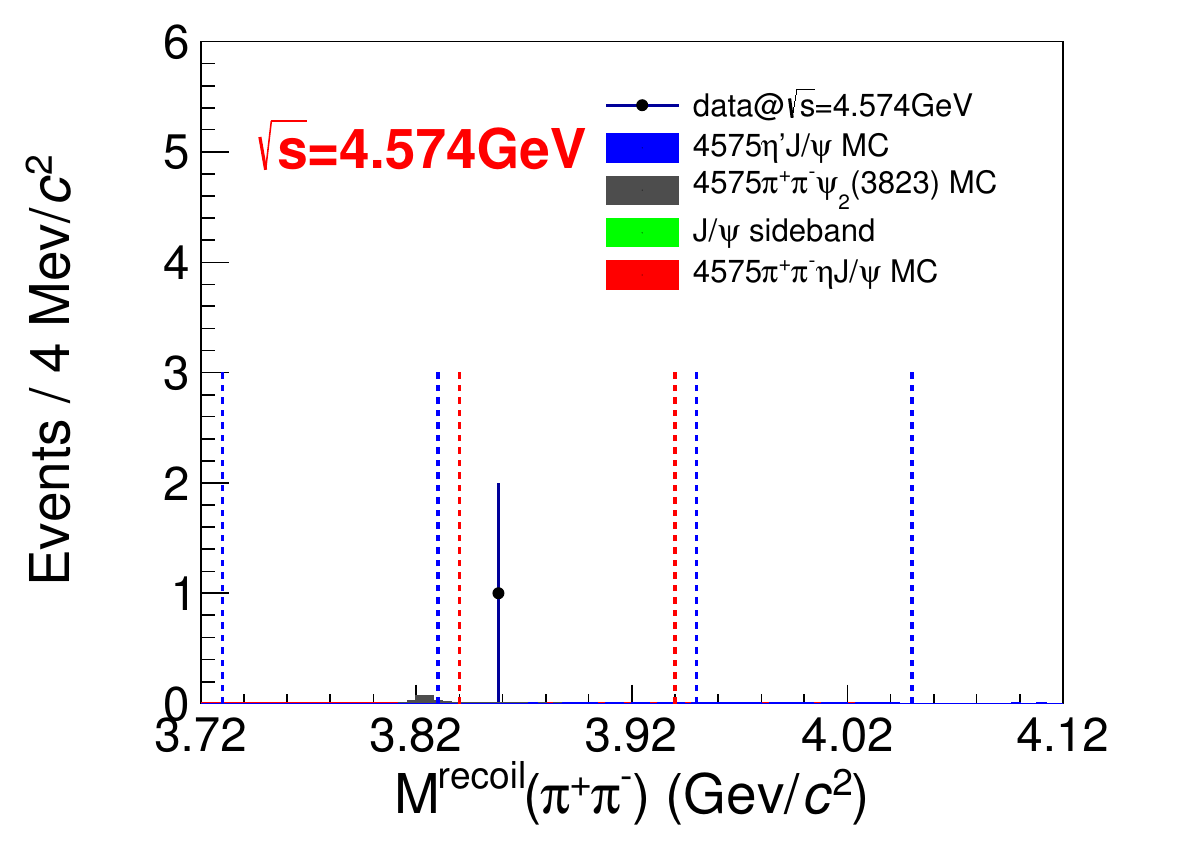}\hspace{5pt}
	\includegraphics[width=0.3\textwidth]{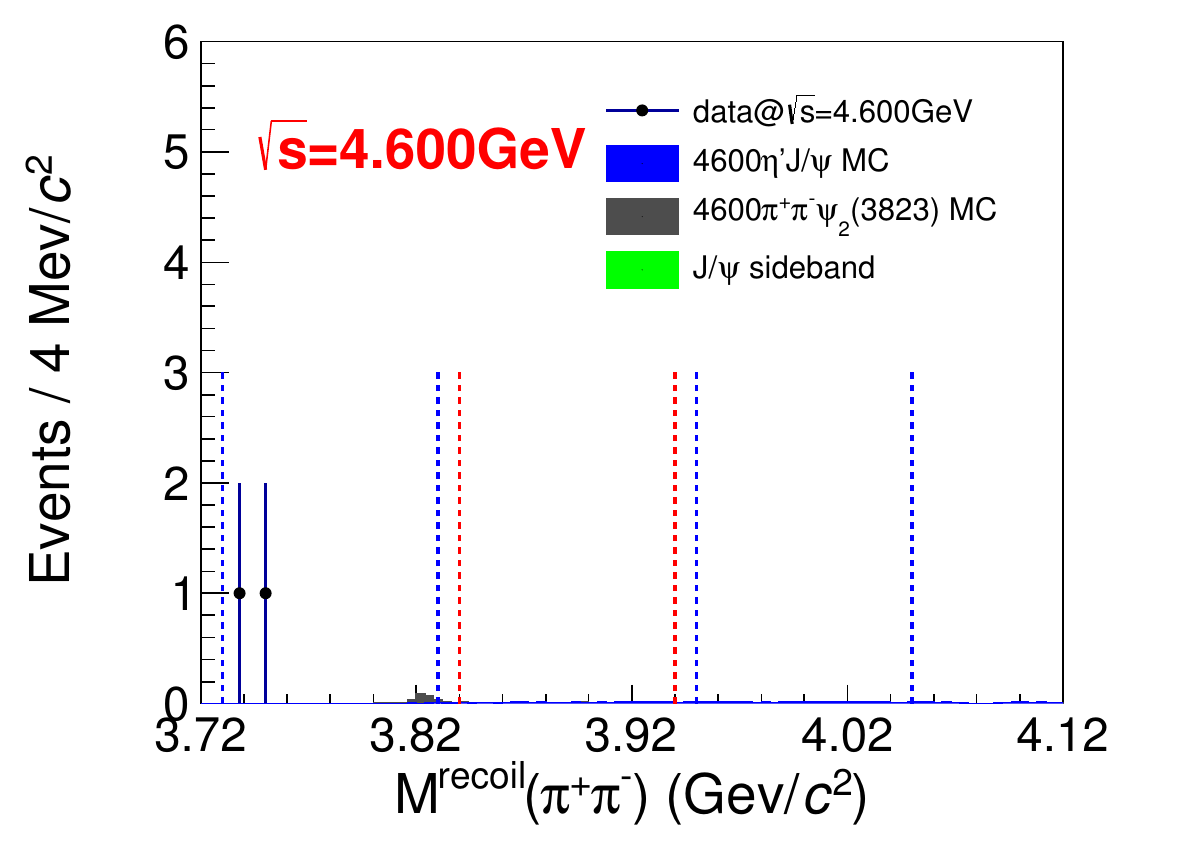}\hspace{5pt}
	\includegraphics[width=0.3\textwidth]{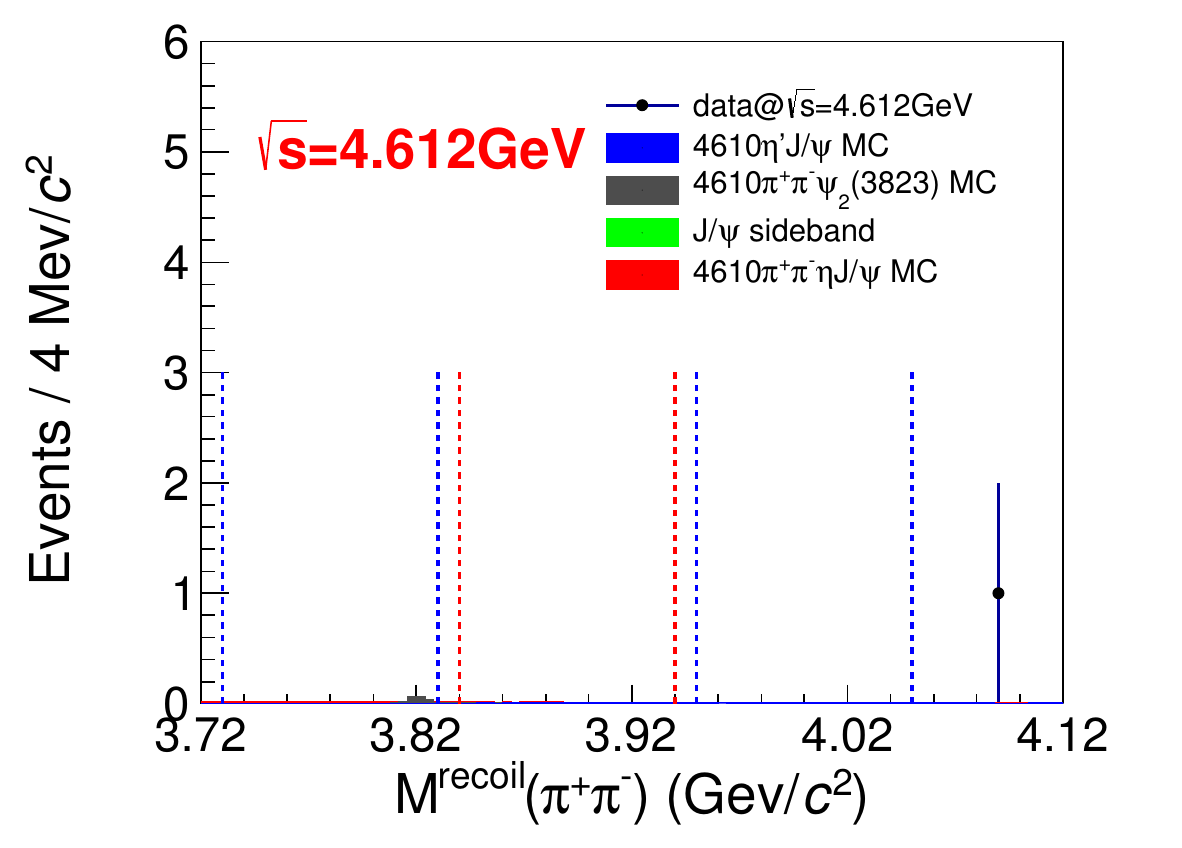}\hspace{5pt}
	\includegraphics[width=0.3\textwidth]{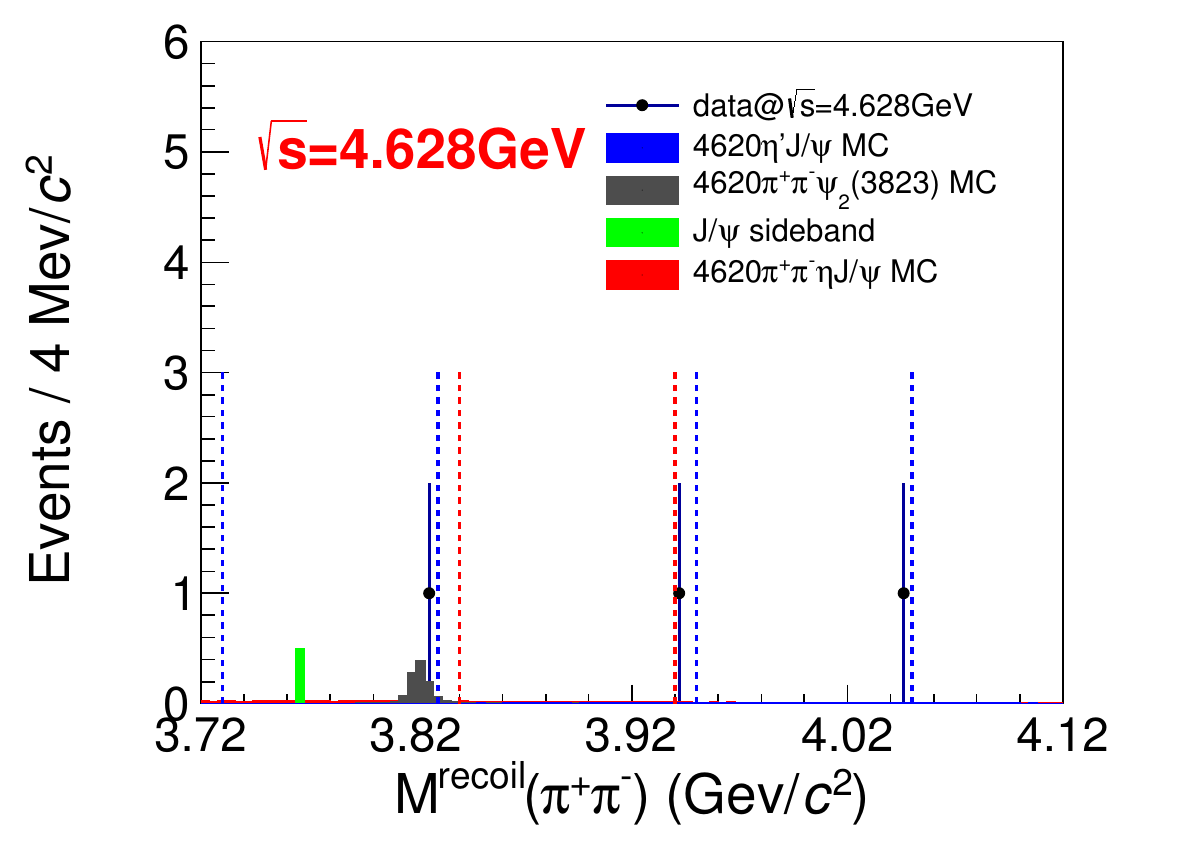}\hspace{5pt}
	\includegraphics[width=0.3\textwidth]{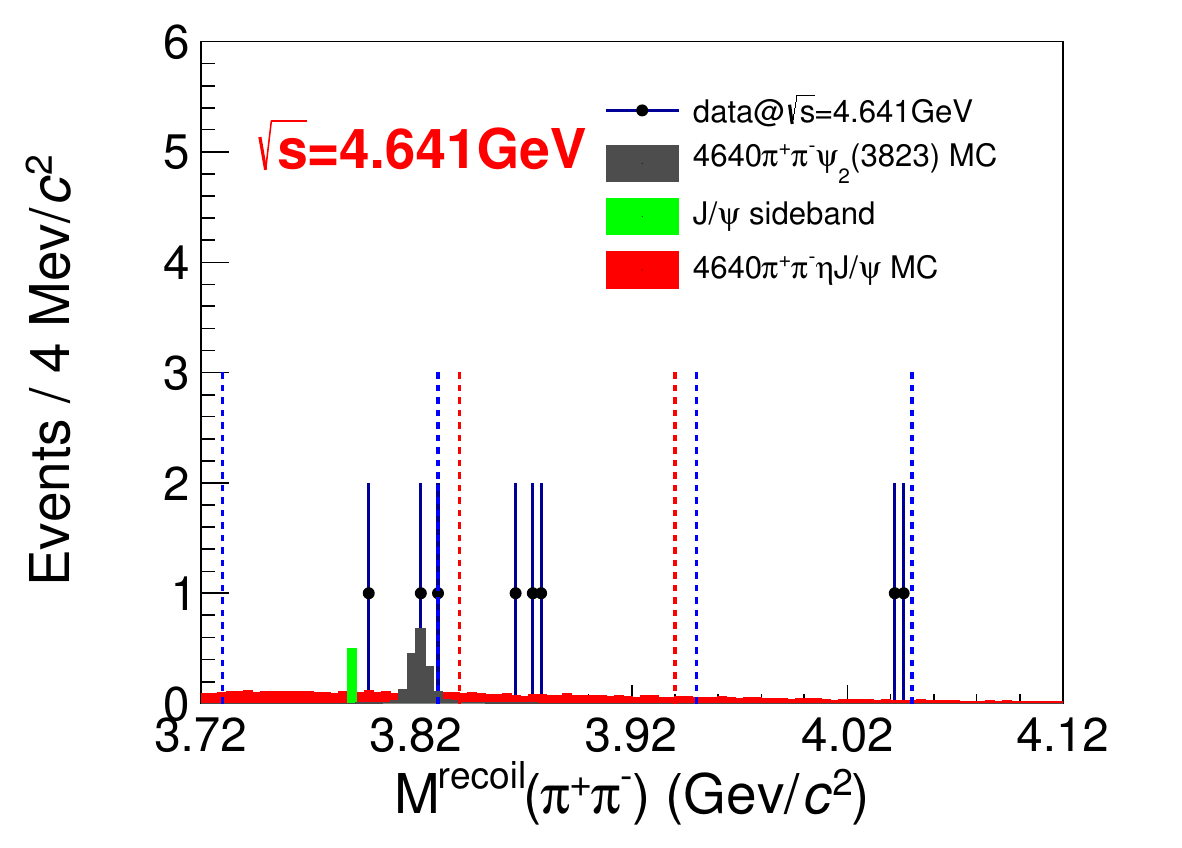}\hspace{5pt}
	\includegraphics[width=0.3\textwidth]{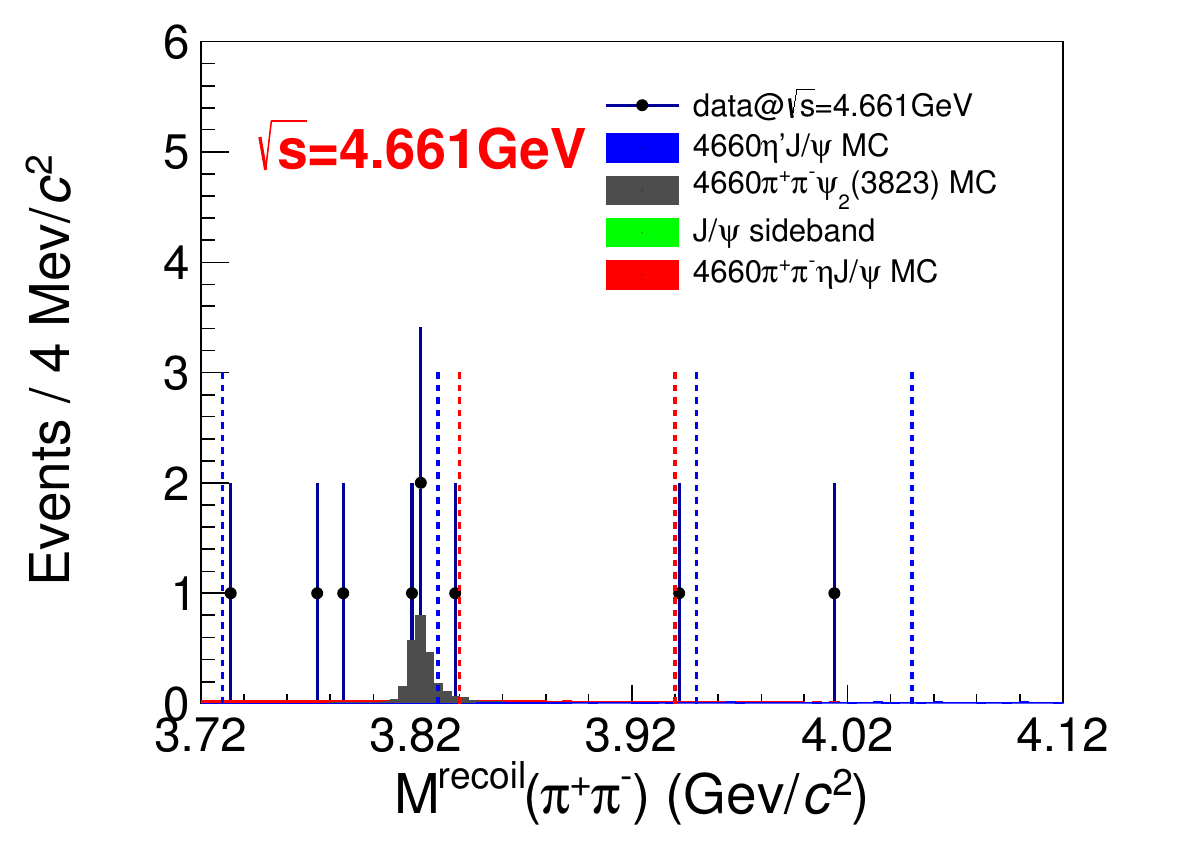}\hspace{5pt}
	\includegraphics[width=0.3\textwidth]{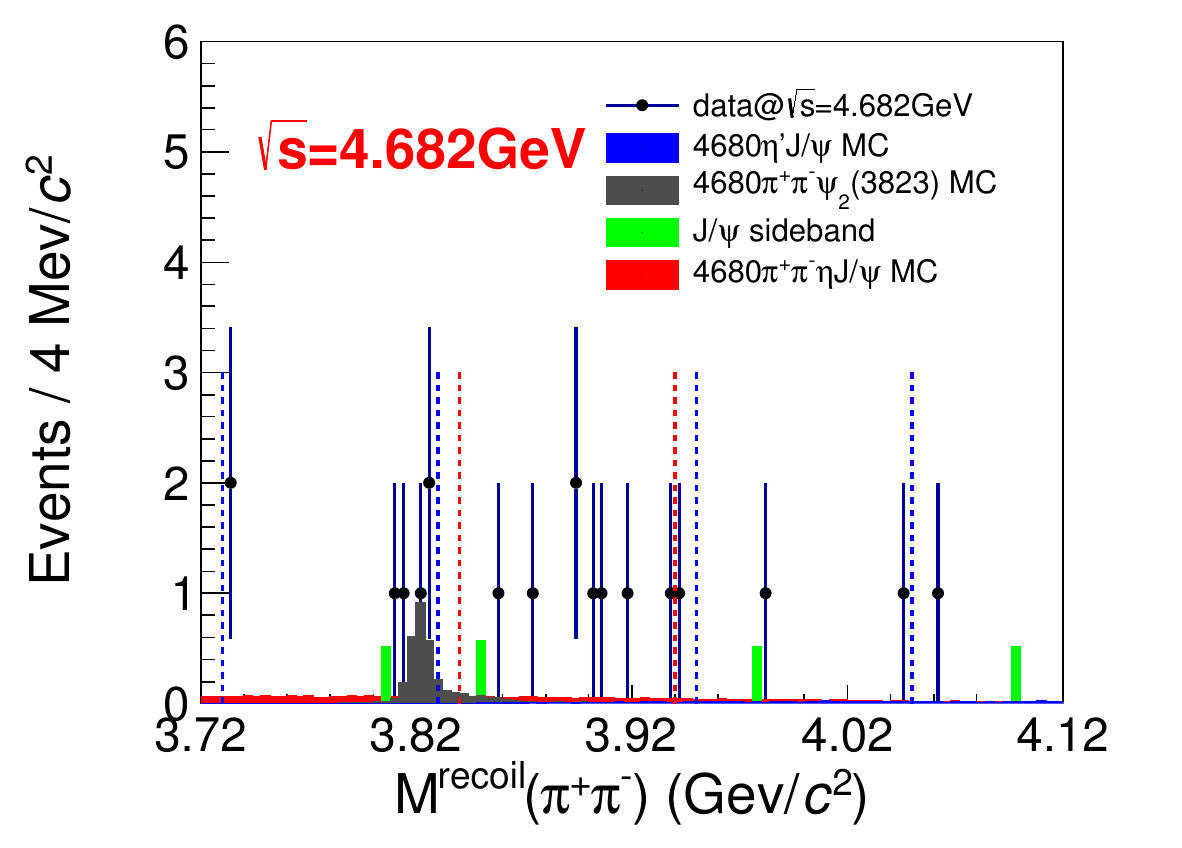}\hspace{5pt}
	\includegraphics[width=0.3\textwidth]{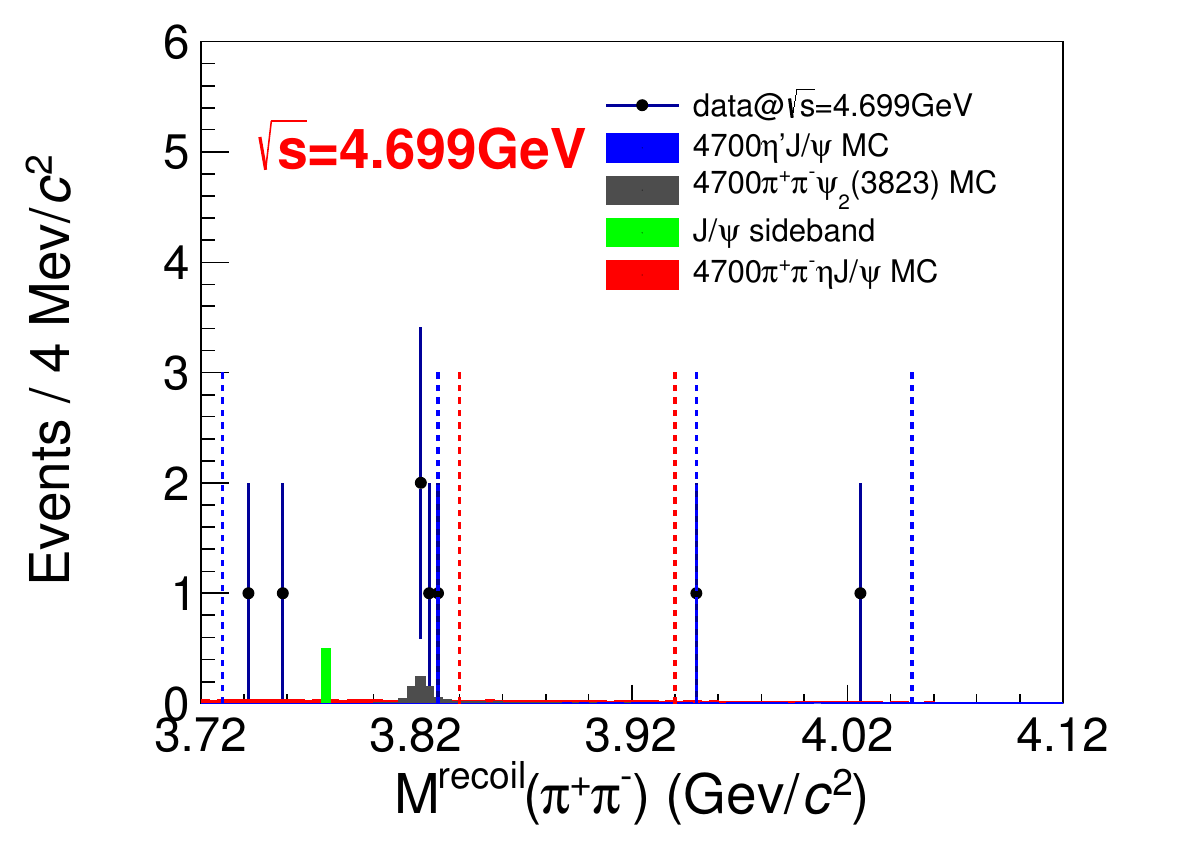}\hspace{5pt}
	\includegraphics[width=0.3\textwidth]{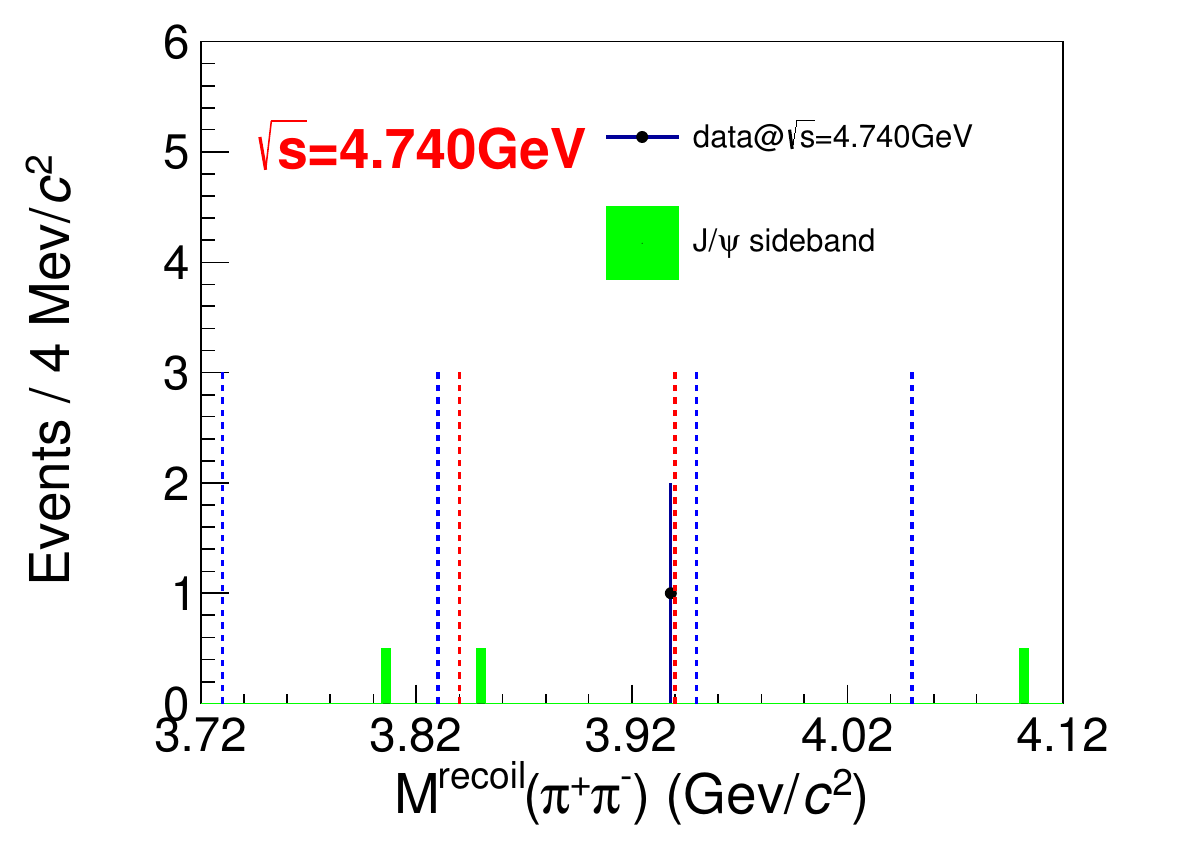}\hspace{5pt}
	\includegraphics[width=0.3\textwidth]{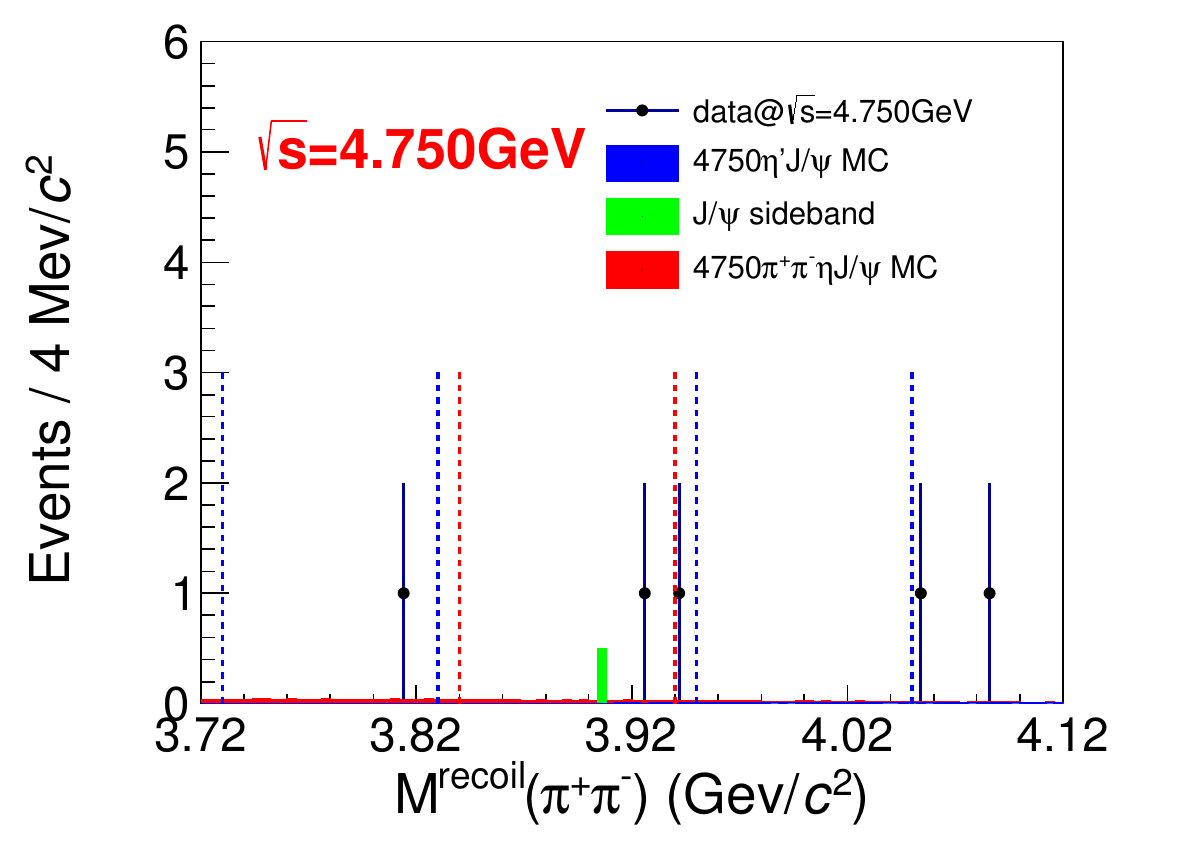}\hspace{5pt}
	\includegraphics[width=0.3\textwidth]{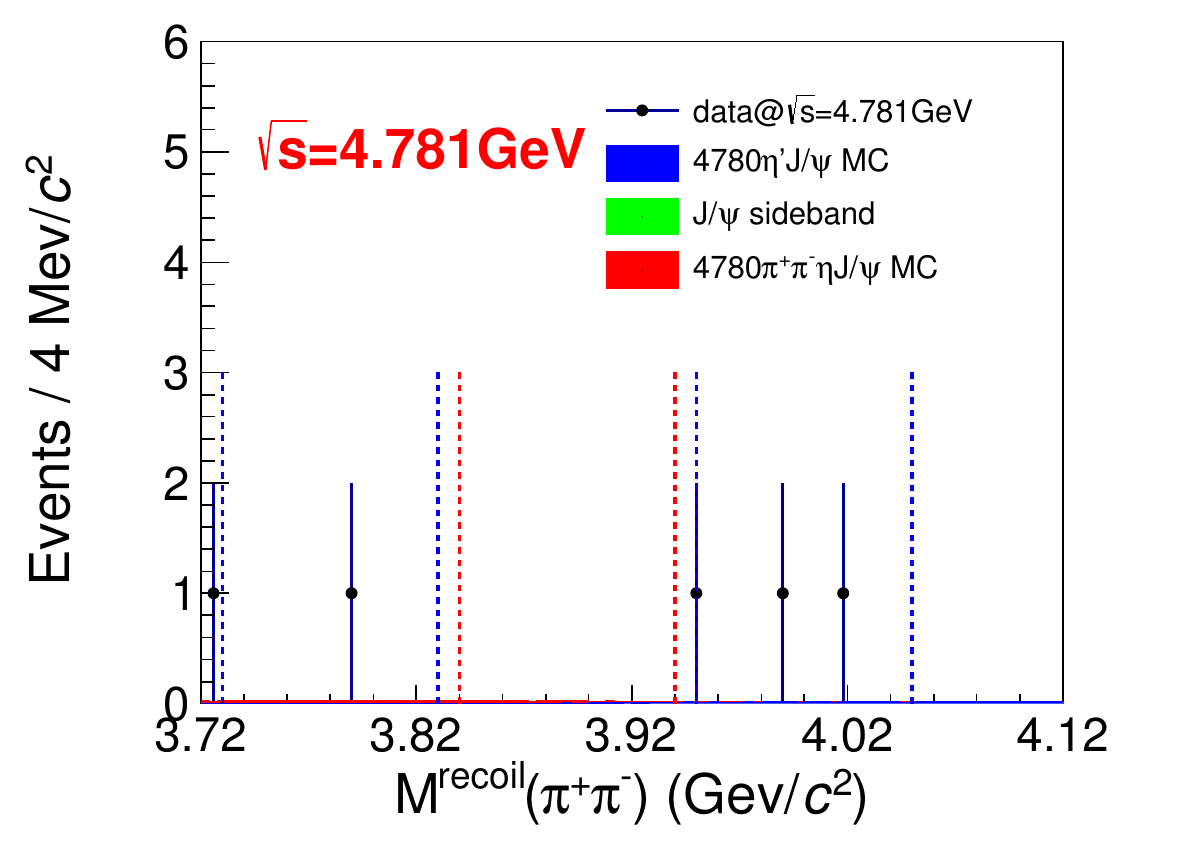}\hspace{5pt}
	\includegraphics[width=0.3\textwidth]{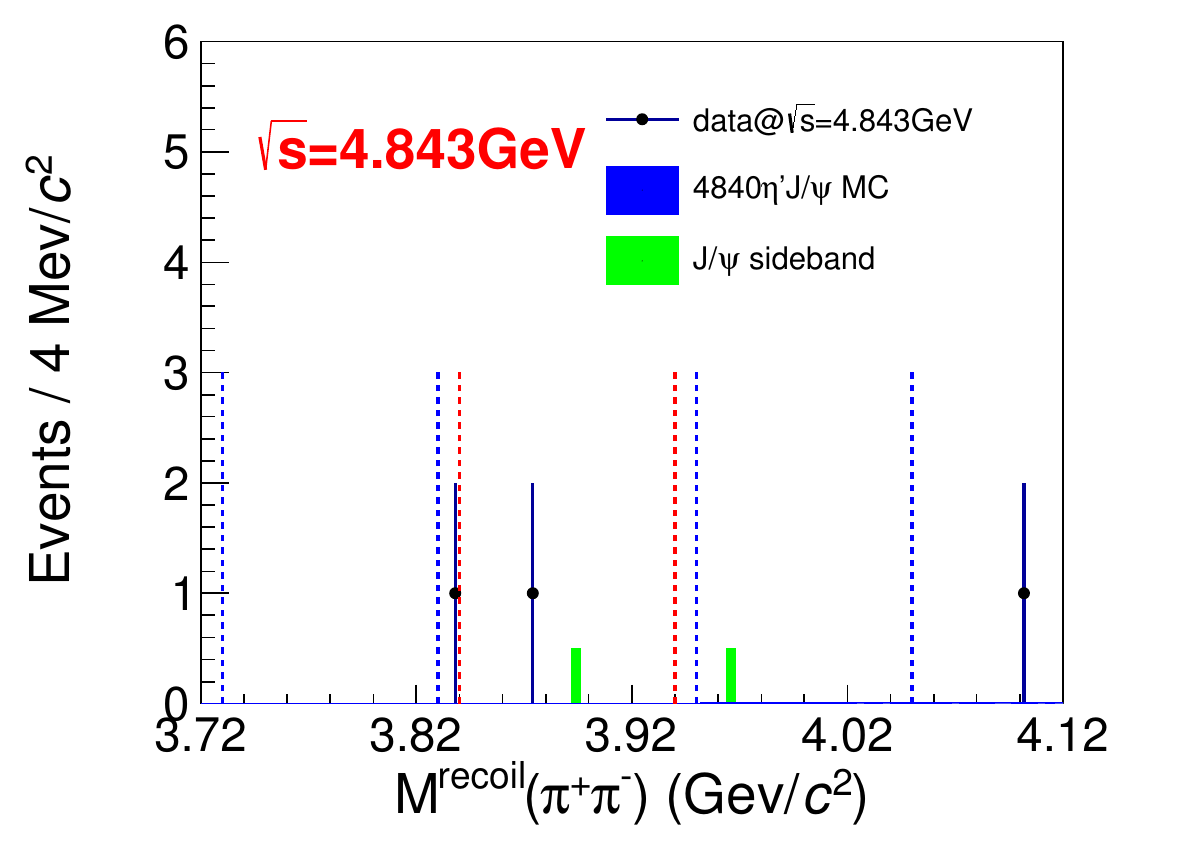}\hspace{5pt}
	\includegraphics[width=0.3\textwidth]{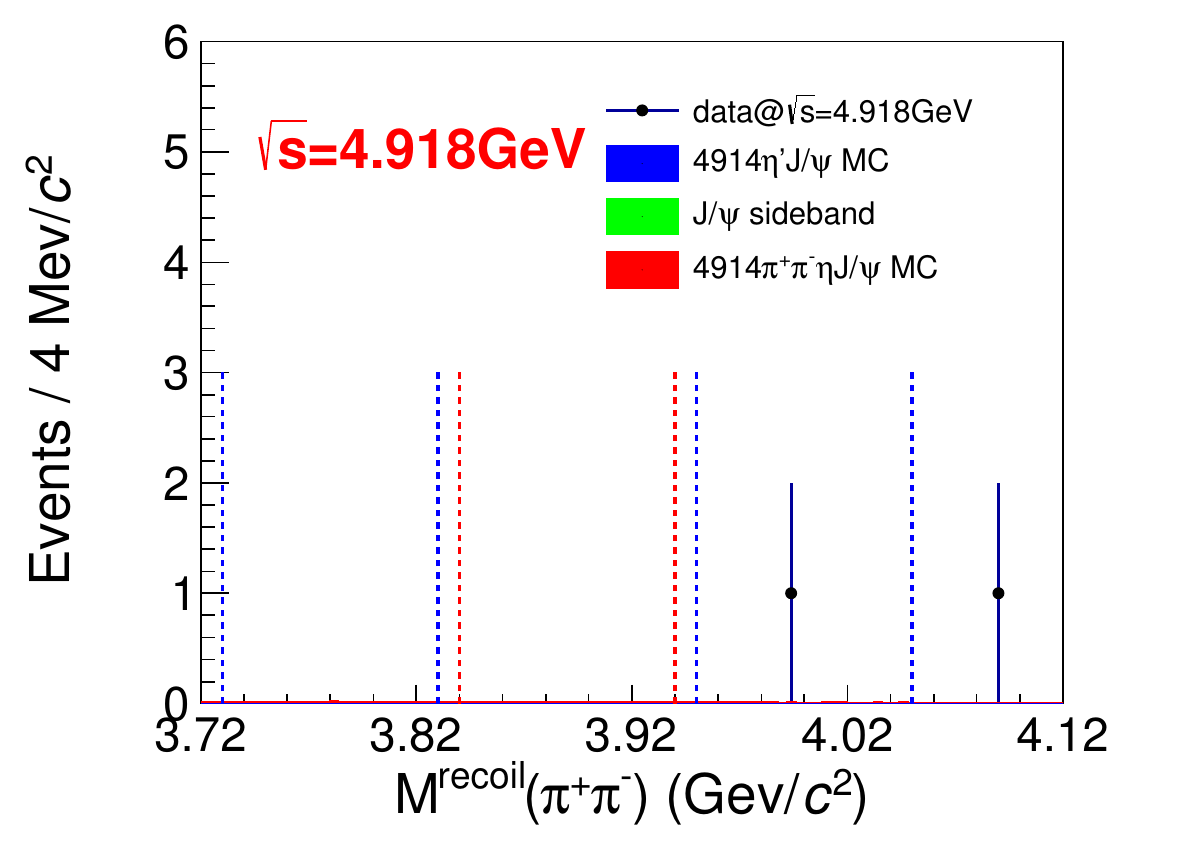}\hspace{5pt}
	\includegraphics[width=0.3\textwidth]{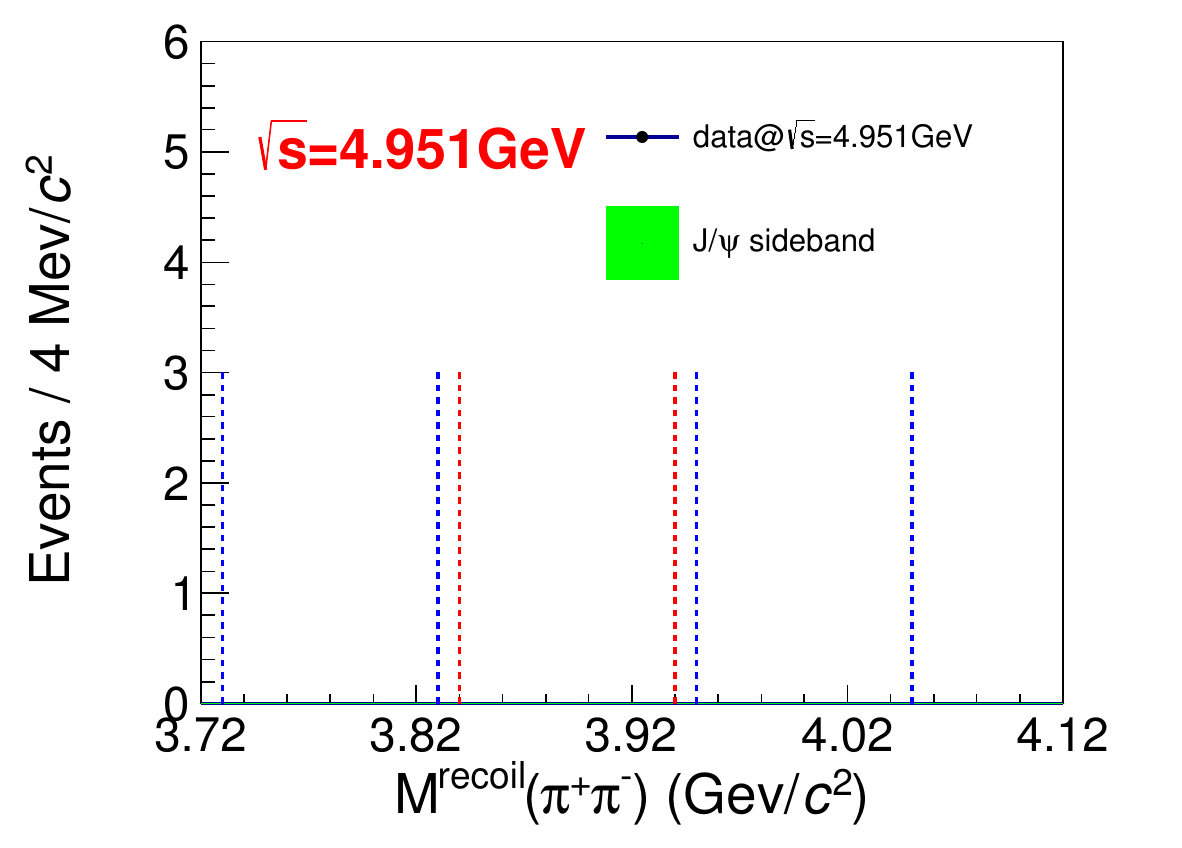}\hspace{5pt}
	\caption{The distributions of $M^{\rm recoil}(\pp)$ at each c.m.~energy from 4.44 GeV to 4.95 GeV. The black dots with error bars are data, the blue histogram is $\eta'\jpsi$ MC, the black histogram is $\pp \p$ MC, the red histogram is $\pp\eta\jpsi$ MC and the green histogram is $\jpsi$ sideband. The red dotted and blue dotted lines represent the signal and sideband regions of $X_5$, respectively.}
	\label{rmpp2}
\end{figure} 

\section{Summary of the results related to the cross section measurement}

\begin{table}[H]
	\caption{The integrated luminosities ($\mathcal{L}_{\rm int}$) of data, in pb$^{-1}$, the ISR correction factor (1+$\delta$), the vacuum polarization factor ($\frac{1}{|1-\Pi^2|}$), the signal efficiency ($\epsilon_{i}$), in $\%$, and the product $\sigma^{\rm Born}[\ee\to\pp X_i] \cdot \mathcal{B}[X_i\to\eta\jpsi]$ at each c.m.~energy, denoted as $(\sigma_{i}\cdot\mathcal{B})$, in pb, where $i$ is 1 to 3, corresponding to $X_1$ to $X_3$. The multiplicative systematic uncertainties have been taken into account.}
	\hspace{15pt}
	\centering
	\begin{tabular}{c c c c c c c c c c} 
		\hline 
		\hline
		$\sqrt{s}$~(GeV) & $\mathcal{L}_{\rm int}$&  $(1+\delta)$ & $\frac{1}{|1-\Pi^2|}$& $\epsilon_{1}$ & $(\sigma_{1}\cdot\mathcal{B})$ & $\epsilon_{2}$ & $(\sigma_{2}\cdot\mathcal{B})$ & $\epsilon_{3}$ & $(\sigma_{3}\cdot\mathcal{B})$\\
		\hline
		4.178 & 3194.5 & 0.66 & 1.055 & 3.8 & 1.25 &    5.8 & 1.33 &    7.4 & 1.37\\
		4.189 & 526.7 & 0.68 & 1.056 & 6.7 & 1.91 &     8.0 & 1.60 &    8.8 & 2.82\\
		4.199 & 526.0 & 0.69 & 1.056 & 9.4 & 1.24 &     10.4 & 2.86 &   10.7 & 3.43\\
		4.209 & 517.1 & 0.71 & 1.057 & 11.7 & 0.93 &    12.1 & 1.16 &   12.4 & 0.51\\
		4.219 & 514.6 & 0.72 & 1.056 & 13.5 & 0.62 &    13.3 & 1.27 &   13.5 & 1.74\\
		4.226 & 1056.4 & 0.72 & 1.056 & 15.5 & 0.75 &   15.1 & 1.96 &   15.6 & 2.27\\
		4.236 & 530.3 & 0.73 & 1.056 & 15.7 & 0.64 &    15.8 & 0.44 &   16.3 & 1.08\\
		4.244 & 538.1 & 0.74 & 1.056 & 16.7 & 1.63 &    16.4 & 1.90 &   16.3 & 2.57\\
		4.257 & 828.4 & 0.75 & 1.054 & 18.0 & 0.94 &    18.1 & 1.58 &   17.6 & 1.68\\
		4.267 & 531.1 & 0.75 & 1.053 & 18.3 & 0.41 &    17.7 & 1.33 &   17.9 & 1.39\\
		4.278 & 175.7 & 0.76 & 1.053 & 17.9 & 2.03 &    18.0 & 1.91 &   17.9 & 1.92\\
		4.288 & 502.4 & 0.76 & 1.053 & 18.8 & 0.81 &    18.7 & 0.35 &   18.6 & 1.95\\
		4.313 & 501.2 & 0.77 & 1.052 & 19.9 & 1.33 &    19.4 & 2.39 &   19.4 & 3.03\\
		4.338 & 505.0 & 0.78 & 1.051 & 20.1 & 1.06 &    20.6 & 0.68 &   19.8 & 1.30\\
		4.358 & 544.0 & 0.79 & 1.051 & 18.5 & 0.54 &    17.8 & 1.31 &   17.6 & 1.33\\
		4.378 & 522.7 & 0.79 & 1.051 & 20.7 & 0.96 &    20.5 & 1.63 &   19.7 & 1.64\\
		4.397 & 507.8 & 0.80 & 1.051 & 19.8 & 1.37 &    19.8 & 1.66 &   19.4 & 3.09\\
		4.416 & 1043.9 & 0.81 & 1.052 & 13.9 & 0.23 &   14.2 & 0.00 &   14.4 & 0.62\\
		4.437 & 569.9 & 0.80 & 1.054 & 17.6 & 1.15 &    17.9 & 4.03 &   18.6 & 3.80\\
		4.467 & 111.1 & 0.82 & 1.055 & 12.9 & 8.08 &    13.3 & 4.64 &   13.5 & 3.73\\
		4.527 & 112.1 & 0.83 & 1.054 & 14.8 & 3.57 &    14.9 & 2.55 &   15.0 & 3.58\\
		4.574 & 48.9 & 0.84 & 1.054 & 16.0 & 13.89 &    16.1 & 13.23 &  15.7 & 14.12\\
		4.600 & 586.9 & 0.85 & 1.055 & 16.1 & 0.61 &    16.3 & 0.26 &   16.4 & 0.16\\
		4.612 & 103.7 & 0.83 & 1.055 & 21.6 & 2.63 &    21.9 & 2.54 &   21.6 & 1.80\\
		4.628 & 521.5 & 0.83 & 1.054 & 21.8 & 0.49 &    21.2 & 0.82 &   21.3 & 0.58\\
		4.641 & 551.6 & 0.83 & 1.054 & 21.7 & 1.64 &    21.7 & 2.21 &   21.8 & 2.00\\
		4.661 & 529.4 & 0.84 & 1.054 & 21.8 & 0.41 &    21.7 & 0.25 &   21.9 & 0.19\\
		4.682 & 1667.4 & 0.84 & 1.054 & 21.8 & 0.31 &   21.9 & 0.95 &   21.4 & 1.05\\
		4.699 & 535.5 & 0.84 & 1.055 & 21.6 & 0.29 &    22.2 & 1.51 &   22.5 & 1.40\\
		4.740 & 163.9 & 0.84 & 1.055 & 22.3 & 1.60 &    22.4 & 0.99 &   21.9 & 1.65\\
		4.750 & 366.6 & 0.84 & 1.055 & 22.0 & 0.54 &    21.9 & 0.49 &   21.9 & 1.22\\
		4.781 & 511.5 & 0.85 & 1.055 & 22.1 & 0.51 &    22.1 & 0.03 &   21.8 & 0.05\\
		4.843 & 525.2 & 0.85 & 1.056 & 22.0 & 0.91 &    22.1 & 1.32 &   22.1 & 1.23\\
		4.918 & 207.8 & 0.86 & 1.056 & 21.6 & 1.27 &    22.0 & 0.94 &   22.2 & 0.61\\
		4.951 & 159.3 & 0.86 & 1.056 & 21.4 & 1.69 &    21.4 & 1.64 &   21.1 & 1.73\\
		\hline 
		\hline
	\end{tabular}
	\label{count_UL_x13}  
\end{table}

\begin{table}[H]
	\caption{The integrated luminosities ($\mathcal{L}_{\rm int}$) of data, in pb$^{-1}$, the ISR correction factor (1+$\delta$), the vacuum polarization factor ($\frac{1}{|1-\Pi^2|}$), the signal efficiency ($\epsilon_{i}$), in $\%$, and the product $\sigma^{\rm Born}[\ee\to\pp X_i] \cdot \mathcal{B}[X_i\to\eta\jpsi]$ at each c.m.~energy, denoted as $(\sigma_{i}\cdot\mathcal{B})$, in pb, where $i$ is 4 to 6, corresponding to $X_4$ to $X_6$. The multiplicative systematic uncertainties have been taken into account.}
	\hspace{15pt}
	\centering
	\begin{tabular}{c c c c c c c c c c} 
		\hline 
		\hline
		$\sqrt{s}$~(GeV) & $\mathcal{L}_{\rm int}$ &  $(1+\delta)$ & $\frac{1}{|1-\Pi^2|}$& $\epsilon_{4}$ & $(\sigma_{4}\cdot\mathcal{B})$ & $\epsilon_{5}$ & $(\sigma_{5}\cdot\mathcal{B})$ & $\epsilon_{6}$ & $(\sigma_{6}\cdot\mathcal{B})$\\
		\hline
		4.178 & 3194.5 & 0.66 & 1.055 & 0.7 & 7.98 &    3.2 & 2.57 &    5.3 & 2.03\\
		4.189 & 526.7 & 0.68 & 1.056 & 1.3 & 11.54 &    4.1 & 4.83 &    5.9 & 2.61\\
		4.199 & 526.0 & 0.69 & 1.056 & 3.6 & 8.50 &     5.7 & 7.05 &    7.5 & 4.76\\
		4.209 & 517.1 & 0.71 & 1.057 & 6.5 & 1.33 &     7.8 & 1.38 &    8.9 & 1.82\\
		4.219 & 514.6 & 0.72 & 1.056 & 8.9 & 0.93 &     9.7 & 1.02 &    10.6 & 1.76\\
		4.226 & 1056.4 & 0.72 & 1.056 & 11.5 & 1.54 &   11.5 & 1.76 &   12.2 & 2.62\\
		4.236 & 530.3 & 0.73 & 1.056 & 13.1 & 0.61 &    12.9 & 0.70 &   13.1 & 0.57\\
		4.244 & 538.1 & 0.74 & 1.056 & 14.1 & 2.22 &    14.1 & 2.48 &   13.8 & 2.36\\
		4.257 & 828.4 & 0.75 & 1.054 & 15.8 & 1.40 &    16.1 & 1.51 &   15.4 & 1.86\\
		4.267 & 531.1 & 0.75 & 1.053 & 15.9 & 0.14 &    16.8 & 0.14 &   15.7 & 1.48\\
		4.278 & 175.7 & 0.76 & 1.053 & 16.1 & 1.93 &    17.2 & 1.98 &   15.9 & 2.17\\
		4.288 & 502.4 & 0.76 & 1.053 & 17.2 & 0.22 &    17.4 & 0.24 &   17.0 & 0.18\\
		4.313 & 501.2 & 0.77 & 1.052 & 18.3 & 2.16 &    18.3 & 2.39 &   18.0 & 2.53\\
		4.338 & 505.0 & 0.78 & 1.051 & 19.5 & 0.83 &    19.2 & 0.93 &   19.2 & 0.68\\
		4.358 & 544.0 & 0.79 & 1.051 & 17.5 & 0.88 &    20.2 & 0.85 &   17.1 & 1.54\\
		4.378 & 522.7 & 0.79 & 1.051 & 20.1 & 1.23 &    19.7 & 1.40 &   19.3 & 1.75\\
		4.397 & 507.8 & 0.80 & 1.051 & 19.7 & 2.79 &    19.6 & 3.11 &   19.3 & 3.19\\
		4.416 & 1043.9 & 0.81 & 1.052 & 14.4 & 0.18 &   19.4 & 0.15 &   14.2 & 0.27\\
		4.437 & 569.9 & 0.80 & 1.054 & 17.8 & 0.76 &    18.2 & 0.85 &   18.9 & 3.82\\
		4.467 & 111.1 & 0.82 & 1.055 & 12.0 & 6.15 &    17.5 & 4.81 &   13.0 & 4.58\\
		4.527 & 112.1 & 0.83 & 1.054 & 13.7 & 2.52 &    19.6 & 2.00 &   14.0 & 2.69\\
		4.574 & 48.9 & 0.84 & 1.054 & 15.5 & 12.28 &    21.1 & 10.11 &  15.4 & 14.40\\
		4.600 & 586.9 & 0.85 & 1.055 & 15.7 & 0.33 &    21.5 & 0.27 &   16.0 & 0.05\\
		4.612 & 103.7 & 0.83 & 1.055 & 21.2 & 2.31 &    21.2 & 2.62 &   20.9 & 1.94\\
		4.628 & 521.5 & 0.83 & 1.054 & 21.2 & 0.37 &    21.1 & 0.42 &   21.1 & 1.28\\
		4.641 & 551.6 & 0.83 & 1.054 & 20.8 & 1.22 &    21.2 & 1.36 &   20.9 & 1.98\\
		4.661 & 529.4 & 0.84 & 1.054 & 21.3 & 0.13 &    21.2 & 0.15 &   21.3 & 0.94\\
		4.682 & 1667.4 & 0.84 & 1.054 & 21.3 & 0.76 &   21.6 & 0.86 &   21.2 & 1.29\\
		4.699 & 535.5 & 0.84 & 1.055 & 21.3 & 0.01 &    21.5 & 0.02 &   21.3 & 2.09\\
		4.740 & 163.9 & 0.84 & 1.055 & 21.6 & 2.63 &    21.6 & 2.98 &   21.5 & 3.13\\
		4.750 & 366.6 & 0.84 & 1.055 & 21.5 & 0.97 &    21.5 & 1.11 &   21.7 & 1.50\\
		4.781 & 511.5 & 0.85 & 1.055 & 21.1 & 0.20 &    21.3 & 0.23 &   21.1 & 0.59\\
		4.843 & 525.2 & 0.85 & 1.056 & 21.4 & 0.81 &    21.9 & 0.91 &   21.5 & 1.27\\
		4.918 & 207.8 & 0.86 & 1.056 & 21.5 & 0.86 &    21.5 & 0.98 &   21.1 & 0.68\\
		4.951 & 159.3 & 0.86 & 1.056 & 21.2 & 1.46 &    21.0 & 1.68 &   21.3 & 1.72\\
		\hline 
		\hline
	\end{tabular}
	\label{count_UL_x46}  
\end{table}

\begin{table}[H]
	\caption{The integrated luminosities ($\mathcal{L}_{\rm int}$) of data, in pb$^{-1}$, the ISR correction factor (1+$\delta$), the vacuum polarization factor ($\frac{1}{|1-\Pi^2|}$), the signal efficiency ($\epsilon_{i}$), in $\%$, and the product $\sigma^{\rm Born}[\ee\to\pp X_i] \cdot \mathcal{B}[X_i\to\eta\jpsi]$ at each c.m.~energy, denoted as $(\sigma_{i}\cdot\mathcal{B})$, in pb, where $i$ is 7 to 9, corresponding to $X_7$ to $X_9$. The multiplicative systematic uncertainties have been taken into account.}
	\hspace{15pt}
	\centering
	\begin{tabular}{c c c c c c c c c c} 
		\hline 
		\hline
		$\sqrt{s}$~(GeV) & $\mathcal{L}_{\rm int}$ & $(1+\delta)$ & $\frac{1}{|1-\Pi^2|}$& $\epsilon_{7}$ & $(\sigma_{7}\cdot\mathcal{B})$ & $\epsilon_{8}$ & $(\sigma_{8}\cdot\mathcal{B})$ & $\epsilon_{9}$ & $(\sigma_{9}\cdot\mathcal{B})$\\
		\hline
		4.199 & 526.0 & 0.69 & 1.056 & 0.6 & 27.93 &    3.1 & 8.68 &    5.4 & 10.64\\
		4.209 & 517.1 & 0.71 & 1.057 & 1.3 & 9.85 &     4.0 & 4.57 &    6.0 & 2.07\\
		4.219 & 514.6 & 0.72 & 1.056 & 3.4 & 3.07 &     5.6 & 2.50 &    7.4 & 1.46\\
		4.226 & 1056.4 & 0.72 & 1.056 & 5.7 & 1.22 &    7.4 & 1.16 &    8.7 & 2.77\\
		4.236 & 530.3 & 0.73 & 1.056 & 8.3 & 1.21 &     8.9 & 1.35 &    9.7 & 1.06\\
		4.244 & 538.1 & 0.74 & 1.056 & 10.0 & 2.44 &    10.6 & 2.65 &   11.0 & 3.18\\
		4.257 & 828.4 & 0.75 & 1.054 & 12.7 & 1.03 &    12.2 & 1.19 &   12.8 & 2.01\\
		4.267 & 531.1 & 0.75 & 1.053 & 14.0 & 0.33 &    13.4 & 0.38 &   13.4 & 0.21\\
		4.278 & 175.7 & 0.76 & 1.053 & 14.1 & 2.20 &    14.3 & 2.40 &   14.3 & 2.43\\
		4.288 & 502.4 & 0.76 & 1.053 & 15.7 & 0.51 &    15.2 & 0.59 &   15.7 & 0.02\\
		4.313 & 501.2 & 0.77 & 1.052 & 17.2 & 2.35 &    17.1 & 2.62 &   16.9 & 2.58\\
		4.338 & 505.0 & 0.78 & 1.051 & 18.4 & 0.92 &    18.0 & 1.04 &   18.2 & 0.97\\
		4.358 & 544.0 & 0.79 & 1.051 & 17.1 & 1.46 &    16.6 & 1.67 &   16.7 & 1.63\\
		4.378 & 522.7 & 0.79 & 1.051 & 18.8 & 1.35 &    18.9 & 1.50 &   19.1 & 1.45\\
		4.397 & 507.8 & 0.80 & 1.051 & 19.7 & 2.38 &    19.0 & 2.76 &   19.0 & 3.07\\
		4.416 & 1043.9 & 0.81 & 1.052 & 14.8 & 0.55 &   14.7 & 0.62 &   14.2 & 0.45\\
		4.437 & 569.9 & 0.80 & 1.054 & 18.6 & 0.74 &    18.6 & 0.84 &   18.8 & 1.26\\
		4.467 & 111.1 & 0.82 & 1.055 & 12.1 & 6.23 &    12.5 & 7.05 &   13.0 & 6.74\\
		4.527 & 112.1 & 0.83 & 1.054 & 12.5 & 2.79 &    13.5 & 3.00 &   13.5 & 2.92\\
		4.574 & 48.9 & 0.84 & 1.054 & 14.1 & 7.27 &     14.8 & 8.01 &   14.6 & 15.37\\
		4.600 & 586.9 & 0.85 & 1.055 & 15.4 & 0.45 &    15.5 & 0.50 &   15.1 & 0.18\\
		4.612 & 103.7 & 0.83 & 1.055 & 20.5 & 2.39 &    20.9 & 2.68 &   20.3 & 2.23\\
		4.628 & 521.5 & 0.83 & 1.054 & 20.5 & 0.79 &    20.6 & 0.88 &   20.3 & 0.90\\
		4.641 & 551.6 & 0.83 & 1.054 & 20.0 & 1.29 &    20.7 & 1.43 &   20.6 & 1.38\\
		4.661 & 529.4 & 0.84 & 1.054 & 20.3 & 0.61 &    20.8 & 0.68 &   20.7 & 0.55\\
		4.682 & 1667.4 & 0.84 & 1.054 & 20.4 & 0.91 &   21.0 & 1.01 &   20.5 & 0.99\\
		4.699 & 535.5 & 0.84 & 1.055 & 20.7 & 0.51 &    20.9 & 0.58 &   21.1 & 0.38\\
		4.740 & 163.9 & 0.84 & 1.055 & 20.8 & 2.71 &    21.4 & 3.03 &   20.7 & 3.26\\
		4.750 & 366.6 & 0.84 & 1.055 & 21.0 & 1.31 &    21.5 & 1.48 &   21.4 & 1.28\\
		4.781 & 511.5 & 0.85 & 1.055 & 20.7 & 0.59 &    21.2 & 0.65 &   20.9 & 0.58\\
		4.843 & 525.2 & 0.85 & 1.056 & 20.7 & 0.50 &    21.6 & 0.54 &   21.2 & 0.53\\
		4.918 & 207.8 & 0.86 & 1.056 & 21.0 & 0.88 &    21.3 & 1.00 &   21.3 & 0.69\\
		4.951 & 159.3 & 0.86 & 1.056 & 20.4 & 1.51 &    20.6 & 1.72 &   20.6 & 1.79\\	
		\hline 
		\hline
	\end{tabular}
	\label{count_UL_x79}  
\end{table}

\section{Systematic uncertainties for the cross section measurement}
\label{sys-xs}
\begin{table}
	\caption{Systematic uncertainties (in $\%$) for the $\sigma^{\rm U.L.}[\ee\to\pp\x] \cdot \mathcal{B}[\x\to\eta\jpsi]$ measurement. Here $\Delta_{1}$, $\Delta_{2}$ and $\Delta_{3}$ represent the systematic uncertainties from 4C kinematic fit, MC model and ISR, and $\gamma^{ISR}\psip$ cut, respectively. The sources marked with "$*$" are shared systematic uncertainties for different data sets.}
	\hspace{15pt}
	\centering
	\begin{tabular}{c|c|c|c|c|c|c|c|c|c|c} 
		\hline 
		\hline
		Data set & $\mathcal{L}_{\rm int}$ & $\gamma^{*}$ & Tracking$^{*}$ & $\mathcal{B}(\eta)^{*}$ & $\mathcal{B}(\jpsi)^{*}$ & $\Delta_{1}$ & $\Delta_{2}$ & $\Delta_{3}^{*}$ & MUC & total \\
		\hline
		4180 & 0.7 & 2.0 & 4.0 & 0.5 & 0.4 & 1.6 & 9.6 & 0.2 & 2.7 & 11.1\\
		4190 & 0.7 & 2.0 & 4.0 & 0.5 & 0.4 & 1.6 & 11.2 & 0.2 & 2.7 & 12.5\\
		4200 & 0.7 & 2.0 & 4.0 & 0.5 & 0.4 & 1.7 & 9.6 & 0.2 & 2.7 & 11.1\\
		4210 & 0.7 & 2.0 & 4.0 & 0.5 & 0.4 & 1.6 & 9.3 & 0.2 & 2.8 & 10.8\\
		4220 & 0.7 & 2.0 & 4.0 & 0.5 & 0.4 & 1.5 & 9.9 & 0.2 & 2.8 & 11.3\\
		4230 & 0.7 & 2.0 & 4.0 & 0.5 & 0.4 & 1.6 & 8.8 & 0.2 & 2.8 & 10.4\\
		4237 & 0.7 & 2.0 & 4.0 & 0.5 & 0.4 & 1.2 & 1.2 & 0.2 & 2.7 & 5.6\\
		4246 & 0.7 & 2.0 & 4.0 & 0.5 & 0.4 & 1.1 & 1.8 & 0.2 & 2.8 & 5.7\\
		4260 & 0.7 & 2.0 & 4.0 & 0.5 & 0.4 & 1.2 & 1.5 & 0.2 & 2.8 & 5.7\\
		4270 & 0.7 & 2.0 & 4.0 & 0.5 & 0.4 & 1.5 & 1.2 & 0.2 & 2.8 & 5.7\\
		4280 & 0.7 & 2.0 & 4.0 & 0.5 & 0.4 & 2.0 & 2.6 & 0.2 & 2.8 & 6.3\\
		4290 & 0.7 & 2.0 & 4.0 & 0.5 & 0.4 & 1.4 & 5.3 & 0.2 & 1.4 & 7.3\\
		4315 & 0.7 & 2.0 & 4.0 & 0.5 & 0.4 & 1.5 & 4.0 & 0.2 & 1.4 & 6.4\\
		4340 & 0.7 & 2.0 & 4.0 & 0.5 & 0.4 & 1.3 & 4.8 & 0.2 & 1.4 & 6.9\\
		4360 & 0.7 & 2.0 & 4.0 & 0.5 & 0.4 & 1.6 & 1.4 & 0.2 & 2.8 & 5.8\\
		4380 & 0.7 & 2.0 & 4.0 & 0.5 & 0.4 & 1.6 & 2.8 & 0.2 & 1.4 & 5.7\\
		4400 & 0.7 & 2.0 & 4.0 & 0.5 & 0.4 & 1.4 & 2.1 & 0.2 & 1.4 & 5.4\\
		4420 & 0.7 & 2.0 & 4.0 & 0.5 & 0.4 & 1.2 & 1.0 & 0.2 & 2.8 & 5.6\\
		4440 & 0.7 & 2.0 & 4.0 & 0.5 & 0.4 & 1.5 & 3.1 & 0.2 & 1.4 & 5.9\\
		4470 & 0.7 & 2.0 & 4.0 & 0.5 & 0.4 & 1.5 & 9.7 & 0.2 & 2.8 & 11.2\\
		4530 & 0.7 & 2.0 & 4.0 & 0.5 & 0.4 & 1.5 & 3.5 & 0.2 & 2.8 & 6.5\\
		4575 & 0.7 & 2.0 & 4.0 & 0.5 & 0.4 & 1.3 & 6.0 & 0.2 & 2.8 & 8.2\\
		4600 & 0.7 & 2.0 & 4.0 & 0.5 & 0.4 & 1.5 & 4.6 & 0.2 & 2.8 & 7.2\\
		4610 & 0.5 & 2.0 & 4.0 & 0.5 & 0.4 & 1.4 & 5.8 & 0.2 & 1.4 & 7.7\\
		4620 & 0.5 & 2.0 & 4.0 & 0.5 & 0.4 & 1.5 & 5.5 & 0.2 & 1.4 & 7.4\\
		4640 & 0.5 & 2.0 & 4.0 & 0.5 & 0.4 & 1.4 & 9.2 & 0.2 & 1.4 & 10.4\\
		4660 & 0.5 & 2.0 & 4.0 & 0.5 & 0.4 & 1.5 & 10.5 & 0.2 & 1.4 & 11.6\\
		4680 & 0.5 & 2.0 & 4.0 & 0.5 & 0.4 & 1.4 & 7.8 & 0.2 & 1.4 & 9.2\\
		4700 & 0.5 & 2.0 & 4.0 & 0.5 & 0.4 & 1.5 & 6.8 & 0.2 & 1.4 & 8.4\\
		4740 & 0.5 & 2.0 & 4.0 & 0.5 & 0.4 & 1.4 & 9.5 & 0.2 & 1.6 & 10.7\\
		4750 & 0.5 & 2.0 & 4.0 & 0.5 & 0.4 & 1.4 & 8.0 & 0.2 & 1.6 & 9.4\\
		4780 & 0.5 & 2.0 & 4.0 & 0.5 & 0.4 & 1.4 & 9.9 & 0.2 & 1.6 & 11.1\\
		4840 & 0.5 & 2.0 & 4.0 & 0.5 & 0.4 & 1.8 & 9.0 & 0.2 & 1.6 & 10.3\\
		4914 & 0.5 & 2.0 & 4.0 & 0.5 & 0.4 & 1.4 & 9.9 & 0.2 & 1.6 & 11.1\\
		4946 & 0.5 & 2.0 & 4.0 & 0.5 & 0.4 & 1.6 & 9.1 & 0.2 & 1.6 & 10.4\\
		\hline 
		\hline
	\end{tabular}
	\label{systematical error for xs}  

\end{table}

\end{document}